\documentclass[nofootinbib,showpacs,preprintnumbers,amsmath,amssymb, prd]{revtex4}
\pdfoutput=1

\usepackage{color}
\usepackage{amsmath,amssymb,graphicx}
\usepackage{epsf}
\usepackage{epstopdf}
\usepackage {amssymb}

\usepackage{epstopdf}

\def\bk{{\bf k}}
\def\bp{{\bf p}}

\def\CL{{\cal L}}

\def\half{\frac{1}{2}}

\usepackage {amssymb}
\newcommand{\nc}{\newcommand}
\nc{\ba}{\begin{eqnarray}}
\nc{\ea}{\end{eqnarray}}

\newcommand{\bea}{\begin{eqnarray}}
\newcommand{\eea}{\end{eqnarray}}
\newcommand{\barr}{\begin{array}}
\newcommand{\earr}{\end{array}}

\begin{document}
\title{Spectroscopy of Masses and Couplings during Inflation}

\author{Razieh Emami$^{~1,2}$}
\email{emami-AT-ipm.ir}
\affiliation{$^1$ School of Physics, Institute for Research in
Fundamental Sciences (IPM), P. O. Box 19395-5531, Tehran, Iran  \\
$^2$ Abdus Salam International Centre for Theoretical Physics\\ Strada Costiera 11, 34151, Trieste, Italy}

\begin{abstract}
In this work, we extend the idea of Quasi Single Field inflation \cite{Chen:2009zp} to the case of multiple isocurvaton fields with masses of order of  Hubble, which are coupled kinetically to the inflaton field and have some interactions among themselves. We consider the effects of these massive modes in both the size and the shape of the bispectrum. We show that the shape of the bispectrum in the squeezed limit is dominated by the lightest field and is the same as in Quasi Single Field inflation. This is a generic feature of multiple isocurvaton fields and is independent of the details of the interactions among the massive fields. When the isocurvaton fields have similar masses, we can potentially distinguish two different shapes in the squeezed limit so that the shape of the bispectrum can act as a particle detector. However, in the presence of hierarchy among the massive fields, the dominant effect is due to the lightest field.

 \end{abstract}

\maketitle


\section{Introduction}
\label{intro}

Inflation is the leading paradigm for early universe and structure formation which is well consistent with recent observations from Planck satellite
\cite{Ade:2013lta, Ade:2013uln}. In simple models of inflation a scalar field rolls down towards the minimum of its potential supporting a long enough period of inflation. In order to solve the flatness and the horizon problem one requires 60 or so number of e-foldings.
The near dS background of the inflationary mechanism usually guarantees that the primordial anisotropies imprinted on  cosmic microwave background (CMB) map are nearly scale-invariant, nearly adiabatic and  nearly Gaussian. These basics predictions of inflation are well
consistent with the cosmological observations.

There have been much interests on non-Gaussianities during past few years, for a review see
\cite{Chen:2010xka, Komatsu:2010hc}. There was no detection of non-Gaussianity by
Planck \cite{Ade:2013ydc} and there are strong upper bound on the amplitude of local-type non-Gaussianities, $f_{NL}^{loc}=2.7 \pm 5.8$ (68 \% CL)   \cite{Ade:2013ydc}. Having this said, the constraints on other shapes of non-Gaussianities,  such as equilateral-type shape, are not as tight and there is still room for large observable non-Gaussianities in these shapes. On the other hand, there are also some constraints coming from the Large Scale Structure (LSS) analysis such as galaxy power spectrum as well as the scale dependence corrections to galaxy bias. As it is shown in \cite{Sefusatti:2012ye} the galaxy halo bias  and CMB analysis are complementray to each other. In the sense that, the halo bias is relatively more sensitive to the isocurvaton mass, since the scale dependent corrections depend on the behavior of curvature bispectrum in the squeezed limit. The sensitivity of CMB on the other hand, is more on the values of $f_{NL}$. So combining both, depending on the amplitude of $f_{NL}$, in principle we can distinguish the Quasi Single Field inflation (QSF) model, \cite{Chen:2009zp}, as well is its extended version, in a significant fraction of mass scales.

While the simplest inflationary models are based on single field, one can easily conceive situations in which more then one fields are responsible for inflation dynamics and generating curvature perturbations. There have been much works on models of multiple field inflation, for a review see \cite{Bassett:2005xm, Wands:2007bd}.  In multiple fields scenarios,  there is one adiabatic direction responsible for  curvature perturbations and several isocurvature directions which can generate entropy perturbations \cite{Gordon:2000hv}.
In these models, it is usually assumed that all fields are light,  i.e. lighter than
$H$, the Hubble expansion rate during inflation.   This is because only the light fields with mass $\lesssim H$ are expected to play important roles. Very heavy fields with mass $m \gg H$ are expected to rapidly rolls down towards their minimum and play no important roles during inflation.  Having this said, it is possible that one encounters situations in which two or more light or semi-heavy fields with some hierarchies of masses  are present during inflation. Therefore, it is an important question as how the hierarchy of masses and couplings show themselves in inflationary predictions such as the power spectrum and the bispectrum.

The effects of an isocurvaton field, a semi-heavy field  with the mass  $m \lesssim 3H/2$,  on inflation bispectrum and trispectrum were studied by Chen and Wang in \cite{Chen:2009zp, Chen:2009we}.  Phenomenologically, this model is very interesting. Because by adjusting the magnitude of the ratio $m/H$, the shape of the bispectrum in
this model  effectively interpolates  between the shape of the bispectrum in
single field and the shape of bispectrum in
multi field inflationary models.  On the other hand, this model is well-motivated  from the theoretical point of view too. Indeed, many models of high energy physics such as string theory and supersymmetry contain many scalar degrees of freedom,  for a review see \cite{HenryTye:2006uv, Cline:2006hu, Burgess:2007pz, McAllister:2007bg, Baumann:2009ni, Baumann:2011nk, Ellis:2013zsa, Halter:2012ola}. In embedding inflation in these models of high energy physics one can naturally encounter several fields with different masses. Also in this setup it is quite natural to have massive fields, with masses at the order of Hubble parameter $H$ or so.


The authors in  \cite{Chen:2009zp} considered the situation in which the isocurvaton field does not affect the power spectrum but can have significant contributions on bispectrum via turning trajectory. To see how this can happen suppose we denote the inflaton field by $\theta$
which is an angle in a two-dimensional field space trajectory while $\sigma$ denotes the
additional semi-heavy field  with the self-coupling potential $V(\sigma)$. At the background level, we can stabilize the semi-heavy field on its own minimum without any effect on the background trajectory. In \cite{Chen:2009zp} it is shown that the amplitude of $f_{NL}$ scales like $ \left(\frac{\dot{\theta}}{H}\right)^3 \left( \frac{V_{, \sigma \sigma \sigma} }{H}\right){P_{\zeta}}^{-\frac{1}{2}}$
in which $ P_{\zeta} \sim 10^{-9}$ is the observed curvature perturbation power spectrum. With large enough self-coupling  $V_{, \sigma \sigma \sigma}$ one can easily saturate the current observational bound on the amplitude of non-Gaussianity.
In addition, it is shown in \cite{Chen:2009zp} that depending on the $\sigma$ field's mass, a new shape of non-Gaussianity is generated which continuously interpolate between the local shape and the equilateral shape. For other works with similar ideas  see
\cite{ Elliston:2013zja, Gao:2013ota, Gong:2013sma, Chen:2012ge, Pi:2012gf,
Assassi:2013gxa, Battefeld:2013xka, Green:2013rd, Noumi:2012vr, Burgess:2012dz, Elliston:2012ab}.

In this work we extend the analysis of \cite{Chen:2009zp} to the case in which two semi-heavy fields are present in the model, denoted by $\sigma_1$ and $\sigma_2$. We show that while the correction to the power spectrum is quite negligible, these semi-heavy fields can have significant contributions in the bispectrum.
It is interesting to study the effects of these massive fields in both the size and the shape of the bispectrum in the squeezed limit. It turns out that, while these fields have non trivial effects in the size of the bispectrum, they will not produce any new shapes in the squeezed limit. This means that in the squeezed limit, the leading shape is the same as in the quasi single field inflation, which scales as $ k_{3}^{-(\frac{3}{2}+ \nu_{i})} $ with $ (k_{3}\ll k_{1} \simeq k_{2})$ and where we have defined $\nu_{i}$ as, $\nu_{i}\equiv \sqrt{\frac{9}{4} - \frac{\widehat{m}_{i}^2}{H^2}}, ( i=1,2)$. This is a generic feature of multiple isocurvaton fields and is independent of the details of the interactions among the massive modes.

The rest of this paper is organized as follows. In Section \ref{setup} we present our model containing the motivation as well as the phenomenological Lagrangian. We study  the perturbation at the quadratic level,  considering the free field action as well as the exchange vertex interactions which are necessary in order to convert the contribution from the isocurvaton fields into the curvature perturbation. In addition we investigate the interaction among the isocurvaton fields at the cubic order in perturbations.
In Section \ref{power} we present our power spectrum analysis. We show that the correction into the power-spectrum, which is controlled by the ratio $\frac{\dot{\theta}_{0}}{H}$, is  small and can be neglected. In Section \ref{bispectrum} we present the bispectrum analysis employing the in-in formalism. As in \cite{Chen:2009zp} we use two different methods in order to calculate the in-in integrals. In Section  \ref{squeezed} we consider the squeezed limit of the bispectrum followed by
our conclusions and discussions in Section \ref{summary}. We relegate many technical analysis  of the in-in integrals  into Appendices.
\section{The Setup}
\label{setup}

Here we present our setup. This is an extension of the work by Chen and Wang  \cite{Chen:2009zp} known as Quasi Single Field Inflation (QSF). Our model contains two iso-curvaton fields (semi-heavy fields) $\sigma_1$ and $\sigma_2$. Following the logic in \cite{Chen:2009zp} we want to study the effects of these iso-curvaton fields on power spectrum and bispectrum.  First we start with the motivation.

In original QSF idea, it is imagined that the light field $\theta$ is moving along a constant radius circular trajectory with the fixed radius $R$ determined by the vacuum expectation
value of the heavy field $\sigma$, $R= \sigma_0$. Here we extend this view to higher dimensional field space geometry.  Consider the model containing a light field moving on the surface of a two-dimensional sphere. To be specific, let us denote the light field by the azimuthal direction $\phi$, while the heavy fields are denoted by the radius $r$ and the other angular direction $\theta$.  In this picture the heavy fields $(r, \theta)$ are stabilized around the background value $(r_0, \theta_0)$, while the light field $\phi$ moves nearly freely on the surface of the sphere. The Lagrangian for the kinetic energy is
\ba
\label{L-kin}
L_{\mathrm{kin}} = \frac{1}{2} \dot r^2 + \frac{r^2}{2}   \dot \theta^2 +
\frac{r^2}{2}  \sin^2 \theta \,  \dot \phi^2 \, .
\ea
Now consider the fluctuations $r(t) = r_0 + \delta r(t), \theta(t) = \theta_0 + \delta \theta(t)$ and
$\phi(t) = \phi_0(t) + \delta \phi(t)$.
Plugging these back into the kinetic Lagrangian yields the quadratic Lagrangian
\ba
\label{L-kin2}
 \Delta L_{\mathrm{kin}}^{(2)} =  \frac{1}{2} \delta \dot r^2 + \frac{r_0^2}{2}   \delta \dot \theta^2
+  \frac{r_0^2}{2}  \sin^2 \theta_0 \delta \dot \phi^2 \, ,
\ea
plus the exchange vertex interactions
\ba
\Delta  L_3 =   \dot \phi_0 \left(  r_0 \sin^2 \theta_0 \, \delta r + r_0^2 \sin 2 \theta_0  \delta \theta  \right) \delta \dot \phi \, .
\ea
This is a multiple field  extension of \cite{Chen:2009zp} with the exchange vertex interactions
$\delta r \delta \dot \phi$ and $\delta \theta \delta \dot \phi$.  Note that from Eq. (\ref{L-kin}) we also obtain other sub-leading interactions, e.g. the exchange vertex between $r$ and $\theta$, which we discard.

Having presented our motivation for the multiple massive fields kinetically coupled to the inflaton field, we proceed with our phenomenological model now. From now on we follow
the notation of  \cite{Chen:2009zp} with $\theta$ being the inflaton field while the heavy fields are denoted by $\sigma_1$ and $\sigma_2$.  The action is
\begin{align}
\CL =& -\half (R^2 + c_1 \sigma_1 + c_2 \sigma_2) g^{\mu\nu} \partial_\mu \theta \partial_\nu \theta
- \half g^{\mu\nu} \partial_\mu \sigma_1 \partial_\nu \sigma_1
- \half g^{\mu\nu} \partial_\mu \sigma_2 \partial_\nu \sigma_2
\nonumber \\
& - V_{\rm sr}(\theta) - V(\sigma_1,\sigma_2)  \, ,
\label{Lagrangian}
\end{align}
in which $V_{\rm sr}(\theta)$ is the slow-roll potential and  $V(\sigma_1,\sigma_2)$ represents the potential between the fields $\sigma_1$ and $\sigma_2$. Also $c_{i}, i=1,2$ are both constants of the order R. \\
At the background level, the Hubble equation and the continuity equation are

\begin{eqnarray}
 3 M_{P}^2 H^2 &=& \frac{1}{2}\widetilde{R}\dot{\theta}_{0}^2 + V_{sr}(\theta) ~,~ \\
-2 M_{P}^2 \dot{H} &=& \widetilde{R}\dot{\theta}_{0}^2  \, ,
\label{Hub-cont}
\end{eqnarray}
where we have absorbed the background fields' values $\sigma_{10}$ and $\sigma_{20}$ into the net turning radius by defining
\ba
\widetilde{R} \equiv( R^2 + c_{1}\sigma_{10} + c_{2}\sigma_{20}).
\ea
Also the equations of motion for $\sigma_{i0}(t) , (i= 0,1) $ and $\theta_{0}(t)$ are
\bea
\sigma_{i0}= const. ~~~~,~~~~ V_{, \sigma_{i}} = \frac{1}{2} c_{i} \dot{\theta _{0}^2} ~~,~~ i= 0,1 ~,~
\label{sigma}
\eea
\bea
\widetilde{R} \, \ddot{\theta_{0}} + 3 \widetilde{R} H \, \dot{\theta_{0}} +\frac{\partial}{\partial \theta} V_{sr } =0.
\label{sigma}
\eea
In this picture $\sigma_{i}$'s have been stabilized around their background values while  the inflaton field $\theta$ is slowly rolling along the potential $ V_{sr }$.

Now we can specify the form of the $V(\sigma_{1} , \sigma_{1})$ up to the third orders in fields perturbations which will be used to calculate the bispectrum
\bea
\delta V(\sigma_1, \sigma_2) =
\half m_1^2 \delta\sigma_1^2 + \half m_2^2 \delta\sigma_2^2 + \half m_{12}^2 \delta\sigma_1 \delta\sigma_2
+ \frac{1}{6} \lambda_1 \delta\sigma_1^3 + \frac{1}{6} \lambda_2 \delta\sigma_2^3
+ \half \lambda_3 \delta\sigma_1 \delta\sigma_2^2
+ \half \lambda_4 \delta\sigma_2 \delta\sigma_1^2 ~.
\label{interactions}
\eea
Furthermore, we chose the mass basis such that the mass matrix is diagonal so we can neglect the term $m_{12}$. In what follows we assume $0\leq  m_i/H \leq 3/2$ so the iso-curvaton fields can be light or semi-heavy. This may have a natural interpretation for the
super-symmetric completion of the theory.

\begin{figure}
\includegraphics[ scale=1]{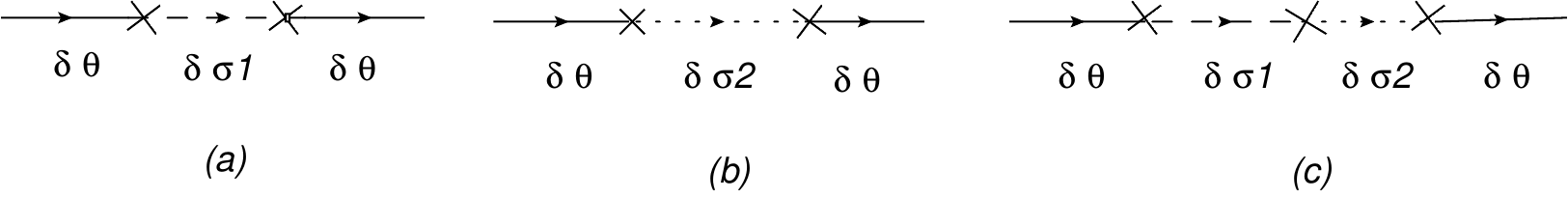}
\medskip
\caption{Feynman diagrams for the correction in the power spectrum. (a) is the correction due to $\sigma_{1}$ , (b) is the correction of $\sigma_{2}$ and (c) describes the sub-leading correction due to the exchange vertex between $\sigma_{1}$ and $\sigma_{2}$. }
\label{Fig:Fdiagram1}
\end{figure}

We choose the \textit{spatially flat gauge} in which the scale factor of the metric is homogeneous,
\bea
h_{ij} = a^2(t)\delta_{ij} \, .
\eea
We expand only the matter part of the lagrangian, given by Eq. (\ref{Lagrangian}), while ignoring the perturbations in the gravitational sector. This is justified, since from Maldacena's
analysis   \cite{Maldacena:2002vr} it is expected that the gravitational back-reactions in bispectrum are slow-roll suppressed. This was specifically demonstrated in \cite{Chen:2009zp} and this is expected to be the case in our setup. For the future references, the slow-roll parameters are
\begin{align}
\epsilon &\equiv  \frac{\widetilde{R}\dot{\theta_{0}}^2}{2 H^2 M_{P}^2} ~,~ \nonumber \\
\eta & \equiv \frac{\dot{\epsilon}}{H \epsilon}~.~
\label{epsilon-eta}
\end{align}
In order for $\epsilon$ to be small, we require
\bea
\left(\frac{\dot{\theta_{0}}}{H}\right)^2 \ll 1 \, .
\eea
Calculating the quadratic and the cubic Lagrangians, we have
\ba
L_{2} &=& \frac{1}{2} a(t)^3 \widetilde{R} \delta \dot{\theta}^2 - \frac{1}{2} a(t) \widetilde{R} (\partial_{i} \delta \theta )^2 + \frac{1}{2}a(t)^3 \delta \dot{\sigma _{1}}^2 - \frac{1}{2} a(t)(\partial_{i} \delta \sigma _{1})^2 + \nonumber\\
 &+&\frac{1}{2}a(t)^3 \delta \dot{\sigma _{2}}^2 - \frac{1}{2} a(t)(\partial_{i} \delta \sigma _{2})^2 - \frac{1}{2} a(t)^3 m_1^2 \delta\sigma_1^2 - \frac{1}{2} a(t)^3 m_2^2 \delta\sigma_2^2
\label{free-lagrangian}
\nonumber\\
\delta L_{2} &=& c_{1}a(t)^3 \dot{\theta_{0}} \delta \dot{\theta} \delta \sigma_{1} + c_{2}a(t)^3 \dot{\theta_{0}} \delta \dot{\theta} \delta \sigma_{2}
\label{2-interaction}
\nonumber\\
\delta L_{3} &=& - \frac{1}{6} a(t)^3 \lambda_1 \delta\sigma_1^3 - \frac{1}{6} a(t)^3 \lambda_2 \delta\sigma_2^3
- \half a(t)^3 \lambda_3 \delta\sigma_1 \delta\sigma_2^2
- \half a(t)^3 \lambda_4 \delta\sigma_2 \delta\sigma_1^2 ~.
\label{3-interaction}
\ea
in which $L_2$ is the quadratic action describing the free field lagrangian. As was mentioned  before  $\theta$ is nearly massless while the isocurvaton fields, $\sigma_{i}, i=1,2$,  can have  masses at the order of Hubble parameter. In addition, $\delta L_{2}$ is the coupling between the fields which describes the exchange vertex between them, as shown in Fig. \ref{Fig:Fdiagram1}. Finally, $\delta L_{3}$ represents the self-interactions between $\sigma_{i}$'s.

Now, we calculate the Hamiltonian of this system. After some simple calculations the Hamiltonian densities are calculated to be
\ba
H_{0} &=& \half a(t)^3 \widetilde{R}\delta \dot{\theta}^2 + \half a(t) \widetilde{R} (\partial_{i} \delta \theta )^2 + \half a(t)^3 \delta \dot{\sigma _{1}}^2 + \half a(t) (\partial_{i} \delta \sigma _{1})^2 + \half a(t)^3 \delta \dot{\sigma _{2}}^2 +
\nonumber \\
& +&\half a(t) (\partial_{i} \delta \sigma _{2})^2 + \half a(t)^3 \widehat{m}_{1}^2 \delta \sigma_{1}^2 + \half a(t)^3 \widehat{m}_{2}^2\delta \sigma_{2}^2 ~~,~~
\label{free-Hamiltonian}
 \\
H_{2} &=& - \dot{\theta_{0}}a(t)^3 \delta \dot{\theta} (c_{1} \delta \sigma_{1} + c_{2} \delta \sigma_{2}) +
\frac{a(t)^3 \dot{\theta_{0}}^2}{\widetilde{R}} c_{1} c_{2} \delta \sigma_{1} \delta \sigma_{2} ~,~
\label{quad-Hamiltonian}
\\
H_{3} &=& + \frac{1}{6} a(t)^3 \lambda_1 \delta\sigma_1^3 + \frac{1}{6} a(t)^3 \lambda_2 \delta\sigma_2^3
+ \half a(t)^3 \lambda_3 \delta\sigma_1 \delta\sigma_2^2
+ \half a(t)^3 \lambda_4 \delta\sigma_2 \delta\sigma_1^2 ~.
\label{cubic-Hamiltonian}
\ea
where
\bea
\widehat{m}_{1}^2 \equiv \left(m_{1}^2 + \frac{\dot{\theta_{0}}^2}{\widetilde{R}}c_{1}^2 \right)  \quad , \quad
\widehat{m}_{2}^2 \equiv \left(m_{2}^2 + \frac{\dot{\theta_{0}}^2}{\widetilde{R}}c_{2}^2 \right)  \, .
\eea
We  quantize the Fourier components of the free fields in the interaction picture denoted by
$\delta \theta^{I} _{k}$ , $\delta \sigma ^{I} _{1k}$ and $\delta \sigma ^{I} _{2k}$ such that \begin{align}
\delta \theta^{I} _{\bk} & = u_{k} a_{\bk} + u^\star _{-\bk} a^\dag_{-\bk} ~~, \\
\delta \sigma ^{I} _{1k} & = v_{k} b_{\bk} + v^\star_{-\bk} b^\dag_{-\bk} ~~, \\
\delta \sigma ^{I} _{2k} & = w_{k} c_{\bk} + w^\star_{-k} c^\dag_{-\bk} ~~.
\end{align}
Here $a_{\bk} (a^\dagger_{\bk})$ , $b_{\bk}(b^\dagger_{\bk})$ and $c_{\bk}(c^{\dagger}_{\bk})$ are the usual annihilation (creation) operators defined for
$\delta \theta^{I} _{k}$ , $\delta \sigma ^{I} _{1k}$ and $\delta \sigma ^{I} _{2k}$ respectively. Since these fields are treated as independent random fields we assume
that the annihilation and creation operators for different fields are independent of each other and they commute with each other. Furthermore, they satisfy the usual commutation relations
\begin{align}
[a_{\bk},a^\dag_{-\bk'}] = (2\pi)^3 \delta ^{3}(\bk+\bk')  \quad , \quad
[b_{\bk},b^\dag_{-\bk'}] = (2\pi)^3 \delta ^{3}(\bk+\bk') \quad , \quad
[c_{\bk},c^\dag_{-\bk'}] = (2\pi)^3 \delta ^{3}(\bk+\bk').
\end{align}
The mode functions $u_{k}$ , $v_{k}$ and $w_{k}$ satisfy the following linear equations of motion obtained from the free Hamiltonian $H_0$ given in Eq. (\ref{free-Hamiltonian})
\begin{align}
u_{k}'' - \frac{2}{\tau}u_{k}' + k^2 u_{k} =0 ~~,\\
v_{k}'' - \frac{2}{\tau}v_{k}' + k^2 v_{k} + \frac{\widehat{m}_{1}^2}{H^2 \tau^2}v_{k} =0 ~~,\\
w_{k}'' - \frac{2}{\tau}w_{k}' + k^2 w_{k} + \frac{\widehat{m}_{2}^2}{H^2 \tau^2}w_{k} =0~~.
\end{align}
Here $\tau$ is the conformal time, $dt\equiv a(t) d\tau $, and the prime denotes the derivative with respect to $\tau$.

The solution for the inflaton mode function, $u_k$,  imposing the Minkowski initial condition for modes deep inside the horizon  is
\begin{align}
u_{k} = \frac{H}{\sqrt{2\widetilde{R}k^3}} (1 + ik \tau) e^{-ik \tau }
\label{u-mode} \, .
\end{align}
Now, as it has been mentioned in  \cite{Chen:2009zp},  there are three different regions in the mass parameter space of $\sigma_{i}$'s, in which $v_k$ and $w_k$  can be either over-damped corresponding to $  (\frac{\widehat{m}_{i}^2}{H^2}) < \frac{9}{4}$, critical with $ (\frac{\widehat{m}_{i}^2}{H^2})= \frac{9}{4} $, or under-damped corresponding to
$ (\frac{\widehat{m}_{i}^2}{H^2})> \frac{9}{4} $. However, due to Boltzmann suppression , as in \cite{Chen:2009zp}, we are only interested in the first two regions,  the over-damped and the critical cases in which the wave functions are given by
\begin{align}
v_{k} &= -i e^{i(\nu_{1} + \half)\frac {\pi}{2}} \frac{\sqrt{\pi}}{2} H (-\tau)^{3/2}H^{(1)}_{\nu_{1}}(-k \tau)  \\
w_{k} &= -i e^{i(\nu_{2} + \half)\frac {\pi}{2}} \frac{\sqrt{\pi}}{2} H (-\tau)^{3/2}H^{(1)}_{\nu_{2}}(-k \tau)
\end{align}
where
\ba
\nu_{i}\equiv \sqrt{\frac{9}{4} - \frac{\widehat{m}_{i}^2}{H^2}} \, .
\ea
As explained before,  the normalization of all of the above fields have been chosen such that deep inside the horizon we recover the Bunch-Davies vacuum
\begin{align}
\sqrt{\widetilde{R}u_{k}} ~~,~~ v_{k} ~~,~~ w_{k} \rightarrow \frac{iH}{\sqrt{2k}}\tau e^{-ik\tau} \, .
\end{align}

For the future reference,  the behavior of the mode functions after the horizon exit,
$k \tau \rightarrow 0$, is also useful
\bea
(v_{k}, w_{k})=
\left\{
\begin{array}{ccc}
-\left(\frac{2^{\nu_{i}-1}}{\sqrt{\pi}}\right) \frac{H}{k^{\nu_{i}}} \Gamma (\nu_{i})\left(-\tau \right)^{-(\nu_{i}- \frac{3}{2})} e^{i(\nu_{i} + \half)\frac {\pi}{2}}\quad , ~~0<\nu_{i} \leq \frac{3}{2}   \\ \\
                                                                               \left(\frac{1}{\sqrt{\pi}}\right)H (-\tau)^{3/2} \ln(-k\tau)e^{i\frac {\pi}{4}} \quad  \qquad , ~~\nu_{i}=0
\end{array}
\right.
\label{alpha_m}
\eea

\section{Power Spectrum}
\label{power}

As we mentioned before, the turning trajectory, especially the exchange vertices between $\sigma_{i}$'s and $\theta$, lead to  corrections in the power spectrum. The terms which are responsible for this corrections are given in Eq. (\ref{quad-Hamiltonian}). The contributions from the first two terms  are very similar to \cite{Chen:2009zp} (with a summation over two isocurvaton fields), as shown in Fig.~\ref{Fig:Fdiagram1}, plots (a) and (b). In addition,  there is another term which describes the exchange vertex between $\delta \sigma_{1}$ and $\delta \sigma_{2}$ as has been shown in plot (c) of  Fig.~\ref{Fig:Fdiagram1}. However, since this last correction is proportional to $ \left(\frac{c_{1}^2 c_{2}^2}{\widetilde{R}^3}\right) \left(\frac{\dot{\theta_{0}}^2}{H}\right)^2$, the ratio between this term and the first two corrections  is proportional to $\left(\frac{c_{1} c_{2}}{\widetilde{R}}\right) \frac{\dot{\theta_{0}}^2}{H^2}$. So for  $c_{1} \sim c_{2}\sim \widetilde{R}$ this correction is very smaller than the first two terms and can be neglected in the following. Therefore,  we are left with the first two terms in Eq. (\ref{quad-Hamiltonian}),  corresponding to plots (a) and (b) in  Fig.~\ref{Fig:Fdiagram1}.

The analysis is very similar to the results of  \cite{Chen:2009zp}
\bea
\langle\zeta_{\bp_{1}}\zeta_{\bp_{2}}\rangle = (2\pi)^{5} \delta^{3}(\bp_{1} +\bp_{2})\frac{1}{2p^{3}}P_{\zeta}
\label{Power spectrum1}
\eea
where $\zeta$ is the curvature perturbation on flat slice and
the power spectrum is
\bea
P_{\zeta} = \frac{H^4}{4\pi^2\widetilde{R} \dot{\theta_{0}}^2}\left( 1+ 2 D_{\nu_{1}} \frac{ c_{1}^2 \dot{\theta_{0}}^2}{\widetilde{R} H^2} + 2 D_{\nu_{2}} \frac{c_{2}^2 \dot{\theta_{0}}^2}{\widetilde{R} H^2} \right)
\label{Power spectrum2}
\eea
 As  in \cite{Chen:2009zp},  $D_{\nu_{i}}$ is defined as
\ba
D_{\nu_{i}}&\equiv& \frac{\pi}{4} Re \left[ \int_{0}^{\infty}dx_{1}\int_{x_{1}}^{\infty}dx_{2} ( x_{1}^{-\frac{1}{2}}H_{\nu_{i}}^{(1)}(x_{1}) e^{ix_{1}}x_{2}^{-\frac{1}{2}}H_{\nu_{i}}^{(2)}(x_{2}) e^{-ix_{2}} \right. \nonumber \\
&&\left. - x_{1}^{-\frac{1}{2}}H_{\nu_{i}}^{(1)}(x_{1}) e^{-ix_{1}}x_{2}^{-\frac{1}{2}}H_{\nu_{i}}^{(2)}(x_{2}) e^{-ix_{2}})  \, \right] \, .
\label{neumerical-power spectrum}
\ea

\begin{figure}
\includegraphics[ scale=0.4]{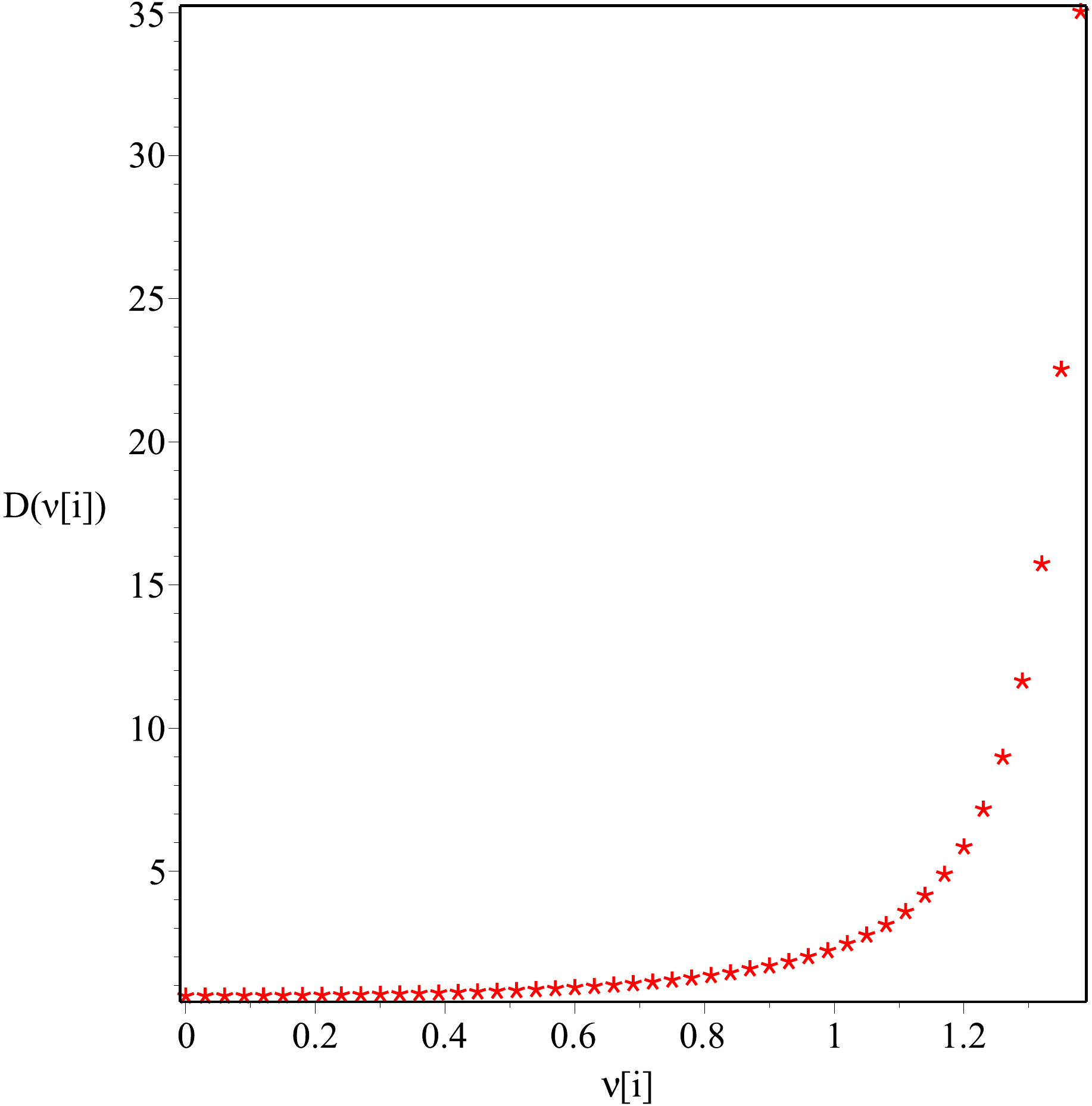}
\medskip
\caption{The behavior of $D(v_{i})$ in terms of $\nu_{i}$, as given in Eq. (\ref{neumerical-power spectrum}).  }
\label{Fig:correctionPower}
\end{figure}

In Fig. \ref{Fig:correctionPower} we have plotted  $D_{\nu_{i}}$ as a function of $\nu_{i}$. As in  \cite{Chen:2009zp}, we see that there is no divergence in the plot. As we see from Eq. (\ref{Power spectrum2}), since the correction to the power spectrum is proportional to the ratio $\left(\frac{\dot{\theta_{0}}}{H}\right)^2 \ll 1$, the correction to the power spectrum in this model is quite negligible.

The spectral index is
\bea
n_{s}-1 \equiv \frac{d \ln {P_{\zeta}}}{d \ln {k}} = -2 \epsilon - \eta + 2 \eta \left( D_{\nu_{1}} \frac{c_{1}^2}{\widetilde{R}}+ D_{\nu_{2}} \frac{c_{2}^2}{\widetilde{R}} \right)\left(\frac{\dot{\theta_{0}}^2}{H^2}\right)
\label{spectral index}
\eea
In which the last two terms refer to the roles of the heavy fields, turning trajectory. As in the power spectrum case, since the ratio $\left(\frac{\dot \theta_{0}}{H}\right)^2$ is very small, the correction due to the heavy fields in the spectral tilt are quite negligible.

\begin{figure}
\includegraphics[ scale=0.2]{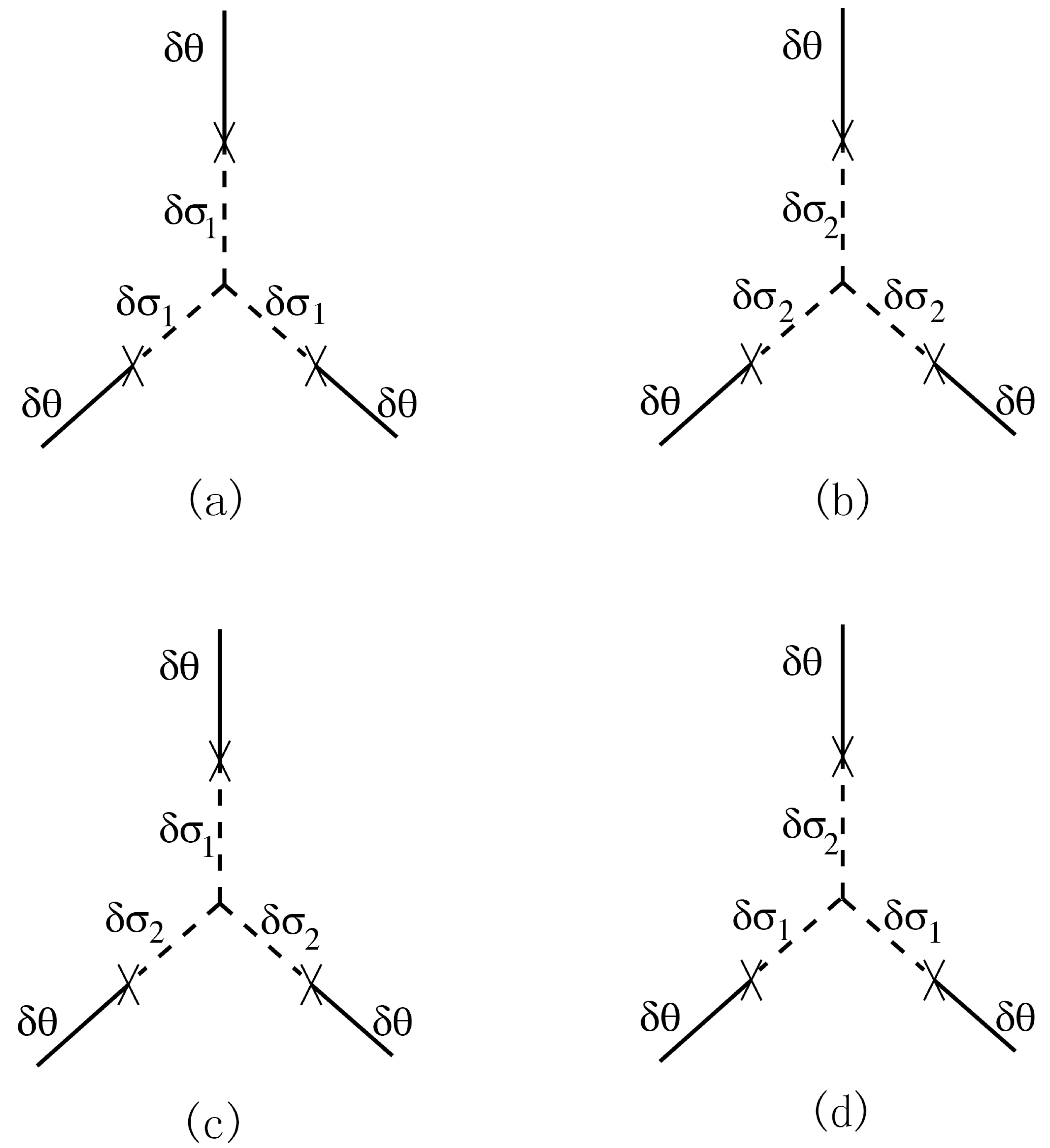}
\medskip
\caption{Leading corrections in the bispectrum. (a) is due to $ H^{3I}_{1}$, (b) is from $ H^{3I}_{2}$, (c) comes from $ H^{3I}_{3}$  and (d) is from $ H^{3I}_{4}$.}
\label{Fig:correctionBispectrum}
\end{figure}

\section{Bispectrum}
\label{bispectrum}

In this section we calculate the leading terms in the bispectrum. As we mentioned before, we
neglect the gravitational back-reactions  which are sub-leading and we only consider the corrections due to the interactions between $\sigma_{i}$ as given by Eq. (\ref{cubic-Hamiltonian}). The corresponding Feynman diagrams  are  shown in Fig. \ref{Fig:correctionBispectrum}.

The cubic Hamiltonian density in Fourier space is
\begin{align}
H^{I}_{3} &= \int d^3x \mathcal{H}^{I}_{3} = a^{3}(t) \int d^3p d^3q ~  \{
 \frac{\lambda_{1}}{6}\delta \sigma^{I}_{1p}(t) \delta \sigma^{I}_{1q}(t) \delta \sigma^{I}_{1(-p-q)}(t) + \frac{\lambda_{2}}{6}\delta \sigma^{I}_{2p}(t) \delta \sigma^{I}_{2q}(t) \delta \sigma^{I}_{2(-p-q)}(t) \nonumber \\
& + \frac{\lambda_{3}}{2}\delta \sigma^{I}_{1p}(t) \delta \sigma^{I}_{2q}(t) \delta \sigma^{I}_{2(-p-q)}(t) +\frac{\lambda_{4}}{2}\delta \sigma^{I}_{2p}(t) \delta \sigma^{I}_{1q}(t) \delta \sigma^{I}_{1(-p-q)(t)} \} \nonumber\\
&\equiv \int d^3p d^3q \left( H^{3I}_{1} + H^{3I}_{2} + H^{3I}_{3} + H^{3I}_{4} \right)
\end{align}
As in the power spectrum case, the transfer vertexes, the first two terms in (\ref{quad-Hamiltonian}), convert the above interactions to curvature perturbations. As it is mentioned in \cite{Chen:2009zp} the three point function for $\delta \theta$ is given by
\begin{align}
\langle\delta \theta ^{3}\rangle ~ \equiv ~ \langle 0 \mid \left[ \overline{T} exp( i\int_{t_{0}}^{t} dt' H_{I}(t')) \right] \delta \theta_{I} ^{3}(t) \left[ T exp( -i\int_{t_{0}}^{t} dt' H_{I}(t')) \right] \mid 0 \rangle \, ,
\label{three point function}
\end{align}
in which the symbols $T$ and $\bar T$  represent the time ordering and anti-time ordering respectively.

We can expand the above equation in two equivalent forms, which are called
in \cite{Chen:2009zp} as the "factorized" and "commutator" forms. Each of these forms has its own advantage and disadvantages. Specially, as it is mentioned in \cite{Chen:2009zp}, in the calculation of the integrals there are some unphysical  divergences in both IR limit, $ \tau \rightarrow 0$, and UV limit, $ \tau \rightarrow -\infty$
which can be neglected in commutator and factorized forms, respectively.
In the following we briefly mention each of them and leave the details to Appendix \ref{AppenB}.

Let us start with the  factorized form.  The expansion of  Eq. (\ref{three point function}) is
\begin{align}
\langle \delta \theta ^{3} \rangle & =  \int_{t_{0}}^{t} d \widetilde {t}_{1} \int_{t_{0}}^{\widetilde{t}_{1}} d\widetilde{t}_{2} \int_{t_{0}}^{t} dt_{1}\int_{t_{0}}^{t_1} dt_{2} \langle H_{I}(\widetilde{t}_{2})H_{I}(\widetilde{t}_{1})\delta \theta_{I} ^{3} H_{I}(t_{1})H_{I}(t_{2}) \rangle \label{factorized1}\\
& -2 {\rm Re} \left[ \int_{t_{0}}^{t} d \widetilde {t}_{1} \int_{t_{0}}^{t} dt_{1} \int_{t_{0}}^{t_{1}} dt_{2}\int_{t_{0}}^{t_2} dt_{3} \langle H_{I}(\widetilde{t}_{1}) \delta \theta_{I} ^{3} H_{I}(t_{1}) H_{I}(t_{2})H_{I}(t_{3}) \rangle\right] \label{factorized2}\\
& +2 {\rm Re} \left[ \int_{t_{0}}^{t} dt_{1} \int_{t_{0}}^{t_{1}} dt_{2} \int_{t_{0}}^{t_{2}} dt_{3}\int_{t_{0}}^{t_3} dt_{4} \langle \delta \theta_{I} ^{3} H_{I}(t_{1}) H_{I}(t_{2}) H_{I}(t_{3})H_{I}(t_{4}) \rangle \right]  \, .
\label{factorized3}
\end{align}
In each term, the self interactions between the $\delta \sigma_{i}$'s can appear in one of four $H_{I}$. For example,  for $H^{3I}_{3}$ interaction we need one transfer vertex of the form  $H^{2}_{1} \equiv - c_{1} a(t)^3 \dot{\theta_{0}} \delta \dot{\theta} \delta \sigma_{1} $ and two of the form $H^{2}_{2} \equiv -c_{2} a(t)^3 \dot{\theta_{0}} \delta \dot{\theta} \delta \sigma_{2} $. After contractions we get,
\begin{align}
&-12 u^{*}_{k_{1}}(0)u_{k_{2}}(0)u_{k_{3}}(0) \nonumber\\
& \times {\rm Re} \bigg{[}\int_{-\infty}^{0} d \widetilde {\tau}_{1} a^{3} c_{1} \dot{\theta_{0}} v^{*}_{k_{1}}(\widetilde {\tau}_{1}) u'_{k_{1}}(\widetilde {\tau}_{1}) \int_{-\infty}^{\widetilde {\tau}_{1}} d \widetilde {\tau}_{2} \frac{\lambda_{1}}{6} a^{4} v_{k_{1}}(\widetilde {\tau}_{2}) v_{k_{2}}(\widetilde {\tau}_{2}) v_{k_{3}}(\widetilde {\tau}_{2}) \nonumber\\
& \times \int_{-\infty}^{0} d \tau_{1} a^{3} c_{1} \dot{\theta_{0}} v^{*}_{k_{2}}(\tau_{1}) u'^{*}_{k_{2}}(\tau_{1})\int_{-\infty}^{\tau_{1}} d \tau_{2} a^{3} c_{1} \dot{\theta_{0}} v^{*}_{k_{3}}(\tau_{2}) u'^{*}_{k_{3}}(\tau_{2})\bigg{]}(2\pi)^{3} \delta ^{3}(k_{1}+k_{2}+k_{3}) + 5 perm.
\end{align}
There are altogether 80 different terms in this format, which are given in details in Appendix \ref{AppenB.1}. \\
Now let us look into the  commutator form.  The expansion of Eq. (\ref{three point function}) in terms of the nested commutators is given by \cite{Weinberg:2005vy}
\begin{align}
\langle\delta \theta^{3}\rangle = \int_{t_{0}}^{t} dt_{1} \int_{t_{0}}^{t_{1}} dt_{2} \int_{t_{0}}^{t_{2}} dt_{3} \int_{t_{0}}^{t_{3}} dt_{4} \langle [H_{I}(t_{4}), [H_{I}(t_{3}),[ H_{I}(t_{2}), [H_{I}(t_{1}),\delta \theta_{I}(t)^{3}] \, ]\, ]\, ] \rangle  \, .
\label{three commutator}
\end{align}
Again we should consider each of $H^{3}_{i}$ separately. There are 24 terms. The details are given in Appendix \ref{AppenB.2}.\\
As mentioned in \cite{Chen:2009zp}, when we calculate the integrals we enter unphysical singularities. To eliminate this spurious singularities we have to perform the integral in the so-called ``mixed form" which is a combination of the factorized form and the commutator form.
In principle one can find the full shape of the bispectrum in the mixed form  in terms of different couplings as well as different masses of $\sigma_{i}$ fields.  However, this involves very complicated analysis which is beyond the scope of this work. In the following we only consider the squeezed limit of the bispectrum which contains interesting physical information.


\section{Squeezed limit of bispectrum}
\label{squeezed}
The goal of this section is the calculation of the bispectrum in the squeezed limit, $k_{3}\ll k_{1}\simeq k_{2}$. The importance of doing analysis in this limit is that, we can analytically estimate the effect of the massive fields in both the size and the shape of the bispectrum. As we will show explicitly, while these massive fields have non-trivial effect in the amplitude of the bispectrum, they will not generate any new shapes as compared with the quasi single field inflation, \cite{Chen:2009zp}.
Before going through the analysis, let us review the case of quasi single field inflation. In quasi single field inflation, we can understand the squeezed limit behavior in the following way,  \cite{Baumann:2009ni}. First, remember that the squeezed limit corresponds to the correlation between a long mode and two short modes. Let us assume that this long mode crosses the horizon at $\tau_{1}$ while short modes cross the horizon some time after, say $\tau_{2}$.  Since the massive modes decay after exiting the horizon, at $\tau_{2}$ amplitude has been reduced as,
\ba
\sigma_{L}(k_{3}\tau_{2}) &=& \sigma_{L}(k_{3}\tau_{1}) \left(\frac{\tau_{2}}{\tau_{1}}\right)^{\left(\frac{3}{2} - \nu \right)}\nonumber\\
& \sim& \sigma_{L}(k_{3}\tau_{1}) \left(\frac{k_{3}}{k_{1}}\right)^{\left(\frac{3}{2} - \nu \right)}
\ea
On the other hand, due to the scale invariance of the theory one expect that bispectrum scales as $\left(\frac{1}{k^6}\right)$.
Combining these two point with each other leads us to the following leading shape,
\ba
\lim_{k_{3}\rightarrow 0} \langle \zeta_{k_{1}}\zeta_{k_{2}}\zeta_{k_{3}}\rangle \simeq \frac{1}{k_{1}^3 k_{3}^3} \left(\frac{k_{3}}{k_{1}}\right)^{\left(\frac{3}{2} - \nu \right)}
\ea
Which as we see is the leading shape of the bispectrum in the squeezed limit.\\
We argue that this result is generic and is due to a couple of points. First, although there is a mixing between the massive modes with inflaton field, the approximate scale invariance of the model is still alive, the Goldstone boson will not be massive. Second, there is not any mixing between massive modes in the free field action, which means that at the zeroth order in the quadratic action, these massive modes do not communicate with each other and freely decay after the horizon crossing. \\
In the following we are going to show the above results in more details.\\
In this limit we can use the small argument of the Hankel functions which is given in the following,
\begin{align}
H^{(1)}_{\nu}\rightarrow -i \frac{2^{\nu} \Gamma(\nu)}{\pi} x^{-\nu} - i \frac{2^{-2+\nu} \Gamma(\nu)}{\pi(-1+\nu)} x^{-\nu+2}+ \left(-i \frac{\cos(\pi \nu)\Gamma(-\nu)}{2^{\nu}\pi} + \frac{1}{2^{\nu}\Gamma(1+\nu)} \right)x^{\nu} + ...
\label{Hankel limit}
\end{align}
In what follows, we only consider the commutator forms, since the results of the factorized form is exactly the same as this form, \cite{Chen:2009zp}. we separate the contributions from three types of terms denoted by A, B and C.
First we look at the A term, (\ref{A terms}).  As we see from this equation, there are four different contributions originating from $H_{i}^{3} , i=1..4$. But as we mentioned in the Appendix \ref{AppenB.2} only two of them are independent, namely $H^{3}_{1}$ and $H^{3}_{3}$ while the contribution from the other terms can be easily calculated. So we can skip them and only consider the independent terms.\\
We just mention one term here and leave the details in the appendix, \ref{AppenB.3}.\\
$(1_{A})A_{H^{3}_{1}}$:  This part is very similar to \cite{Chen:2009zp}, so we summarize the calculations and express the final result. With the definition $x_{i} \equiv k_{1} \tau_{i}$ the contribution from $A_{H^{3}_{1}}$ is
\begin{align}
\delta \theta ^{3} (A_{H^{3}_{1}}) &= - \frac{\dot{\theta_{0}}^{3}}{2^{7}} \frac{\lambda_{1}c_{1}^{3}}{H\widetilde{R}^{3}}\frac{\pi^{3}}{k_{1}^{4}k_{2}k_{3}} \nonumber \\
& \times {\rm Re} \bigg{[} i\int_{-\infty}^{0} dx_{1}\int_{-\infty}^{x_{1}} dx_{2}\int_{-\infty}^{x_{2}} dx_{3}\int_{-\infty}^{x_{3}} dx_{4} (-x_{1})^{-\frac{1}{2}} (-x_{2})^{\frac{1}{2}}(-x_{3})^{-\frac{1}{2}}(-x_{4})^{-\frac{1}{2}} \nonumber\\
& \times \sin(-x_{1})\left( H^{(1)}_{\nu_{1}}(-x_{1})H^{(2)}_{\nu_{1}}(-x_{2}) - c.c. \right) \left( H^{(2)}_{\nu_{1}}(-\frac{k_{3}}{k_{1}}x_{2})H^{(1)}_{\nu_{1}}(-\frac{k_{3}}{k_{1}}x_{4})e^{-i\frac{k_{3}}{k_{1}}x_{4}} - c.c. \right) \nonumber\\
& \times H^{(1)}_{\nu_{1}}(-\frac{k_{2}}{k_{1}}x_{2})H^{(2)}_{\nu_{1}}(-\frac{k_{2}}{k_{1}}x_{3})e^{i\frac{k_{2}}{k_{1}}x_{3}}\bigg{]} \, .
\end{align}
Now as in \cite{Chen:2009zp}, the term $H^{(2)}_{\nu_{1}}(-\frac{k_{3}}{k_{1}}x_{2})$ in the 3rd line  can be approximated in the small $-k_{3}/k_{1}x_{2}$ limit. However, the term $H^{(1)}_{\nu_{1}}(-\frac{k_{3}}{k_{1}}x_{4})$ in the 3rd line can not be approximated. So by redefining $y_{4} \equiv k_{3}/k_{1}x_{4}$, we get
\begin{align}
\delta \theta ^{3} (A_{H^{3}_{1}}) &= - \frac{\dot{\theta_{0}}^{3}}{2^{5-\nu_{1}}} \frac{\lambda_{1}c_{1}^{3}}{H\widetilde{R}^{3}}\frac{\pi^{2}\Gamma{(\nu_{1})}}{k_{1}^{\frac{7}{2}-\nu_{1}}k_{2}k_{3}^{\frac{3}{2}+\nu_{1}}} \nonumber \\
& \times \int_{-\infty}^{0} dx_{1}\int_{-\infty}^{x_{1}} dx_{2}\int_{-\infty}^{x_{2}} dx_{3} (-x_{1})^{-\frac{1}{2}} (-x_{2})^{\frac{1}{2}-\nu_{1}}(-x_{3})^{-\frac{1}{2}} \nonumber\\
& \times \sin(-x_{1})Im\left( H^{(1)}_{\nu_{1}}(-x_{1})H^{(2)}_{\nu_{1}}(-x_{2})\right) {\rm Im}\left( H^{(1)}_{\nu_{1}}(-x_{2})H^{(2)}_{\nu_{1}}(-x_{3})e^{ix_{3}} \right) \nonumber\\
& \times \int_{-\infty}^{0} dy_{4} (-y_{4})^{-\frac{1}{2}} {\rm Re}\left(  H^{(1)}_{\nu_{1}}(-y_{4})e^{-iy_{4}}\right) \, .
\end{align}
The scaling behavior of the above term is
\ba
N_{1}\equiv  \left(k_{1}^{\frac{7}{2}-\nu_{1}}k_{2}k_{3}^{\frac{3}{2}+\nu_{1}}\right)^{-1}
\ea

\begin{figure}
\begin{center}
\includegraphics[ scale=.8]{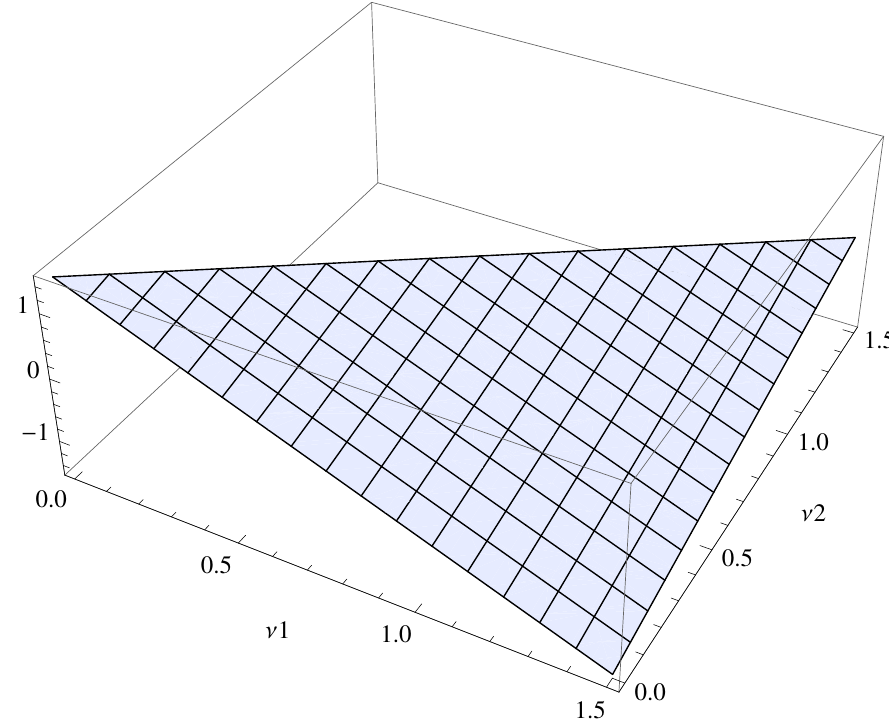}
\end{center}
\medskip
\caption{ The plot of $(3/2 + \nu_{2} -2\nu_{1})$ for $\nu_{2}<\nu_{1}$. }
\label{Fig:Fdiagram3}
\end{figure}

We next look at the term with the permutation $k_{1}\leftrightarrow k_{3} $. With the same redefinition of $x_{i} (i=1, 2, 3, 4)$, we get
\begin{align}
\delta \theta ^{3} (A_{H^{3}_{1}}) &= - \frac{\dot{\theta_{0}}^{3}}{2^{7}} \frac{\lambda_{1}c_{1}^{3}}{H\widetilde{R}^{3}}\frac{\pi^{3}}{k_{1}^{4}k_{2}k_{3}} \nonumber \\
& \times {\rm Re} \bigg{[} i\int_{-\infty}^{0} dx_{1}\int_{-\infty}^{x_{1}} dx_{2}\int_{-\infty}^{x_{2}} dx_{3}\int_{-\infty}^{x_{3}} dx_{4} (-x_{1})^{-\frac{1}{2}} (-x_{2})^{\frac{1}{2}}(-x_{3})^{-\frac{1}{2}}(-x_{4})^{-\frac{1}{2}} \nonumber\\
& \times \sin(-\frac{k_{3}}{k_{1}}x_{1})\left( H^{(1)}_{\nu_{1}}(-\frac{k_{3}}{k_{1}}x_{1})H^{(2)}_{\nu_{1}}(-\frac{k_{3}}{k_{1}}x_{2}) - c.c. \right) \left( H^{(2)}_{\nu_{1}}(-x_{2})H^{(1)}_{\nu_{1}}(-x_{4})e^{-ix_{4}} - c.c. \right) \nonumber\\
& \times H^{(1)}_{\nu_{1}}(-\frac{k_{2}}{k_{1}}x_{2})H^{(2)}_{\nu_{1}}(-\frac{k_{2}}{k_{1}}x_{3})e^{i\frac{k_{2}}{k_{1}}x_{3}}\bigg{]}
\end{align}
Again as in \cite{Chen:2009zp}, in the 3rd line the first three functions can be approximated in the small $-(k_{3}/k_{1})x_{i}\, , (i= 1,2)$ limit,  and in order to get a non-zero result, one of the Hankel functions should be expanded to $\mathcal{O}(x^{\nu_{1}}_{i})$ . By this approximation, the scaling behavior of the above term is obtained to be
\begin{align}
M\equiv\frac{1}{k_{1}^{5}k_{2}}
\end{align}
Since $\frac{M}{N_{1}} = \frac{K_{3}}{k_{1}}^{3/2+\nu_{1}}$, $M$ is much smaller than $N_{1}$.\\
Then we look at the term with the permutation $k_{2}\leftrightarrow k_{3} $
\begin{align}
\delta \theta ^{3} (A_{H^{3}_{1}}) &= - \frac{\dot{\theta_{0}}^{3}}{2^{7}} \frac{\lambda_{1}c_{1}^{3}}{H\widetilde{R}^{3}}\frac{\pi^{3}}{k_{1}^{4}k_{2}k_{3}} \nonumber \\
& \times {\rm Re} \bigg{[} i\int_{-\infty}^{0} dx_{1}\int_{-\infty}^{x_{1}} dx_{2}\int_{-\infty}^{x_{2}} dx_{3}\int_{-\infty}^{x_{3}} dx_{4} (-x_{1})^{-\frac{1}{2}} (-x_{2})^{\frac{1}{2}}(-x_{3})^{-\frac{1}{2}}(-x_{4})^{-\frac{1}{2}} \nonumber\\
& \times \sin(-x_{1})\left( H^{(1)}_{\nu_{1}}(-x_{1})H^{(2)}_{\nu_{1}}(-x_{2}) - c.c. \right) \left( H^{(2)}_{\nu_{1}}(-\frac{k_{2}}{k_{1}}x_{2})H^{(1)}_{\nu_{1}}(-\frac{k_{2}}{k_{1}}x_{4})e^{-i\frac{k_{2}}{k_{1}}x_{4}} - c.c. \right) \nonumber\\
& \times H^{(1)}_{\nu_{1}}(-\frac{k_{3}}{k_{1}}x_{2})H^{(2)}_{\nu_{1}}(-\frac{k_{3}}{k_{1}}x_{3})e^{i\frac{k_{3}}{k_{1}}x_{3}}\bigg{]}
\end{align}
Again in the 4th line, we can approximate the three functions in the small
$-x_{i}k_{3}/k_{1}, \,  (i= 2,3)$ limit. For $\nu_{1}> \frac{1}{2}$, we use the leading term for the two Hankel functions and sub-leading term for the exponential function. Then the result is
\begin{align}
\delta \theta ^{3} (A_{H^{3}_{1}}) &= - \frac{\dot{\theta_{0}}^{3}}{2^{5-2\nu_{1}}} \frac{\lambda_{1}c_{1}^{3}}{H\widetilde{R}^{3}}\frac{\pi}{k_{1}^{5-2\nu_{1}}k_{2}k_{3}^{2\nu_{1}}}(\Gamma{(\nu_{1})})^{2} \nonumber \\
& \times \int_{-\infty}^{0} dx_{1}\int_{-\infty}^{x_{1}} dx_{2}\int_{-\infty}^{x_{2}} dx_{3}\int_{-\infty}^{x_{3}} dx_{4} (-x_{1})^{-\frac{1}{2}} (-x_{2})^{\frac{1}{2}-\nu_{1}}(-x_{3})^{\frac{1}{2}-\nu_{1}}(-x_{4})^{-\frac{1}{2}} \nonumber\\
& \times \sin(-x_{1}){\rm Im}\left( H^{(1)}_{\nu_{1}}(-x_{1})H^{(2)}_{\nu_{1}}(-x_{2})\right) {\rm Im}\left( H^{(2)}_{\nu_{1}}(-x_{2})H^{(1)}_{\nu_{1}}(-x_{4})e^{-ix_{4}} \right)
\end{align}
The scaling behavior of this term is $P_{1} \equiv \left(k_{1}^{5-2\nu_{1}}k_{2}k_{3}^{2\nu_{1}}\right)^{-1} $ and since  $\frac{P_{1}}{N_{1}}= \frac{k_{3}}{k_{1}}^{3/2 - \nu_{1}}$, it was argued in \cite{Chen:2009zp} that $P_{1}$ is negligible compared to $N_{1}$. However, in our case we have two different indices $\nu_{1}$ and $\nu_{2}$ and as we can see from Fig. \ref{Fig:Fdiagram3}, it is possible to choose some parts of the parameter space such that $P_{1}$ becomes larger than $N_{2}\equiv \left(k_{1}^{\frac{7}{2}-\nu_{2}}k_{2}k_{3}^{\frac{3}{2}+\nu_{2}}\right)^{-1}$. This happens when in the ratio $\frac{P_{1}}{N_{2}}= \frac{k_{3}}{k_{1}}^{3/2 + \nu_{2} -2\nu_{1}}$ the expression $3/2 + \nu_{2} -2\nu_{1}$  becomes negative.  From Fig.\ref{Fig:Fdiagram3} we see that this actually happens for some values of $\nu_{i}$ in the parameter space. So we should consider  this term as well as $N_{1}$ and $N_{2}$.

On the other hand, for $\nu_{1} < \frac{1}{2} $ one of the Hankel functions should be expanded to $\mathcal{O}(x^{\nu_{1}}_{i})$ and by using the leading term for the exponential function, the scaling behavior is
\begin{align}
Q\equiv \frac{1}{k_{1}^{4}k_{2}k_{3}}
\end{align}
Since $\frac{Q}{N_{1}}= \frac{k_{3}}{k_{1}}^{\frac{1}{2} + \nu_{1}}$, $Q$ is negligible compared to $N_{1}$.

Finally,  the other permutation $k_{1}\leftrightarrow k_{2}$ gives each term a factor of 2.

The final result can be obtained by summing up all of the above contributions from A, B and C terms.
\begin{align}
\langle\zeta(k_{1})\zeta(k_{2})\zeta(k_{3})\rangle |_{k_{3} \ll k_{1}=k_{2}} = \frac{H^{2}}{\widetilde{R}^{3}} \bigg{[} &\frac{S_{1}(\nu_{1} , \nu_{2})}{k_{1}^{\frac{7}{2}-\nu_{1}}k_{2}k_{3}^{\frac{3}{2} + \nu_{1}}} +  \frac{S_{2}(\nu_{1} , \nu_{2})}{k_{1}^{\frac{7}{2}-\nu_{2}}k_{2}k_{3}^{\frac{3}{2} + \nu_{2}}} \nonumber \\
& + \frac{S_{3}(\nu_{1} , \nu_{2})} {k_{1}^{5-2\nu_{1}}k_{2}k_{3}^{2\nu_{1}}} + \frac{S_{4}(\nu_{1} , \nu_{2})} {k_{1}^{5-2\nu_{2}}k_{2}k_{3}^{2\nu_{2}}}   \bigg{]} (2\pi)^{3} \delta^{3}(\sum_{i} \bk_{i})
\label{squeezed limit}
\end{align}
This is the main result of this Section. \\
Although the full analysis has been given in Appendix (\ref{AppenB.3}), it is worth to elaborate on how different terms are obtained, especially it is important to connect them to the Feynman diagrams in Fig.\ref{Fig:correctionBispectrum}. \\
As it can be seen from Appendix (\ref{AppenB.3}), there are different contributions from the $H^{3I}_{i}$s in the above shapes. \\
$\bullet$ $S_{1}(\nu_{1},\nu_{2})$ and $S_{3}(\nu_{1},\nu_{2})$ come from
$H^{3I}_{1}$, $H^{3I}_{3}$ and $H^{3I}_{4}$ which are given in Fig.(\ref{Fig:correctionBispectrum}a), Fig.(\ref{Fig:correctionBispectrum}c) and Fig.(\ref{Fig:correctionBispectrum}d) respectively. \\
$\bullet$ $S_{2}(\nu_{1},\nu_{2})$ and $S_{4}(\nu_{1},\nu_{2})$ come from $H^{3I}_{2}$, $H^{3I}_{3}$ and $H^{3I}_{4}$ which are presented in Fig.(\ref{Fig:correctionBispectrum}b), Fig.(\ref{Fig:correctionBispectrum}c) and
Fig.(\ref{Fig:correctionBispectrum}d) respectively.\\
Since $S_{1}(\nu_{1},\nu_{2})$ is related to the leading shape, we express its form here, while leave the details of $S_{3}(\nu_{1},\nu_{2})$ in the appendix, \ref{AppenB.4}.
\ba
S_{1}(\nu_{1} , \nu_{2}) &\equiv& \frac{\pi^{2}\Gamma{(\nu_{1})}}{2^{4-\nu_{1}}} \int_{-\infty}^{0} dx_{1}\int_{-\infty}^{x_{1}} dx_{2}\int_{-\infty}^{x_{2}} dx_{3}\nonumber \\
&& \times \bigg{[} \bigg{(}  (-x_{1})^{-\frac{1}{2}} (-x_{2})^{\frac{1}{2}-\nu_{1}}(-x_{3})^{-\frac{1}{2}}
\sin(-x_{1})  \bigg{(}  \lambda_{1}c_{1}^{3} {\rm Im}\left( H^{(1)}_{\nu_{1}}(-x_{1})H^{(2)}_{\nu_{1}}(-x_{2})\right) \nonumber \\
&& \times {\rm Im}\left( H^{(1)}_{\nu_{1}}(-x_{2})H^{(2)}_{\nu_{1}}(-x_{3})e^{ix_{3}}\right) + \lambda_{4}c_{1}^{2}c_{2}{\rm Im}\left( H^{(1)}_{\nu_{2}}(-x_{1})H^{(2)}_{\nu_{2}}(-x_{2})\right) \nonumber\\
&&\times {\rm Im}\left( H^{(1)}_{\nu_{1}}(-x_{2})H^{(2)}_{\nu_{1}}(-x_{3})e^{ix_{3}} \right) + \lambda_{4}c_{1}^{2}c_{2}{\rm Im}\left( H^{(1)}_{\nu_{1}}(-x_{1})H^{(2)}_{\nu_{1}}(-x_{2})\right) \nonumber\\
&&\times {\rm Im}\left( H^{(1)}_{\nu_{2}}(-x_{2})H^{(2)}_{\nu_{2}}(-x_{3})e^{ix_{3}} \right) + \lambda_{3}c_{1}c_{2}^{2}{\rm Im}\left( H^{(1)}_{\nu_{2}}(-x_{1})H^{(2)}_{\nu_{2}}(-x_{2})\right) \nonumber \\
&& \times {\rm Im}\left( H^{(1)}_{\nu_{2}}(-x_{2})H^{(2)}_{\nu_{2}}(-x_{3})e^{ix_{3}} \right)
\bigg{)}+ (-x_{1})^{-\frac{1}{2}}(-x_{2})^{-\frac{1}{2}}(-x_{3})^{\frac{1}{2}-\nu_{1}} \nonumber\\
&& \times \sin(-x_{1})\sin(-x_{2}) \bigg{(}\lambda_{1}c_{1}^{3} {\rm Im}\left( H^{(2)}_{\nu_{1}}(-x_{1})H^{(2)}_{\nu_{1}}(-x_{2}) \left(H^{(1)}_{\nu_{1}}(-x_{3})\right)^{2}\right) \nonumber \\
&&+\lambda_{4}c_{1}^{2}c_{2}{\rm Im}\left( H^{(2)}_{\nu_{2}}(-x_{1})H^{(1)}_{\nu_{2}}(-x_{3})H^{(2)}_{\nu_{1}}(-x_{2})H^{(1)}_{\nu_{1}}(-x_{3})\right)\nonumber\\
&&+\lambda_{4}c_{1}^{2}c_{2}{\rm Im}\left( H^{(2)}_{\nu_{1}}(-x_{1})H^{(1)}_{\nu_{1}}(-x_{3})H^{(2)}_{\nu_{2}}(-x_{2})H^{(1)}_{\nu_{2}}(-x_{3})\right)\nonumber\\
&&+\lambda_{3}c_{1}c_{2}^{2}{\rm Im}\left( H^{(2)}_{\nu_{2}}(-x_{1})H^{(2)}_{\nu_{2}}(-x_{2})(H^{(1)}_{\nu_{2}}(-x_{3}))^{2}\right)
\bigg{)}  \bigg{)} \nonumber\\
&& \times \int_{-\infty}^{0} dy_{4} (-y_{4})^{-\frac{1}{2}}{\rm Re}\left(  H^{(1)}_{\nu_{1}}(-y_{4})e^{-iy_{4}}\right) \bigg{]}
\ea
Eq. (\ref{squeezed limit}) is one of our main results in this paper. There are different shapes in the bispectrum, which depending on the value of $\nu_{1}$ and $\nu_{2}$ i.e.  different values of the isocurvatons mass and the coupling constants for the interaction among $\sigma_{i}$ fields, can be dominant or subdominant. It is an interesting question to study the full dependence of the bispectrum shape on the model parameters. This requires extensive analysis which is beyond the scope of this work. Having this said, we still can obtain important
information from our current results as follows. \\

$\blacklozenge$ First of all, as it was mentioned in \cite{Chen:2009zp}, the origins of non-Gaussianity in this model are the self-interactions between $\delta \sigma_{i}$'s which scales like $k_{3}^{-2\nu_{i}}$, \cite{Chen:2009zp}. However,  it has to convert to the curvature perturbation via the exchange vertex. While converting to the curvature bispectrum, for a typical value of $\nu_{i}$, the leading shape will be changed from its original contribution,$k_{3}^{-2\nu_{i}}$, to $k_{3}^{-\left(\frac{3}{2} + \nu_{i}\right)}$, which is more like the local shape, $k_{3}^{-3}$. However, as sub-leading term compared to this leading contribution, there is still $k_{3}^{-2\nu_{i}}$ shape which means that in the conversion procedure there is not any change in the shape of the bispectrum. Therefore,  we conclude that the projection effect from $\delta \sigma_{i}$ bispectrum to that of curvature bispectrum can change both of the amplitude and the shape of bispectrum. In every case the amplitude will be changed. However, while for the leading contribution we have a change in the shape of three point function, there is still another sub-leading term which does not contain any change in its shape, as compared to the original contribution of $\sigma_{i}$ bispectrum. \\

$\blacklozenge$ The second point is that, as it is mentioned in \cite{Chen:2009zp}, these shapes are between the equilateral-type shape, $k_{3}^{-1}$, and the local-type shape, $k_{3}^{-3}$.  For  small values of $\nu_{i}$ they are more like equilateral shape while for the larger values of $\nu_i$ they go to the local shape. It is physically understandable, since for the small values of $\nu_{i}$ i.e. for higher values of the isocurvatons' masses, the isocurvaton wave function decays rapidly after horizon crossing which means that effectively we are in the single field regime.  On the other hand,  for the higher values of the $\nu_{i}$, or in another word for small values for the isocurvatons' masses, the wave functions  decay slowly after horizon crossing and we are in multiple field regime with local shape.\\

$\blacklozenge$ Third, for the spectroscopy of different masses, it is worth to consider two different cases in the following, \\

$~~\bigstar$ A) Suppose that we are in the regime in which both isocurvaton fields have masses of the same order. In this case, the first two terms in Eq. \ref{squeezed limit} will dominate and effectively we will have,
\begin{align}
\langle\zeta(k_{1})\zeta(k_{2})\zeta(k_{3})\rangle |_{k_{3} \ll k_{1}=k_{2}} = \frac{H^{2}}{\widetilde{R}^{3}} \bigg{[} &\frac{S_{1}(\nu_{1} , \nu_{2})}{k_{1}^{\frac{7}{2}-\nu_{1}}k_{2}k_{3}^{\frac{3}{2} + \nu_{1}}} +  \frac{S_{2}(\nu_{1} , \nu_{2})}{k_{1}^{\frac{7}{2}-\nu_{2}}k_{2}k_{3}^{\frac{3}{2} + \nu_{2}}} \bigg{]} (2\pi)^{3} \delta^{3}(\sum_{i} \bk_{i})
\label{squeezed limit1}
\end{align}

$~~\bigstar$ B) In this case, we assume that there is a hierarchy between different mass scales, e.g. $ m_{\sigma2} \gg m_{\sigma1}$ which means that $\nu_{1} \gg \nu_{2} $. So as it is shown in Fig. \ref{Fig:Fdiagram3}, the leading scaling dependence of $\nu_{2}$ becomes smaller than the subleading shape coming from $\sigma_{1}$. As a result the shape is dominated with the lightest field and we can neglect the contribution from the heavy field,
\begin{align}
\langle\zeta(k_{1})\zeta(k_{2})\zeta(k_{3})\rangle |_{k_{3} \ll k_{1}=k_{2}} = \frac{H^{2}}{\widetilde{R}^{3}} \bigg{[} &\frac{S_{1}(\nu_{1} , \nu_{2})}{k_{1}^{\frac{7}{2}-\nu_{1}}k_{2}k_{3}^{\frac{3}{2} + \nu_{1}}} + \frac{S_{3}(\nu_{1} , \nu_{2})} {k_{1}^{5-2\nu_{1}}k_{2}k_{3}^{2\nu_{1}}} \bigg{]} (2\pi)^{3} \delta^{3}(\sum_{i} \bk_{i})
\label{squeezed limit2}
\end{align}
in which we have kept the contribution from the subleading term to specify that potentially one can detect this shape in the squeezed limit, although it is subleading.\\

$\blacklozenge$ Fourth, let us look at the order of magnitude of $S_{i}(\nu_{1}, \nu_{2}), i=1,..,4$. In our case, since $S_{3}(\nu_{1}, \nu_{2})$ and $S_{4}(\nu_{1}, \nu_{2})$ contain four layer integrals the detail calculation of these terms are tremendous and are left to the future work. However, compared to \cite{Chen:2009zp} we can estimate their order of magnitudes. We have checked that as we vary $\nu_{i}$ from zero to $\frac{3}{2}$, the
values of $S_{3}(\nu_{1}, \nu_{2})$ and $S_{4}(\nu_{1}, \nu_{2})$
change from zero to around 30, in the limit in which all of the coupling constants are of the same order.  \\

$\blacklozenge$ Fifth, although these massive modes have non-trivial effects in the amplitude of the bispectrum, they will not produce any new shapes in this limit. In the sense that they are completely separated at the level of the shape and there will not be any mixing between their masses at this level. As we showed, it is a generic feature of this kind of models and is due to the scale invariance of this model and the fact that at the zeroth order in show roll these massive fields do not communicate with each other and decay  after horizon crossing.\\

\begin{figure}
\includegraphics[ scale=1]{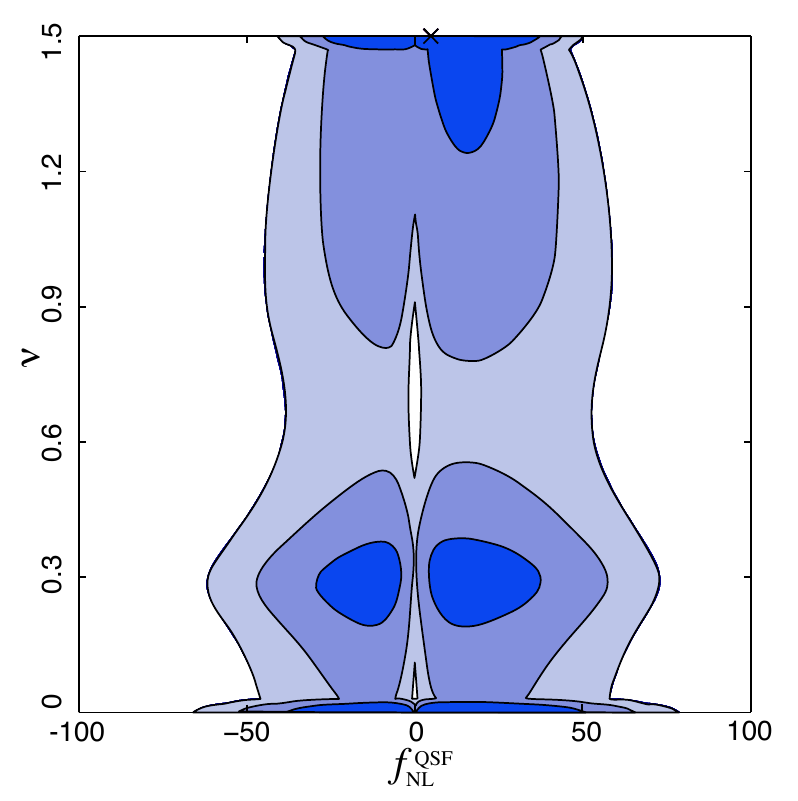}
\medskip
\caption { $68\%$, $95\%$, and $99.7\%$ confidence intervals for $\nu$ and
$f_\mathrm{NL}^{\rm QSI}$ in original quasi-single field model borrowed from PLANCK   \cite{Ade:2013ydc}.
There is not any preferred value for $\nu$ with all values allowed at $3\sigma$.}
\label{Fig:QSF}
\end{figure}

$\blacklozenge$ The last point is the comparison of this model with the PLANCK data. Because  there was no detection of non-Gaussianity by PLANCK,
there is not any preference for the values of  $\nu_{i}$,  see Fig. \ref{Fig:QSF} but more strong constraints on the amplitude of $f_{NL}$. However,
it seems that there are still some rooms for the equilateral shapes to work after PLANCK.  Naively speaking this means that
our model has a  better fit to the data for the smaller values of $\nu_i$ i.e. for larger values of the isocurvatons' masses. In addition, there are also some constraints from the LSS analysis \cite{Sefusatti:2012ye, Liguori:2010hx, Dalal:2007cu, Desjacques:2010jw}. Interestingly the halo bias is more sensitive to the scale dependence of the bispectrum in the squeezed limit while the CMB analysis are more based on the Bispectrum amplitude. So  these different measurements can act as a complementary to each other in order to distinguish QSF models among other models.

\section{Summary and Discussions}
\label{summary}
In this work we have extended the analysis of \cite{Chen:2009zp} to the model containing
two semi-heavy isocurvaton fields. These kind of models are well motivated in the UV completion of the string theory and supersymmetry, when naturally we can have heavy fields with masses of the order of the Hubble constant. In this model, due to the turning trajectory,
the perturbations from the isocurvaton fields are converted to the curvature perturbation which can have non-trivial effects in both the power spectrum as well as the bispectrum. The correction in the power spectrum is negligible. However, because of
their strong self-interactions, their contributions in the amplitude and the shape of the bispectrum of the curvature perturbation are significant.

The origin of the non-Gaussianity comes from isocurvatons' bispectrum. However they have to convert to the curvature bispectrum via the turning trajectory. While converting to the curvature bispectrum, the leading shape also goes more to local-type  shape. The amplitude of $f_{NL}$ is controlled with the strength of these interactions as well as the velocity of the turning trajectory. On the other hand, adding new massive field does not produce any new shape in the squeezed limit which is due to the scale invariance of the model and the fact that free propagating isocurvaton modes do not communicate with each other. However, squeezed limit can act as a particle detector to recognize the particles with masses of the same order.

It is worth to mention that, while the squeezed limit of the bispectrum is dominated by the lightest isocurvaton, it does not mean that the entire bispectrum signal should be dominated by the lightest isocurvaton field. It would be interesting to consider the analysis of the full shape of bispectrum. This is however beyond the scope of this paper and is left for the future work.

From the observational point of view, there are strong upper bound on the amplitude of the local non-Gaussianity from  PLANCK. This means that the strength of the self-interactions between isocurvaton fields as well as the velocity of the turning trajectory can not be large. However, still there is room for the equilateral shape to be detected in the future observations such as  in large scale structure surveys which are more sensitive tot he scale dependence of the Bispectrum curvature perturbation.

\section*{Acknowledgments}
I am very grateful to Xingang Chen and Hassan Firouzjahi for initiating this work and for many insightful discussions during this project. The author also thanks a lot Paolo Creminelli  and Emiliano Sefusatti for their very useful comments on the draft and also on the final results. I also thank  ICTP for its hospitality during
the completion of this work under the ``Sandwich Training Educational Program'' (STEP) fellowship.

\appendix

\section{The full Lagrangian up to third order}\label{AppenA}
In this Appendix we derive the full third order action in the spatially flat gauge. We also show that (\ref{3-interaction}) is the leading order interaction. We set $M_{P}=1$ in this Appendix.

The full action is
\bea
S = S_{g} + S_{m} \, ,
\label{full-action}
\eea
where
\bea
S_{g} = \half \int d^4 x \sqrt{-g} \mathcal{R}
\label{s-g}
\eea
is the gravitational action  and
\bea
S_{m} = \int d^4 x \mathcal{L}
\label{s-m}
\eea
represents the matter part os the action which is given by Eq. (\ref{Lagrangian}). Using the ADM formalism
\bea
ds^2 = -N^2 dt^2 + h_{ij}(dx^{i} + N^{i}dt)(dx^{j} + N^{j}dt)~,
\label{ds2}
\eea
the action becomes
\bea
S = \half \int dt dx^{3} \sqrt{h}N (R^{(3)} + 2\mathcal{L}) + \half \int dt dx^{3} \sqrt{h} N^{-1} (E_{ij} E^{ij} - E^2) \, .
\label{ADM-action}
\eea
Here the index of $N^{i}$ is lowered by the three-dimensional metric $h_{ij}$ and $R^{(3)}$ is the three-dimensional Rici scalar constructed from $h_{ij}$. The definitions of the extrinsic curvature $E_{ij}$ and its trace $E$ are
\bea
E_{ij}&=& \half (\dot{h}_{ij}-\nabla_{i}N_{j}- \nabla_{j}N_{i})~, \nonumber \\
E&=& E_{ij} H^{ij}
\label{def of E}
\eea

We choose the following spatially flat gauge:
\bea
h_{ij} = a^2(t) \delta_{ij} ~~~~,~~~~ \theta = \theta_{0} (t) + \delta\theta ~~~~,~~~~ \sigma_{i} = \sigma_{i0}(t) + \delta\sigma_{i}
\eea
For the constant turn case $\sigma_{i0}(t) = constant$.

The constraint equations for the Lagrangian multipliers $N$ and $N_{i}$ are
\bea
R^{(3)} + 2 \mathcal{L}_{m} + 2N \frac{\partial\mathcal{L}_{m}}{\partial N} + \frac{1}{N^2} (E_{ij} E^{ij} - E^2) &=0~, \label{lagrange equation1} \\
\nabla_{i} (N^{-1}(E^{ij} - E h^{ij})) + N \frac{\partial\mathcal{L}_{m}}{\partial N_{j}}  &=0 \, .
\label{lagrange equation2}
\eea
Now we expand $N^{i}$ and $N$ up to first order
\bea
N = 1+ \alpha_{1} ~~~~,~~~~ N_{i} = \partial_{i} \psi + \widetilde{N}^{1}_{i} ~~~~,~~~~ \partial_{i}\widetilde{N}^{1}_{i} = 0 ~~~~,
\eea
Plugging them into (\ref{lagrange equation1}) and (\ref{lagrange equation2}), the solutions with proper boundary conditions are
\bea
\alpha_{1} &=& \frac{\widetilde{R}\dot{\theta}_{0} \delta \theta}{2H} \quad , \quad  \widetilde{N}^{(1)}_{i} =0 ~, \\
\partial^{2} \psi &=& \frac{a^2}{2H} \left((-6H^2 +\widetilde{R}\dot{\theta}_{0}^2)\alpha_{1} -\widetilde{R}\dot{\theta}_{0} \delta \dot{\theta}-\frac{\dot{\theta}_{0}^2}{2}(c_{1}\delta \sigma_{1} + c_{2}\delta \sigma_{2})- V'_{sr}(\theta_{0})\delta \theta \right) \, .
\eea
Now we plug these solutions into the action and expand up to third order in perturbations. The first order terms give the equation of motion for $\theta_{0}(t)$ and $\sigma_{0}(t)$.  The quadratic action is
\begin{align}
L_{2} &= \frac{1}{2} a(t)^3 \widetilde{R} \delta \dot{\theta}^2 - \frac{1}{2} a(t) \widetilde{R} (\partial_{i} \delta \theta )^2 + \frac{1}{2}a(t)^3 \delta \dot{\sigma} _{1}^2 - \frac{1}{2} a(t)(\partial_{i} \delta \sigma _{1})^2 + +\frac{1}{2}a(t)^3 \delta \dot{\sigma} _{2}^2 \nonumber\\
& - \frac{1}{2} a(t)(\partial_{i} \delta \sigma _{2})^2 - \frac{1}{2} a(t)^3 m_1^2 \delta\sigma_1^2 - \frac{1}{2} a(t)^3 m_2^2 \delta\sigma_2^2- \frac{1}{2}V''_{sr}\delta \theta^2 + c_{1}a(t)^3 \dot{\theta}_{0} \delta \dot{\theta} \delta \sigma_{1} \nonumber\\
& - c_{1} a(t)^3 H \epsilon \dot{\theta}_{0} \delta \theta \delta \sigma_{1} + c_{2}a(t)^3 \dot{\theta}_{0} \delta \dot{\theta} \delta \sigma_{2} - c_{2} a(t)^3 H \epsilon \dot{\theta}_{0} \delta \theta \delta \sigma_{2} + \widetilde{R}H^2 (3\epsilon + \epsilon\eta - \epsilon^{2})\delta \theta^2.
\end{align}
The third order action is
\begin{align}
\frac{L_{3}}{a^{3}} &= -\frac{1}{2}\widetilde{R} \alpha_{1} \delta \dot{\theta}^2 + \widetilde{R}\dot{\theta}_{0} \alpha_{1}^2 \delta \dot{\theta} - \frac{1}{2}\widetilde{R}\dot{\theta}_{0}^2 \alpha_{1}^3 + \frac{1}{2}c_{1}\delta \sigma_{1}\delta \dot{\theta}^2 - c_{1} \dot{\theta}_{0} \alpha_{1} \delta \sigma_{1}\delta \dot{\theta}+ \frac{1}{2} c_{1} \dot{\theta}_{0}^2 \alpha_{1}^2 \delta \sigma_{1} \nonumber\\
&+ \frac{1}{2}c_{2}\delta \sigma_{2}\delta \dot{\theta}^2 - c_{2} \dot{\theta}_{0} \alpha_{1} \delta \sigma_{2}\delta \dot{\theta}+ \frac{1}{2} c_{2} \dot{\theta}_{0}^2 \alpha_{1}^2 \delta \sigma_{2}-a^{-2}\partial_{i} \psi \partial_{i} \delta \theta (\widetilde{R}\delta \dot{\theta} - \widetilde{R}\dot{\theta}_{0} \alpha_{1} + c_{1} \dot{\theta}_{0}\delta \sigma_{1} +  c_{2} \dot{\theta}_{0}\delta \sigma_{2}) \nonumber\\
&- \frac{1}{2}a^{-2}(\partial_{i} \delta \theta )^2(\widetilde{R}\alpha_{1}+ c_{1}\delta \sigma_{1}+ c_{2}\delta \sigma_{2})- \frac{1}{2}\alpha_{1} \delta \dot{\sigma}_{1}^2 - a^{-2}\partial_{i} \psi \partial_{i}\delta \sigma_{1} \delta \dot{\sigma}_{1} - \frac{1}{2}a^{-2}\alpha_{1}(\partial_{i}\delta \sigma_{1})^2 \nonumber\\
&- \frac{1}{2}\alpha_{1} \delta \dot{\sigma}_{2}^2- a^{-2}\partial_{i} \psi \partial_{i}\delta \sigma_{2} \delta \dot{\sigma}_{2} -\frac{1}{2}a^{-2}\alpha_{1}(\partial_{i}\delta \sigma_{2})^2 -\frac{1}{2} \alpha_{1} m_1^2 \delta\sigma_1^2 -\frac{1}{2} \alpha_{1} m_2^2 \delta\sigma_2^2 - \frac{1}{6} \lambda_1 \delta\sigma_1^3  \nonumber\\
& - \frac{1}{6} \lambda_2 \delta\sigma_2^3- \frac{1}{2} \lambda_3 \delta\sigma_1 \delta\sigma_2^2- \frac{1}{2} \lambda_4 \delta\sigma_2 \delta\sigma_1^2 - \frac{1}{2}V''_{sr} \alpha_{1} \delta \theta^2 - \frac{1}{6}V'''_{sr} \delta \theta^3 + 3H^2\alpha_{1}^3 \nonumber\\
&- \frac{1}{2}a^{-4}\alpha_{1}(\partial_{i}\partial_{j}\psi\partial_{i}\partial_{j}\psi - (\partial^2 \psi)^2) + 2a^{-2} H \alpha_{1}^2 \partial^2 \psi.
\end{align}
As we see the terms coming from the gravity back-reaction are smaller than the matter effects.
\section{Details of integrals in various forms}
\label{AppenB}

In this Appendix we give the details of the in-in integrals in the factorized and commutator forms.
\subsection{All terms in the factorized form}
\label{AppenB.1}

Here we calculate the independent contributions  to the bispectrum in  the factorized form. Since $H^{3}_{2}$ can be obtained by changing $(c_{1},\lambda_{1},v_{k_{i}})$ to $(c_{2},\lambda_{2}, w_{k_{i}})$ we skip it. Also the terms $H^{3}_{4}$ is not independent of $H^{3}_{3}$ so we skip it too.

Different terms related to Eq.(\ref{factorized1}) and $H^{3}_{1}$ are:
\begin{align}
(1)=&-12 u^{*}_{k_{1}}(0)u_{k_{2}}(0)u_{k_{3}}(0) \nonumber\\
& \times {\rm Re} \bigg{[}\int_{-\infty}^{0} d \widetilde {\tau}_{1} a^{3} c_{1} \dot{\theta}_{0} v^{*}_{k_{1}}(\widetilde {\tau}_{1}) u'_{k_{1}}(\widetilde {\tau}_{1}) \int_{-\infty}^{\widetilde {\tau}_{1}} d \widetilde {\tau}_{2} \frac{\lambda_{1}}{6} a^{4} v_{k_{1}}(\widetilde {\tau}_{2}) v_{k_{2}}(\widetilde {\tau}_{2}) v_{k_{3}}(\widetilde {\tau}_{2}) \nonumber\\
& \times \int_{-\infty}^{0} d \tau_{1} a^{3} c_{1} \dot{\theta}_{0} v^{*}_{k_{2}}(\tau_{1}) u'^{*}_{k_{2}}(\tau_{1})\int_{-\infty}^{\tau_{1}} d \tau_{2} a^{3} c_{1} \dot{\theta}_{0} v^{*}_{k_{3}}(\tau_{2}) u'^{*}_{k_{3}}(\tau_{2})\bigg{]}(2\pi)^{3} \delta ^{3}(\sum_{i}k_{i}) + 5 perm.
\end{align}
\begin{align}
(2)=&-12 u^{*}_{k_{1}}(0)u_{k_{2}}(0)u_{k_{3}}(0) \nonumber\\
& \times {\rm Re} \bigg{[}\int_{-\infty}^{0} d \widetilde {\tau}_{1} \frac{\lambda_{1}}{6} a^{4} v^{*}_{k_{1}}(\widetilde {\tau}_{1}) v_{k_{2}}(\widetilde {\tau}_{1}) v_{k_{3}}(\widetilde {\tau}_{1}) \int_{-\infty}^{\widetilde {\tau}_{1}} d \widetilde {\tau}_{2} a^{3} c_{1} \dot{\theta}_{0} v_{k_{1}}(\widetilde {\tau}_{2}) u'_{k_{1}}(\widetilde {\tau}_{2}) \nonumber\\
& \times \int_{-\infty}^{0} d \tau_{1} a^{3} c_{1} \dot{\theta}_{0} v^{*}_{k_{2}}(\tau_{1}) u'^{*}_{k_{2}}(\tau_{1})\int_{-\infty}^{\tau_{1}} d \tau_{2} a^{3} c_{1} \dot{\theta}_{0} v^{*}_{k_{3}}(\tau_{2}) u'^{*}_{k_{3}}(\tau_{2})\bigg{]}(2\pi)^{3} \delta ^{3}(\sum_{i}k_{i}) + 5 perm.
\end{align}

Different terms related to Eq.(\ref{factorized1}) and $H^{3}_{3}$ are:
\begin{align}
(3)=&-4 u^{*}_{k_{1}}(0)u_{k_{2}}(0)u_{k_{3}}(0) \nonumber\\
& \times {\rm Re} \bigg{[}\int_{-\infty}^{0} d \widetilde {\tau}_{1} a^{3} c_{1} \dot{\theta}_{0} v^{*}_{k_{1}}(\widetilde {\tau}_{1}) u'_{k_{1}}(\widetilde {\tau}_{1}) \int_{-\infty}^{\widetilde {\tau}_{1}} d \widetilde {\tau}_{2} \frac{\lambda_{3}}{2} a^{4} v_{k_{1}}(\widetilde {\tau}_{2}) w_{k_{2}}(\widetilde {\tau}_{2}) w_{k_{3}}(\widetilde {\tau}_{2}) \nonumber\\
& \times \int_{-\infty}^{0} d \tau_{1} a^{3} c_{2} \dot{\theta}_{0} w^{*}_{k_{2}}(\tau_{1}) u'^{*}_{k_{2}}(\tau_{1})\int_{-\infty}^{\tau_{1}} d \tau_{2} a^{3} c_{2} \dot{\theta}_{0} w^{*}_{k_{3}}(\tau_{2}) u'^{*}_{k_{3}}(\tau_{2})\bigg{]}(2\pi)^{3} \delta ^{3}(\sum_{i}k_{i}) + 5 perm.
\end{align}
\begin{align}
(4)=&-4 u_{k_{1}}(0)u^{*}_{k_{2}}(0)u_{k_{3}}(0) \nonumber\\
& \times {\rm Re} \bigg{[}\int_{-\infty}^{0} d \widetilde {\tau}_{1} a^{3} c_{2} \dot{\theta}_{0} w^{*}_{k_{2}}(\widetilde {\tau}_{1}) u'_{k_{2}}(\widetilde {\tau}_{1}) \int_{-\infty}^{\widetilde {\tau}_{1}} d \widetilde {\tau}_{2} \frac{\lambda_{3}}{2} a^{4} v_{k_{1}}(\widetilde {\tau}_{2}) w_{k_{2}}(\widetilde {\tau}_{2}) w_{k_{3}}(\widetilde {\tau}_{2}) \nonumber\\
& \times \int_{-\infty}^{0} d \tau_{1} a^{3} c_{1} \dot{\theta}_{0} v^{*}_{k_{1}}(\tau_{1}) u'^{*}_{k_{1}}(\tau_{1})\int_{-\infty}^{\tau_{1}} d \tau_{2} a^{3} c_{2} \dot{\theta}_{0} w^{*}_{k_{3}}(\tau_{2}) u'^{*}_{k_{3}}(\tau_{2})\bigg{]}(2\pi)^{3} \delta ^{3}(\sum_{i}k_{i}) + 5 perm.
\end{align}
\begin{align}
(5)=&-4 u_{k_{1}}(0)u^{*}_{k_{2}}(0)u_{k_{3}}(0) \nonumber\\
& \times {\rm Re} \bigg{[}\int_{-\infty}^{0} d \widetilde {\tau}_{1} a^{3} c_{2} \dot{\theta}_{0} w^{*}_{k_{2}}(\widetilde {\tau}_{1}) u'_{k_{2}}(\widetilde {\tau}_{1}) \int_{-\infty}^{\widetilde {\tau}_{1}} d \widetilde {\tau}_{2} \frac{\lambda_{3}}{2} a^{4} v_{k_{1}}(\widetilde {\tau}_{2}) w_{k_{2}}(\widetilde {\tau}_{2}) w_{k_{3}}(\widetilde {\tau}_{2}) \nonumber\\
& \times \int_{-\infty}^{0} d \tau_{1} a^{3} c_{2} \dot{\theta}_{0} w^{*}_{k_{3}}(\tau_{1}) u'^{*}_{k_{3}}(\tau_{1})\int_{-\infty}^{\tau_{1}} d \tau_{2} a^{3} c_{1} \dot{\theta}_{0} v^{*}_{k_{1}}(\tau_{2}) u'^{*}_{k_{1}}(\tau_{2})\bigg{]}(2\pi)^{3} \delta ^{3}(\sum_{i}k_{i}) + 5 perm.
\end{align}
\begin{align}
(6)=&-4 u^{*}_{k_{1}}(0)u_{k_{2}}(0)u_{k_{3}}(0) \nonumber\\
& \times {\rm Re} \bigg{[}\int_{-\infty}^{0} d \widetilde {\tau}_{1} \frac{\lambda_{3}}{2} a^{4} v^{*}_{k_{1}}(\widetilde {\tau}_{1}) w_{k_{2}}(\widetilde {\tau}_{1}) w_{k_{3}}(\widetilde {\tau}_{1}) \int_{-\infty}^{\widetilde {\tau}_{1}} d \widetilde {\tau}_{2} a^{3} c_{1} \dot{\theta}_{0} v_{k_{1}}(\widetilde {\tau}_{2}) u'_{k_{1}}(\widetilde {\tau}_{2}) \nonumber\\
& \times \int_{-\infty}^{0} d \tau_{1} a^{3} c_{2} \dot{\theta}_{0} w^{*}_{k_{2}}(\tau_{1}) u'^{*}_{k_{2}}(\tau_{1})\int_{-\infty}^{\tau_{1}} d \tau_{2} a^{3} c_{2} \dot{\theta}_{0} w^{*}_{k_{3}}(\tau_{2}) u'^{*}_{k_{3}}(\tau_{2})\bigg{]}(2\pi)^{3} \delta ^{3}(\sum_{i}k_{i}) + 5 perm.
\end{align}
\begin{align}
(7)=&-4 u_{k_{1}}(0)u^{*}_{k_{2}}(0)u_{k_{3}}(0) \nonumber\\
& \times {\rm Re} \bigg{[}\int_{-\infty}^{0} d \widetilde {\tau}_{1} \frac{\lambda_{3}}{2} a^{4} v_{k_{1}}(\widetilde {\tau}_{1}) w^{*}_{k_{2}}(\widetilde {\tau}_{1}) w_{k_{3}}(\widetilde {\tau}_{1}) \int_{-\infty}^{\widetilde {\tau}_{1}} d \widetilde {\tau}_{2} a^{3} c_{2} \dot{\theta}_{0} w_{k_{2}}(\widetilde {\tau}_{2}) u'_{k_{2}}(\widetilde {\tau}_{2}) \nonumber\\
& \times \int_{-\infty}^{0} d \tau_{1} a^{3} c_{1} \dot{\theta}_{0} v^{*}_{k_{1}}(\tau_{1}) u'^{*}_{k_{1}}(\tau_{1})\int_{-\infty}^{\tau_{1}} d \tau_{2} a^{3} c_{2} \dot{\theta}_{0} w^{*}_{k_{3}}(\tau_{2}) u'^{*}_{k_{3}}(\tau_{2})\bigg{]}(2\pi)^{3} \delta ^{3}(\sum_{i}k_{i}) + 5 perm.
\end{align}
\begin{align}
(8)=&-4 u_{k_{1}}(0)u^{*}_{k_{2}}(0)u_{k_{3}}(0) \nonumber\\
& \times {\rm Re} \bigg{[}\int_{-\infty}^{0} d \widetilde {\tau}_{1} \frac{\lambda_{3}}{2} a^{4} v_{k_{1}}(\widetilde {\tau}_{1}) w^{*}_{k_{2}}(\widetilde {\tau}_{1}) w_{k_{3}}(\widetilde {\tau}_{1}) \int_{-\infty}^{\widetilde {\tau}_{1}} d \widetilde {\tau}_{2} a^{3} c_{2} \dot{\theta}_{0} w_{k_{2}}(\widetilde {\tau}_{2}) u'_{k_{2}}(\widetilde {\tau}_{2}) \nonumber\\
& \times \int_{-\infty}^{0} d \tau_{1} a^{3} c_{2} \dot{\theta}_{0} w^{*}_{k_{3}}(\tau_{1}) u'^{*}_{k_{3}}(\tau_{1})\int_{-\infty}^{\tau_{1}} d \tau_{2} a^{3} c_{1} \dot{\theta}_{0} v^{*}_{k_{1}}(\tau_{2}) u'^{*}_{k_{1}}(\tau_{2})\bigg{]}(2\pi)^{3} \delta ^{3}(\sum_{i}k_{i}) + 5 perm.
\end{align}

Different terms related to Eq.(\ref{factorized2}) and $H^{3}_{1}$ are:
\begin{align}
(9)=&12 u_{k_{1}}(0)u_{k_{2}}(0)u_{k_{3}}(0) \nonumber\\
& \times {\rm Re} \bigg{[}\int_{-\infty}^{0} d \widetilde {\tau}_{1} \frac{\lambda_{1}}{6} a^{4} v_{k_{1}}(\widetilde {\tau}_{1}) v_{k_{2}}(\widetilde {\tau}_{1}) v_{k_{3}}(\widetilde {\tau}_{1}) \int_{-\infty}^{0} d\tau_{1} a^{3} c_{1} \dot{\theta}_{0} v^{*}_{k_{1}}(\tau_{1}) u'^{*}_{k_{1}}(\tau_{1}) \nonumber\\
& \times \int_{-\infty}^{\tau_{1}} d \tau_{2} a^{3} c_{1} \dot{\theta}_{0} v^{*}_{k_{2}}(\tau_{2}) u'^{*}_{k_{2}}(\tau_{2})\int_{-\infty}^{\tau_{2}} d \tau_{3} a^{3} c_{1} \dot{\theta}_{0} v^{*}_{k_{3}}(\tau_{3}) u'^{*}_{k_{3}}(\tau_{3})\bigg{]}(2\pi)^{3} \delta ^{3}(\sum_{i}k_{i}) + 5 perm.
\end{align}
\begin{align}
(10)=&12 u^{*}_{k_{1}}(0)u_{k_{2}}(0)u_{k_{3}}(0) \nonumber\\
& \times {\rm Re} \bigg{[}\int_{-\infty}^{0} d \widetilde {\tau}_{1} a^{3} c_{1} \dot{\theta}_{0} v_{k_{1}}(\widetilde {\tau}_{1}) u'_{k_{1}}(\widetilde {\tau}_{1}) \int_{-\infty}^{0} d\tau_{1} \frac{\lambda_{1}}{6} a^{4} v^{*}_{k_{1}}(\tau_{1}) v_{k_{2}}(\tau_{1}) v_{k_{3}}(\tau_{1}) \nonumber\\
& \times \int_{-\infty}^{\tau_{1}} d \tau_{2} a^{3} c_{1} \dot{\theta}_{0} v^{*}_{k_{2}}(\tau_{2}) u'^{*}_{k_{2}}(\tau_{2})\int_{-\infty}^{\tau_{2}} d \tau_{3} a^{3} c_{1} \dot{\theta}_{0} v^{*}_{k_{3}}(\tau_{3}) u'^{*}_{k_{3}}(\tau_{3})\bigg{]}(2\pi)^{3} \delta ^{3}(\sum_{i}k_{i}) + 5 perm.
\end{align}
\begin{align}
(11)=&12 u^{*}_{k_{1}}(0)u_{k_{2}}(0)u_{k_{3}}(0) \nonumber\\
& \times {\rm Re} \bigg{[}\int_{-\infty}^{0} d \widetilde {\tau}_{1} a^{3} c_{1} \dot{\theta}_{0} v_{k_{1}}(\widetilde {\tau}_{1}) u'_{k_{1}}(\widetilde {\tau}_{1}) \int_{-\infty}^{0} d\tau_{1}a^{3} c_{1} \dot{\theta}_{0} v_{k_{2}}(\tau_{1}) u'^{*}_{k_{2}}(\tau_{1})\nonumber\\
& \times \int_{-\infty}^{\tau_{1}} d \tau_{2} \frac{\lambda_{1}}{6} a^{4} v^{*}_{k_{1}}(\tau_{2}) v^{*}_{k_{2}}(\tau_{2}) v_{k_{3}}(\tau_{2}) \int_{-\infty}^{\tau_{2}} d \tau_{3} a^{3} c_{1} \dot{\theta}_{0} v^{*}_{k_{3}}(\tau_{3}) u'^{*}_{k_{3}}(\tau_{3})\bigg{]}(2\pi)^{3} \delta ^{3}(\sum_{i}k_{i}) + 5 perm.
\end{align}
\begin{align}
(12)=&12 u^{*}_{k_{1}}(0)u_{k_{2}}(0)u_{k_{3}}(0) \nonumber\\
& \times {\rm Re} \bigg{[}\int_{-\infty}^{0} d \widetilde {\tau}_{1} a^{3} c_{1} \dot{\theta}_{0} v_{k_{1}}(\widetilde {\tau}_{1}) u'_{k_{1}}(\widetilde {\tau}_{1}) \int_{-\infty}^{0} d\tau_{1}a^{3} c_{1} \dot{\theta}_{0} v_{k_{2}}(\tau_{1}) u'^{*}_{k_{2}}(\tau_{1})\nonumber\\
& \times \int_{-\infty}^{\tau_{1}} d \tau_{2} a^{3} c_{1} \dot{\theta}_{0} v_{k_{3}}(\tau_{2}) u'^{*}_{k_{3}}(\tau_{2})
\int_{-\infty}^{\tau_{2}} d \tau_{3} \frac{\lambda_{1}}{6} a^{4} v^{*}_{k_{1}}(\tau_{3}) v^{*}_{k_{2}}(\tau_{3}) v^{*}_{k_{3}}(\tau_{3}) \bigg{]}(2\pi)^{3} \delta ^{3}(\sum_{i}k_{i}) + 5 perm.
\end{align}

Different terms related to Eq.(\ref{factorized2}) and $H^{3}_{3}$ are:
\begin{align}
(13)=&4 u_{k_{1}}(0)u_{k_{2}}(0)u_{k_{3}}(0) \nonumber\\
& \times {\rm Re} \bigg{[}\int_{-\infty}^{0} d \widetilde {\tau}_{1} \frac{\lambda_{3}}{2} a^{4} v_{k_{1}}(\widetilde {\tau}_{1}) w_{k_{2}}(\widetilde {\tau}_{1}) w_{k_{3}}(\widetilde {\tau}_{1}) \int_{-\infty}^{0} d\tau_{1} a^{3} c_{1} \dot{\theta}_{0} v^{*}_{k_{1}}(\tau_{1}) u'^{*}_{k_{1}}(\tau_{1}) \nonumber\\
& \times \int_{-\infty}^{\tau_{1}} d \tau_{2} a^{3} c_{2} \dot{\theta}_{0} w^{*}_{k_{2}}(\tau_{2}) u'^{*}_{k_{2}}(\tau_{2})\int_{-\infty}^{\tau_{2}} d \tau_{3} a^{3} c_{2} \dot{\theta}_{0} w^{*}_{k_{3}}(\tau_{3}) u'^{*}_{k_{3}}(\tau_{3})\bigg{]}(2\pi)^{3} \delta ^{3}(\sum_{i}k_{i}) + 5 perm.
\end{align}
\begin{align}
(14)=&4 u_{k_{1}}(0)u_{k_{2}}(0)u_{k_{3}}(0) \nonumber\\
& \times {\rm Re} \bigg{[}\int_{-\infty}^{0} d \widetilde {\tau}_{1} \frac{\lambda_{3}}{2} a^{4} v_{k_{1}}(\widetilde {\tau}_{1}) w_{k_{2}}(\widetilde {\tau}_{1}) w_{k_{3}}(\widetilde {\tau}_{1}) \int_{-\infty}^{0} d\tau_{1} a^{3} c_{2} \dot{\theta}_{0} w^{*}_{k_{2}}(\tau_{1}) u'^{*}_{k_{2}}(\tau_{1}) \nonumber\\
& \times \int_{-\infty}^{\tau_{1}} d \tau_{2} a^{3} c_{1} \dot{\theta}_{0} v^{*}_{k_{1}}(\tau_{2}) u'^{*}_{k_{1}}(\tau_{2})\int_{-\infty}^{\tau_{2}} d \tau_{3} a^{3} c_{2} \dot{\theta}_{0} w^{*}_{k_{3}}(\tau_{3}) u'^{*}_{k_{3}}(\tau_{3})\bigg{]}(2\pi)^{3} \delta ^{3}(\sum_{i}k_{i}) + 5 perm.
\end{align}
\begin{align}
(15)=&4 u_{k_{1}}(0)u_{k_{2}}(0)u_{k_{3}}(0) \nonumber\\
& \times {\rm Re} \bigg{[}\int_{-\infty}^{0} d \widetilde {\tau}_{1} \frac{\lambda_{3}}{2} a^{4} v_{k_{1}}(\widetilde {\tau}_{1}) w_{k_{2}}(\widetilde {\tau}_{1}) w_{k_{3}}(\widetilde {\tau}_{1}) \int_{-\infty}^{0} d\tau_{1} a^{3} c_{2} \dot{\theta}_{0} w^{*}_{k_{2}}(\tau_{1}) u'^{*}_{k_{2}}(\tau_{1}) \nonumber\\
& \times \int_{-\infty}^{\tau_{1}} d \tau_{2} a^{3} c_{2} \dot{\theta}_{0} w^{*}_{k_{3}}(\tau_{2}) u'^{*}_{k_{3}}(\tau_{2})\int_{-\infty}^{\tau_{2}} d \tau_{3} a^{3} c_{1} \dot{\theta}_{0} v^{*}_{k_{1}}(\tau_{3}) u'^{*}_{k_{1}}(\tau_{3})\bigg{]}(2\pi)^{3} \delta ^{3}(\sum_{i}k_{i}) + 5 perm.
\end{align}
\begin{align}
(16)=&4 u^{*}_{k_{1}}(0)u_{k_{2}}(0)u_{k_{3}}(0) \nonumber\\
& \times {\rm Re} \bigg{[}\int_{-\infty}^{0} d \widetilde {\tau}_{1} a^{3} c_{1} \dot{\theta}_{0} v_{k_{1}}(\widetilde {\tau}_{1}) u'_{k_{1}}(\widetilde {\tau}_{1}) \int_{-\infty}^{0} d\tau_{1} \frac{\lambda_{3}}{2} a^{4} v^{*}_{k_{1}}(\tau_{1}) w_{k_{2}}(\tau_{1}) w_{k_{3}}(\tau_{1}) \nonumber\\
& \times \int_{-\infty}^{\tau_{1}} d \tau_{2} a^{3} c_{2} \dot{\theta}_{0} w^{*}_{k_{2}}(\tau_{2}) u'^{*}_{k_{2}}(\tau_{2})\int_{-\infty}^{\tau_{2}} d \tau_{3} a^{3} c_{2} \dot{\theta}_{0} w^{*}_{k_{3}}(\tau_{3}) u'^{*}_{k_{3}}(\tau_{3})\bigg{]}(2\pi)^{3} \delta ^{3}(\sum_{i}k_{i}) + 5 perm.
\end{align}
\begin{align}
(17)=&4 u_{k_{1}}(0)u^{*}_{k_{2}}(0)u_{k_{3}}(0) \nonumber\\
& \times {\rm Re} \bigg{[}\int_{-\infty}^{0} d \widetilde {\tau}_{1} a^{3} c_{2} \dot{\theta}_{0} w_{k_{2}}(\widetilde {\tau}_{1}) u'_{k_{2}}(\widetilde {\tau}_{1}) \int_{-\infty}^{0} d\tau_{1} \frac{\lambda_{3}}{2} a^{4} v_{k_{1}}(\tau_{1}) w^{*}_{k_{2}}(\tau_{1}) w_{k_{3}}(\tau_{1}) \nonumber\\
& \times \int_{-\infty}^{\tau_{1}} d \tau_{2} a^{3} c_{1} \dot{\theta}_{0} v^{*}_{k_{1}}(\tau_{2}) u'^{*}_{k_{1}}(\tau_{2})\int_{-\infty}^{\tau_{2}} d \tau_{3} a^{3} c_{2} \dot{\theta}_{0} w^{*}_{k_{3}}(\tau_{3}) u'^{*}_{k_{3}}(\tau_{3})\bigg{]}(2\pi)^{3} \delta ^{3}(\sum_{i}k_{i}) + 5 perm.
\end{align}
\begin{align}
(18)=&4 u_{k_{1}}(0)u^{*}_{k_{2}}(0)u_{k_{3}}(0) \nonumber\\
& \times {\rm Re} \bigg{[}\int_{-\infty}^{0} d \widetilde {\tau}_{1} a^{3} c_{2} \dot{\theta}_{0} w_{k_{2}}(\widetilde {\tau}_{1}) u'_{k_{2}}(\widetilde {\tau}_{1}) \int_{-\infty}^{0} d\tau_{1} \frac{\lambda_{3}}{2} a^{4} v_{k_{1}}(\tau_{1}) w^{*}_{k_{2}}(\tau_{1}) w_{k_{3}}(\tau_{1}) \nonumber\\
& \times \int_{-\infty}^{\tau_{1}} d \tau_{2} a^{3} c_{2} \dot{\theta}_{0} w^{*}_{k_{3}}(\tau_{2}) u'^{*}_{k_{3}}(\tau_{2})\int_{-\infty}^{\tau_{2}} d \tau_{3} a^{3} c_{1} \dot{\theta}_{0} v^{*}_{k_{1}}(\tau_{3}) u'^{*}_{k_{1}}(\tau_{3})\bigg{]}(2\pi)^{3} \delta ^{3}(\sum_{i}k_{i}) + 5 perm.
\end{align}
\begin{align}
(19)=&4 u^{*}_{k_{1}}(0)u_{k_{2}}(0)u_{k_{3}}(0) \nonumber\\
& \times {\rm Re} \bigg{[}\int_{-\infty}^{0} d \widetilde {\tau}_{1} a^{3} c_{1} \dot{\theta}_{0} v_{k_{1}}(\widetilde {\tau}_{1}) u'_{k_{1}}(\widetilde {\tau}_{1}) \int_{-\infty}^{0} d\tau_{1}a^{3} c_{2} \dot{\theta}_{0} w_{k_{2}}(\tau_{1}) u'^{*}_{k_{2}}(\tau_{1})\nonumber\\
& \times \int_{-\infty}^{\tau_{1}} d \tau_{2} \frac{\lambda_{3}}{2} a^{4} v^{*}_{k_{1}}(\tau_{2}) w^{*}_{k_{2}}(\tau_{2}) w_{k_{3}}(\tau_{2}) \int_{-\infty}^{\tau_{2}} d \tau_{3} a^{3} c_{2} \dot{\theta}_{0} w^{*}_{k_{3}}(\tau_{3}) u'^{*}_{k_{3}}(\tau_{3})\bigg{]}(2\pi)^{3} \delta^{3}(\sum_{i}k_{i}) + 5 perm.
\end{align}
\begin{align}
(20)=&4 u_{k_{1}}(0)u^{*}_{k_{2}}(0)u_{k_{3}}(0) \nonumber\\
& \times {\rm Re} \bigg{[}\int_{-\infty}^{0} d \widetilde {\tau}_{1} a^{3} c_{2} \dot{\theta}_{0} w_{k_{2}}(\widetilde {\tau}_{1}) u'_{k_{2}}(\widetilde {\tau}_{1}) \int_{-\infty}^{0} d\tau_{1}a^{3} c_{1} \dot{\theta}_{0} v_{k_{1}}(\tau_{1}) u'^{*}_{k_{1}}(\tau_{1})\nonumber\\
& \times \int_{-\infty}^{\tau_{1}} d \tau_{2} \frac{\lambda_{3}}{2} a^{4} v^{*}_{k_{1}}(\tau_{2}) w^{*}_{k_{2}}(\tau_{2}) w_{k_{3}}(\tau_{2}) \int_{-\infty}^{\tau_{2}} d \tau_{3} a^{3} c_{2} \dot{\theta}_{0} w^{*}_{k_{3}}(\tau_{3}) u'^{*}_{k_{3}}(\tau_{3})\bigg{]}(2\pi)^{3} \delta^{3}(\sum_{i}k_{i}) + 5 perm.
\end{align}
\begin{align}
(21)=&4 u_{k_{1}}(0)u^{*}_{k_{2}}(0)u_{k_{3}}(0) \nonumber\\
& \times {\rm Re} \bigg{[}\int_{-\infty}^{0} d \widetilde {\tau}_{1} a^{3} c_{2} \dot{\theta}_{0} w_{k_{2}}(\widetilde {\tau}_{1}) u'_{k_{2}}(\widetilde {\tau}_{1}) \int_{-\infty}^{0} d\tau_{1}a^{3} c_{2} \dot{\theta}_{0} w_{k_{3}}(\tau_{1}) u'^{*}_{k_{3}}(\tau_{1})\nonumber\\
& \times \int_{-\infty}^{\tau_{1}} d \tau_{2} \frac{\lambda_{3}}{2} a^{4} v_{k_{1}}(\tau_{2}) w^{*}_{k_{2}}(\tau_{2}) w^{*}_{k_{3}}(\tau_{2}) \int_{-\infty}^{\tau_{2}} d \tau_{3} a^{3} c_{1} \dot{\theta}_{0} v^{*}_{k_{1}}(\tau_{3}) u'^{*}_{k_{1}}(\tau_{3})\bigg{]}(2\pi)^{3} \delta^{3}(\sum_{i}k_{i}) + 5 perm.
\end{align}
\begin{align}
(22)=&4 u^{*}_{k_{1}}(0)u_{k_{2}}(0)u_{k_{3}}(0) \nonumber\\
& \times {\rm Re} \bigg{[}\int_{-\infty}^{0} d \widetilde {\tau}_{1} a^{3} c_{1} \dot{\theta}_{0} v_{k_{1}}(\widetilde {\tau}_{1}) u'_{k_{1}}(\widetilde {\tau}_{1}) \int_{-\infty}^{0} d\tau_{1}a^{3} c_{2} \dot{\theta}_{0} w_{k_{2}}(\tau_{1}) u'^{*}_{k_{2}}(\tau_{1})\nonumber\\
& \times \int_{-\infty}^{\tau_{1}} d \tau_{2} a^{3} c_{2} \dot{\theta}_{0} w_{k_{3}}(\tau_{2}) u'^{*}_{k_{3}}(\tau_{2})
\int_{-\infty}^{\tau_{2}} d \tau_{3} \frac{\lambda_{3}}{2} a^{4} v^{*}_{k_{1}}(\tau_{3}) w^{*}_{k_{2}}(\tau_{3}) w^{*}_{k_{3}}(\tau_{3}) \bigg{]}(2\pi)^{3} \delta ^{3}(\sum_{i}k_{i}) + 5 perm.
\end{align}
\begin{align}
(23)=&4 u_{k_{1}}(0)u^{*}_{k_{2}}(0)u_{k_{3}}(0) \nonumber\\
& \times {\rm Re} \bigg{[}\int_{-\infty}^{0} d \widetilde {\tau}_{1} a^{3} c_{2} \dot{\theta}_{0} w_{k_{2}}(\widetilde {\tau}_{1}) u'_{k_{2}}(\widetilde {\tau}_{1}) \int_{-\infty}^{0} d\tau_{1}a^{3} c_{1} \dot{\theta}_{0} v_{k_{1}}(\tau_{1}) u'^{*}_{k_{1}}(\tau_{1})\nonumber\\
& \times \int_{-\infty}^{\tau_{1}} d \tau_{2} a^{3} c_{2} \dot{\theta}_{0} w_{k_{3}}(\tau_{2}) u'^{*}_{k_{3}}(\tau_{2})
\int_{-\infty}^{\tau_{2}} d \tau_{3} \frac{\lambda_{3}}{2} a^{4} v^{*}_{k_{1}}(\tau_{3}) w^{*}_{k_{2}}(\tau_{3}) w^{*}_{k_{3}}(\tau_{3}) \bigg{]}(2\pi)^{3} \delta ^{3}(\sum_{i}k_{i}) + 5 perm.
\end{align}
\begin{align}
(24)=&4 u_{k_{1}}(0)u^{*}_{k_{2}}(0)u_{k_{3}}(0) \nonumber\\
& \times {\rm Re} \bigg{[}\int_{-\infty}^{0} d \widetilde {\tau}_{1} a^{3} c_{2} \dot{\theta}_{0} w_{k_{2}}(\widetilde {\tau}_{1}) u'_{k_{2}}(\widetilde {\tau}_{1}) \int_{-\infty}^{0} d\tau_{1}a^{3} c_{2} \dot{\theta}_{0} w_{k_{3}}(\tau_{1}) u'^{*}_{k_{3}}(\tau_{1})\nonumber\\
& \times \int_{-\infty}^{\tau_{1}} d \tau_{2} a^{3} c_{1} \dot{\theta}_{0} v_{k_{1}}(\tau_{2}) u'^{*}_{k_{1}}(\tau_{2})
\int_{-\infty}^{\tau_{2}} d \tau_{3} \frac{\lambda_{3}}{2} a^{4} v^{*}_{k_{1}}(\tau_{3}) w^{*}_{k_{2}}(\tau_{3}) w^{*}_{k_{3}}(\tau_{3}) \bigg{]}(2\pi)^{3} \delta ^{3}(\sum_{i}k_{i}) + 5 perm.
\end{align}

Different terms related to Eq.(\ref{factorized3}) and $H^{3}_{1}$ are:
\begin{align}
(25)=&-12 u_{k_{1}}(0)u_{k_{2}}(0)u_{k_{3}}(0) \nonumber\\
& \times {\rm Re} \bigg{[}\int_{-\infty}^{0} d \tau_{1} \frac{\lambda_{1}}{6} a^{4} v_{k_{1}}(\tau_{1}) v_{k_{2}}(\tau_{1}) v_{k_{3}}(\tau_{1}) \int_{-\infty}^{\tau_{1}} d\tau_{2} a^{3} c_{1} \dot{\theta}_{0} v^{*}_{k_{1}}(\tau_{2}) u'^{*}_{k_{1}}(\tau_{2}) \nonumber\\
& \times \int_{-\infty}^{\tau_{2}} d \tau_{3} a^{3} c_{1} \dot{\theta}_{0} v^{*}_{k_{2}}(\tau_{3}) u'^{*}_{k_{2}}(\tau_{3})\int_{-\infty}^{\tau_{3}} d \tau_{4} a^{3} c_{1} \dot{\theta}_{0} v^{*}_{k_{3}}(\tau_{4}) u'^{*}_{k_{3}}(\tau_{4})\bigg{]}(2\pi)^{3} \delta ^{3}(\sum_{i}k_{i}) + 5 perm.
\end{align}
\begin{align}
(26)=&-12 u_{k_{1}}(0)u_{k_{2}}(0)u_{k_{3}}(0) \nonumber\\
& \times {\rm Re} \bigg{[}\int_{-\infty}^{0} d \tau_{1} a^{3} c_{1} \dot{\theta}_{0} v_{k_{1}}(\tau_{1}) u'^{*}_{k_{1}}(\tau_{1}) \int_{-\infty}^{\tau_{1}} d\tau_{2} \frac{\lambda_{1}}{6} a^{4} v^{*}_{k_{1}}(\tau_{2}) v_{k_{2}}(\tau_{2}) v_{k_{3}}(\tau_{2}) \nonumber\\
& \times \int_{-\infty}^{\tau_{2}} d \tau_{3} a^{3} c_{1} \dot{\theta}_{0} v^{*}_{k_{2}}(\tau_{3}) u'^{*}_{k_{2}}(\tau_{3})\int_{-\infty}^{\tau_{3}} d \tau_{4} a^{3} c_{1} \dot{\theta}_{0} v^{*}_{k_{3}}(\tau_{4}) u'^{*}_{k_{3}}(\tau_{4})\bigg{]}(2\pi)^{3} \delta ^{3}(\sum_{i}k_{i}) + 5 perm.
\end{align}
\begin{align}
(27)=&-12 u_{k_{1}}(0)u_{k_{2}}(0)u_{k_{3}}(0) \nonumber\\
& \times {\rm Re} \bigg{[}\int_{-\infty}^{0} d \tau_{1} a^{3} c_{1} \dot{\theta}_{0} v_{k_{1}}(\tau_{1}) u'^{*}_{k_{1}}(\tau_{1}) \int_{-\infty}^{\tau_{1}} d\tau_{2}a^{3} c_{1} \dot{\theta}_{0} v_{k_{2}}(\tau_{2}) u'^{*}_{k_{2}}(\tau_{2})\nonumber\\
& \times \int_{-\infty}^{\tau_{2}} d \tau_{2} \frac{\lambda_{1}}{6} a^{4} v^{*}_{k_{1}}(\tau_{3}) v^{*}_{k_{2}}(\tau_{3}) v_{k_{3}}(\tau_{3}) \int_{-\infty}^{\tau_{3}} d \tau_{4} a^{3} c_{1} \dot{\theta}_{0} v^{*}_{k_{3}}(\tau_{4}) u'^{*}_{k_{3}}(\tau_{4})\bigg{]}(2\pi)^{3} \delta ^{3}(\sum_{i}k_{i}) + 5 perm.
\end{align}
\begin{align}
(28)=&-12 u_{k_{1}}(0)u_{k_{2}}(0)u_{k_{3}}(0) \nonumber\\
& \times {\rm Re} \bigg{[}\int_{-\infty}^{0} d \tau_{1} a^{3} c_{1} \dot{\theta}_{0} v_{k_{1}}(\tau_{1}) u'^{*}_{k_{1}}(\tau_{1}) \int_{-\infty}^{\tau_{1}} d\tau_{2}a^{3} c_{1} \dot{\theta}_{0} v_{k_{2}}(\tau_{2}) u'^{*}_{k_{2}}(\tau_{2})\nonumber\\
& \times \int_{-\infty}^{\tau_{2}} d \tau_{3} a^{3} c_{1} \dot{\theta}_{0} v_{k_{3}}(\tau_{3}) u'^{*}_{k_{3}}(\tau_{3})
\int_{-\infty}^{\tau_{3}} d \tau_{4} \frac{\lambda_{1}}{6} a^{4} v^{*}_{k_{1}}(\tau_{4}) v^{*}_{k_{2}}(\tau_{4}) v^{*}_{k_{3}}(\tau_{4}) \bigg{]}(2\pi)^{3} \delta ^{3}(\sum_{i}k_{i}) + 5 perm.
\end{align}

Different terms related to Eq.(\ref{factorized3}) and $H^{3}_{3}$ are:
\begin{align}
(29)=&-4 u_{k_{1}}(0)u_{k_{2}}(0)u_{k_{3}}(0) \nonumber\\
& \times {\rm Re} \bigg{[}\int_{-\infty}^{0} d \tau_{1} \frac{\lambda_{3}}{2} a^{4} v_{k_{1}}(\tau_{1}) w_{k_{2}}(\tau_{1}) w_{k_{3}}(\tau_{1}) \int_{-\infty}^{\tau_{1}} d\tau_{2} a^{3} c_{1} \dot{\theta}_{0} v^{*}_{k_{1}}(\tau_{2}) u'^{*}_{k_{1}}(\tau_{2}) \nonumber\\
& \times \int_{-\infty}^{\tau_{2}} d \tau_{3} a^{3} c_{2} \dot{\theta}_{0} w^{*}_{k_{2}}(\tau_{3}) u'^{*}_{k_{2}}(\tau_{3})\int_{-\infty}^{\tau_{3}} d \tau_{4} a^{3} c_{2} \dot{\theta}_{0} w^{*}_{k_{3}}(\tau_{4}) u'^{*}_{k_{3}}(\tau_{4})\bigg{]}(2\pi)^{3} \delta ^{3}(\sum_{i}k_{i}) + 5 perm.
\end{align}
\begin{align}
(30)=&-4 u_{k_{1}}(0)u_{k_{2}}(0)u_{k_{3}}(0) \nonumber\\
& \times {\rm Re} \bigg{[}\int_{-\infty}^{0} d \tau_{1} \frac{\lambda_{3}}{2} a^{4} v_{k_{1}}(\tau_{1}) w_{k_{2}}(\tau_{1}) w_{k_{3}}(\tau_{1}) \int_{-\infty}^{\tau_{1}} d\tau_{2} a^{3} c_{2} \dot{\theta}_{0} w^{*}_{k_{2}}(\tau_{2}) u'^{*}_{k_{2}}(\tau_{2}) \nonumber\\
& \times \int_{-\infty}^{\tau_{2}} d \tau_{3} a^{3} c_{1} \dot{\theta}_{0} v^{*}_{k_{1}}(\tau_{3}) u'^{*}_{k_{1}}(\tau_{3})\int_{-\infty}^{\tau_{3}} d \tau_{4} a^{3} c_{2} \dot{\theta}_{0} w^{*}_{k_{3}}(\tau_{4}) u'^{*}_{k_{3}}(\tau_{4})\bigg{]}(2\pi)^{3} \delta ^{3}(\sum_{i}k_{i}) + 5 perm.
\end{align}
\begin{align}
(31)=&-4 u_{k_{1}}(0)u_{k_{2}}(0)u_{k_{3}}(0) \nonumber\\
& \times {\rm Re} \bigg{[}\int_{-\infty}^{0} d \tau_{1} \frac{\lambda_{3}}{2} a^{4} v_{k_{1}}(\tau_{1}) w_{k_{2}}(\tau_{1}) w_{k_{3}}(\tau_{1}) \int_{-\infty}^{\tau_{1}} d\tau_{2} a^{3} c_{2} \dot{\theta}_{0} w^{*}_{k_{2}}(\tau_{2}) u'^{*}_{k_{2}}(\tau_{2}) \nonumber\\
& \times \int_{-\infty}^{\tau_{2}} d \tau_{3} a^{3} c_{2} \dot{\theta}_{0} w^{*}_{k_{3}}(\tau_{3}) u'^{*}_{k_{3}}(\tau_{3})\int_{-\infty}^{\tau_{3}} d \tau_{4} a^{3} c_{1} \dot{\theta}_{0} v^{*}_{k_{1}}(\tau_{4}) u'^{*}_{k_{1}}(\tau_{4})\bigg{]}(2\pi)^{3} \delta ^{3}(\sum_{i}k_{i}) + 5 perm.
\end{align}
\begin{align}
(32)=&-4 u_{k_{1}}(0)u_{k_{2}}(0)u_{k_{3}}(0) \nonumber\\
& \times {\rm Re} \bigg{[}\int_{-\infty}^{0} d \tau_{1} a^{3} c_{1} \dot{\theta}_{0} v_{k_{1}}(\tau_{1}) u'^{*}_{k_{1}}(\tau_{1}) \int_{-\infty}^{\tau_{1}} d\tau_{2} \frac{\lambda_{3}}{2} a^{4} v^{*}_{k_{1}}(\tau_{2}) w_{k_{2}}(\tau_{2}) w_{k_{3}}(\tau_{2}) \nonumber\\
& \times \int_{-\infty}^{\tau_{2}} d \tau_{3} a^{3} c_{2} \dot{\theta}_{0} w^{*}_{k_{2}}(\tau_{3}) u'^{*}_{k_{2}}(\tau_{3})\int_{-\infty}^{\tau_{3}} d \tau_{4} a^{3} c_{2} \dot{\theta}_{0} w^{*}_{k_{3}}(\tau_{4}) u'^{*}_{k_{3}}(\tau_{4})\bigg{]}(2\pi)^{3} \delta ^{3}(\sum_{i}k_{i}) + 5 perm.
\end{align}
\begin{align}
(33)=&-4 u_{k_{1}}(0)u_{k_{2}}(0)u_{k_{3}}(0) \nonumber\\
& \times {\rm Re} \bigg{[}\int_{-\infty}^{0} d \tau_{1} a^{3} c_{2} \dot{\theta}_{0} w_{k_{2}}(\tau_{1}) u'^{*}_{k_{2}}(\tau_{1}) \int_{-\infty}^{\tau_{1}} d\tau_{2} \frac{\lambda_{3}}{2} a^{4} v_{k_{1}}(\tau_{2}) w^{*}_{k_{2}}(\tau_{2}) w_{k_{3}}(\tau_{2}) \nonumber\\
& \times \int_{-\infty}^{\tau_{2}} d \tau_{3} a^{3} c_{1} \dot{\theta}_{0} v^{*}_{k_{1}}(\tau_{3}) u'^{*}_{k_{1}}(\tau_{3})\int_{-\infty}^{\tau_{3}} d \tau_{4} a^{3} c_{2} \dot{\theta}_{0} w^{*}_{k_{3}}(\tau_{4}) u'^{*}_{k_{3}}(\tau_{4})\bigg{]}(2\pi)^{3} \delta ^{3}(\sum_{i}k_{i}) + 5 perm.
\end{align}
\begin{align}
(34)=&-4 u_{k_{1}}(0)u_{k_{2}}(0)u_{k_{3}}(0) \nonumber\\
& \times {\rm Re} \bigg{[}\int_{-\infty}^{0} d \tau_{1} a^{3} c_{2} \dot{\theta}_{0} w_{k_{2}}(\tau_{1}) u'^{*}_{k_{2}}(\tau_{1}) \int_{-\infty}^{\tau_{1}} d\tau_{2} \frac{\lambda_{3}}{2} a^{4} v_{k_{1}}(\tau_{2}) w^{*}_{k_{2}}(\tau_{2}) w_{k_{3}}(\tau_{2}) \nonumber\\
& \times \int_{-\infty}^{\tau_{2}} d \tau_{3} a^{3} c_{2} \dot{\theta}_{0} w^{*}_{k_{3}}(\tau_{3}) u'^{*}_{k_{3}}(\tau_{3})\int_{-\infty}^{\tau_{3}} d \tau_{4} a^{3} c_{1} \dot{\theta}_{0} v^{*}_{k_{1}}(\tau_{4}) u'^{*}_{k_{1}}(\tau_{4})\bigg{]}(2\pi)^{3} \delta ^{3}(\sum_{i}k_{i}) + 5 perm.
\end{align}
\begin{align}
(35)=&-4 u_{k_{1}}(0)u_{k_{2}}(0)u_{k_{3}}(0) \nonumber\\
& \times {\rm Re} \bigg{[}\int_{-\infty}^{0} d \tau_{1} a^{3} c_{1} \dot{\theta}_{0} v_{k_{1}}(\tau_{1}) u'^{*}_{k_{1}}(\tau_{1}) \int_{-\infty}^{\tau_{1}} d\tau_{2}a^{3} c_{2} \dot{\theta}_{0} w_{k_{2}}(\tau_{2}) u'^{*}_{k_{2}}(\tau_{2})\nonumber\\
& \times \int_{-\infty}^{\tau_{2}} d \tau_{2} \frac{\lambda_{3}}{2} a^{4} v^{*}_{k_{1}}(\tau_{3}) w^{*}_{k_{2}}(\tau_{3}) w_{k_{3}}(\tau_{3}) \int_{-\infty}^{\tau_{3}} d \tau_{4} a^{3} c_{2} \dot{\theta}_{0} w^{*}_{k_{3}}(\tau_{4}) u'^{*}_{k_{3}}(\tau_{4})\bigg{]}(2\pi)^{3} \delta ^{3}(\sum_{i}k_{i}) + 5 perm.
\end{align}
\begin{align}
(36)=&-4 u_{k_{1}}(0)u_{k_{2}}(0)u_{k_{3}}(0) \nonumber\\
& \times {\rm Re} \bigg{[}\int_{-\infty}^{0} d \tau_{1} a^{3} c_{2} \dot{\theta}_{0} w_{k_{2}}(\tau_{1}) u'^{*}_{k_{2}}(\tau_{1}) \int_{-\infty}^{\tau_{1}} d\tau_{2}a^{3} c_{1} \dot{\theta}_{0} v_{k_{1}}(\tau_{2}) u'^{*}_{k_{1}}(\tau_{2})\nonumber\\
& \times \int_{-\infty}^{\tau_{2}} d \tau_{2} \frac{\lambda_{3}}{2} a^{4} v^{*}_{k_{1}}(\tau_{3}) w^{*}_{k_{2}}(\tau_{3}) w_{k_{3}}(\tau_{3}) \int_{-\infty}^{\tau_{3}} d \tau_{4} a^{3} c_{2} \dot{\theta}_{0} w^{*}_{k_{3}}(\tau_{4}) u'^{*}_{k_{3}}(\tau_{4})\bigg{]}(2\pi)^{3} \delta ^{3}(\sum_{i}k_{i}) + 5 perm.
\end{align}
\begin{align}
(37)=&-4 u_{k_{1}}(0)u_{k_{2}}(0)u_{k_{3}}(0) \nonumber\\
& \times {\rm Re} \bigg{[}\int_{-\infty}^{0} d \tau_{1} a^{3} c_{2} \dot{\theta}_{0} w_{k_{2}}(\tau_{1}) u'^{*}_{k_{2}}(\tau_{1}) \int_{-\infty}^{\tau_{1}} d\tau_{2}a^{3} c_{2} \dot{\theta}_{0} w_{k_{3}}(\tau_{2}) u'^{*}_{k_{3}}(\tau_{2})\nonumber\\
& \times \int_{-\infty}^{\tau_{2}} d \tau_{2} \frac{\lambda_{3}}{2} a^{4} v_{k_{1}}(\tau_{3}) w^{*}_{k_{2}}(\tau_{3}) w^{*}_{k_{3}}(\tau_{3}) \int_{-\infty}^{\tau_{3}} d \tau_{4} a^{3} c_{1} \dot{\theta}_{0} v^{*}_{k_{1}}(\tau_{4}) u'^{*}_{k_{1}}(\tau_{4})\bigg{]}(2\pi)^{3} \delta ^{3}(\sum_{i}k_{i}) + 5 perm.
\end{align}
\begin{align}
(38)=&-4 u_{k_{1}}(0)u_{k_{2}}(0)u_{k_{3}}(0) \nonumber\\
& \times {\rm Re} \bigg{[}\int_{-\infty}^{0} d \tau_{1} a^{3} c_{1} \dot{\theta}_{0} v_{k_{1}}(\tau_{1}) u'^{*}_{k_{1}}(\tau_{1}) \int_{-\infty}^{\tau_{1}} d\tau_{2}a^{3} c_{2} \dot{\theta}_{0} w_{k_{2}}(\tau_{2}) u'^{*}_{k_{2}}(\tau_{2})\nonumber\\
& \times \int_{-\infty}^{\tau_{2}} d \tau_{3} a^{3} c_{2} \dot{\theta}_{0} w_{k_{3}}(\tau_{3}) u'^{*}_{k_{3}}(\tau_{3})
\int_{-\infty}^{\tau_{3}} d \tau_{4} \frac{\lambda_{3}}{2} a^{4} v^{*}_{k_{1}}(\tau_{4}) w^{*}_{k_{2}}(\tau_{4}) w^{*}_{k_{3}}(\tau_{4}) \bigg{]}(2\pi)^{3} \delta ^{3}(\sum_{i}k_{i}) + 5 perm.
\end{align}
\begin{align}
(39)=&-4 u_{k_{1}}(0)u_{k_{2}}(0)u_{k_{3}}(0) \nonumber\\
& \times {\rm Re} \bigg{[}\int_{-\infty}^{0} d \tau_{1} a^{3} c_{2} \dot{\theta}_{0} w_{k_{2}}(\tau_{1}) u'^{*}_{k_{2}}(\tau_{1}) \int_{-\infty}^{\tau_{1}} d\tau_{2}a^{3} c_{1} \dot{\theta}_{0} v_{k_{1}}(\tau_{2}) u'^{*}_{k_{1}}(\tau_{2})\nonumber\\
& \times \int_{-\infty}^{\tau_{2}} d \tau_{3} a^{3} c_{2} \dot{\theta}_{0} w_{k_{3}}(\tau_{3}) u'^{*}_{k_{3}}(\tau_{3})
\int_{-\infty}^{\tau_{3}} d \tau_{4} \frac{\lambda_{3}}{2} a^{4} v^{*}_{k_{1}}(\tau_{4}) w^{*}_{k_{2}}(\tau_{4}) w^{*}_{k_{3}}(\tau_{4}) \bigg{]}(2\pi)^{3} \delta ^{3}(\sum_{i}k_{i}) + 5 perm.
\end{align}
\begin{align}
(40)=&-4 u_{k_{1}}(0)u_{k_{2}}(0)u_{k_{3}}(0) \nonumber\\
& \times {\rm Re} \bigg{[}\int_{-\infty}^{0} d \tau_{1} a^{3} c_{2} \dot{\theta}_{0} w_{k_{2}}(\tau_{1}) u'^{*}_{k_{2}}(\tau_{1}) \int_{-\infty}^{\tau_{1}} d\tau_{2}a^{3} c_{2} \dot{\theta}_{0} w_{k_{3}}(\tau_{2}) u'^{*}_{k_{3}}(\tau_{2})\nonumber\\
& \times \int_{-\infty}^{\tau_{2}} d \tau_{3} a^{3} c_{1} \dot{\theta}_{0} v_{k_{1}}(\tau_{3}) u'^{*}_{k_{1}}(\tau_{3})
\int_{-\infty}^{\tau_{3}} d \tau_{4} \frac{\lambda_{3}}{2} a^{4} v^{*}_{k_{1}}(\tau_{4}) w^{*}_{k_{2}}(\tau_{4}) w^{*}_{k_{3}}(\tau_{4}) \bigg{]}(2\pi)^{3} \delta ^{3}(\sum_{i}k_{i}) + 5 perm.
\end{align}
The three point function is the sum of the above 40 terms as,
\begin{align}
\langle\zeta^3\rangle = - \left(\frac{H}{\dot \theta_{0}}\right)^3 \sum_{i=1}^{40} \left( i\right)
\end{align}
\subsection{All of terms in the commutator form}\label{AppenB.2}

Here we present the form of the integrals in commutator form. Again we only calculate the independent parts of bispectrum in this form.

Different terms related to Eq. (\ref{three commutator}) are:
\begin{align}
\langle\delta \theta^{3}\rangle &= 2 \dot{\theta}_{0}^{3} u_{k_{1}}(0)u_{k_{2}}(0)u_{k_{3}}(0) \nonumber\\
& \times {\rm Re} \bigg{[} \int_{-\infty}^{0} d \tau_{1}\int_{-\infty}^{\tau_{1}} d\tau_{2}\times \int_{-\infty}^{\tau_{2}} d \tau_{3}\int_{-\infty}^{\tau_{3}} d \tau_{4} \prod_{i=1}^{4}a^{3}(\tau_{i}) \nonumber\\
&\times \left( a(\tau_{2})A  + a(\tau_{3})B + a(\tau_{4})C \right) \bigg{]} (2\pi)^{3} \delta ^{3}(\sum_{i}k_{i}) + 5 perm. \, ,
\end{align}
in which we have defined
\begin{align}
A &\equiv A_{H^{3}_{1}}  + A_{H^{3}_{2}} + A_{H^{3}_{3}} +  A_{H^{3}_{4}} \label{A terms}\\
B&\equiv B_{H^{3}_{1}} + B_{H^{3}_{2}} +  B_{H^{3}_{3}} + B_{H^{3}_{4}} \label{B terms}\\
C&\equiv C_{H^{3}_{1}} + C_{H^{3}_{2}} + C_{H^{3}_{3}} + C_{H^{3}_{4}}  \label{c terms} \, .
\end{align}
As in the factorized form, we  only calculate  the independent parts of bispectra in the commutator method. Therefore, in the following, we specify the contribution of different terms of $H^{3}_{i}, i=1,3$ in $A$ ,$B$ and $C$:

\begin{align}
A_{H^{3}_{1}} &= \lambda_{1} c_{1}^{3} (u'_{k_{1}}(\tau_{1})-c.c.)(v_{k_{1}}(\tau_{1})v^{*}_{k_{1}}(\tau_{2})- c.c.)(v^{*}_{k_{3}}(\tau_{4})v_{k_{3}}(\tau_{2})u'^{*}_{k_{3}}(\tau_{4})-c.c.)\nonumber\\
&~~v_{k_{2}}(\tau_{2})v^{*}_{k_{2}}(\tau_{3})u'^{*}_{k_{2}}(\tau_{3})
\end{align}
As in the factorized case, we have three terms for $A_{H^{3}_{3}}$, related to the different possible positions for $H^{2}_{i}$
\begin{align}
A_{H^{3}_{31}} &= \lambda_{3} c_{1} c_{2}^{2} (u'_{k_{1}}(\tau_{1})-c.c.)(v_{k_{1}}(\tau_{1})v^{*}_{k_{1}}(\tau_{2})- c.c.)(w^{*}_{k_{3}}(\tau_{4})w_{k_{3}}(\tau_{2})u'^{*}_{k_{3}}(\tau_{4})-c.c.)\nonumber\\
&~~w_{k_{2}}(\tau_{2})w^{*}_{k_{2}}(\tau_{3})u'^{*}_{k_{2}}(\tau_{3})
\end{align}
\begin{align}
A_{H^{3}_{32}} &= \lambda_{3} c_{1} c_{2}^{2} (u'_{k_{1}}(\tau_{1})-c.c.)(w_{k_{1}}(\tau_{1})w^{*}_{k_{1}}(\tau_{2})- c.c.)(w^{*}_{k_{3}}(\tau_{4})w_{k_{3}}(\tau_{2})u'^{*}_{k_{3}}(\tau_{4})-c.c.)\nonumber\\
&~~v_{k_{2}}(\tau_{2})v^{*}_{k_{2}}(\tau_{3})u'^{*}_{k_{2}}(\tau_{3})
\end{align}
\begin{align}
A_{H^{3}_{33}} &= \lambda_{3} c_{1} c_{2}^{2} (u'_{k_{1}}(\tau_{1})-c.c.)(w_{k_{1}}(\tau_{1})w^{*}_{k_{1}}(\tau_{2})- c.c.)(v^{*}_{k_{3}}(\tau_{4})v_{k_{3}}(\tau_{2})u'^{*}_{k_{3}}(\tau_{4})-c.c.)\nonumber\\
&~~w_{k_{2}}(\tau_{2})w^{*}_{k_{2}}(\tau_{3})u'^{*}_{k_{2}}(\tau_{3}) \, .
\end{align}

Also for $B$ we have:
\begin{align}
B_{H^{3}_{1}} &= \lambda_{1} c_{1}^{3} (u'_{k_{1}}(\tau_{1})-c.c.)(u'_{k_{2}}(\tau_{2})-c.c.)
(v^{*}_{k_{1}}(\tau_{1})v^{*}_{k_{2}}(\tau_{2})v_{k_{1}}(\tau_{3})v_{k_{2}}(\tau_{3})- c.c.)\nonumber\\
&~~v_{k_{3}}(\tau_{3})v^{*}_{k_{3}}(\tau_{4})u'^{*}_{k_{3}}(\tau_{4})
\end{align}
\begin{align}
B_{H^{3}_{31}} &= \lambda_{3} c_{1} c_{2}^{2} (u'_{k_{1}}(\tau_{1})-c.c.)(u'_{k_{2}}(\tau_{2})-c.c.)
(v^{*}_{k_{1}}(\tau_{1})v_{k_{1}}(\tau_{3})w^{*}_{k_{2}}(\tau_{2})w_{k_{2}}(\tau_{3})- c.c.)\nonumber\\
&~~w_{k_{3}}(\tau_{3})w^{*}_{k_{3}}(\tau_{4})u'^{*}_{k_{3}}(\tau_{4})
\end{align}
\begin{align}
B_{H^{3}_{32}} &= \lambda_{3} c_{1} c_{2}^{2} (u'_{k_{1}}(\tau_{1})-c.c.)(u'_{k_{2}}(\tau_{2})-c.c.)
(w^{*}_{k_{1}}(\tau_{1})w_{k_{1}}(\tau_{3})v^{*}_{k_{2}}(\tau_{2})v_{k_{2}}(\tau_{3})- c.c.)\nonumber\\
&~~w_{k_{3}}(\tau_{3})w^{*}_{k_{3}}(\tau_{4})u'^{*}_{k_{3}}(\tau_{4})
\end{align}
\begin{align}
B_{H^{3}_{33}} &= \lambda_{3} c_{1} c_{2}^{2} (u'_{k_{1}}(\tau_{1})-c.c.)(u'_{k_{2}}(\tau_{2})-c.c.)
(w^{*}_{k_{1}}(\tau_{1})w_{k_{1}}(\tau_{3})w^{*}_{k_{2}}(\tau_{2})w_{k_{2}}(\tau_{3})- c.c.)\nonumber\\
&~~v_{k_{3}}(\tau_{3})v^{*}_{k_{3}}(\tau_{4})u'^{*}_{k_{3}}(\tau_{4}) \, .
\end{align}

Finally; for $C$ we have:
\begin{align}
C_{H^{3}_{1}} &= -\lambda_{1} c_{1}^{3} (u'_{k_{1}}(\tau_{1})-c.c.)(u'_{k_{2}}(\tau_{2})-c.c.)
(u'_{k_{3}}(\tau_{3})-c.c.)(v_{k_{1}}(\tau_{4})v_{k_{2}}(\tau_{4})v_{k_{3}}(\tau_{4})\nonumber\\
&~~~v^{*}_{k_{1}}(\tau_{1})v^{*}_{k_{2}}(\tau_{2})v^{*}_{k_{3}}(\tau_{3}))
\end{align}
\begin{align}
C_{H^{3}_{31}} &= -\lambda_{3} c_{1} c_{2}^{2} (u'_{k_{1}}(\tau_{1})-c.c.)(u'_{k_{2}}(\tau_{2})-c.c.)
(u'_{k_{3}}(\tau_{3})-c.c.)(v_{k_{1}}(\tau_{4})v^{*}_{k_{1}}(\tau_{1})w_{k_{2}}(\tau_{4})\nonumber\\
&~~~w^{*}_{k_{2}}(\tau_{2})w_{k_{3}}(\tau_{4})w^{*}_{k_{3}}(\tau_{3}))
\end{align}
\begin{align}
C_{H^{3}_{32}} &= -\lambda_{3} c_{1} c_{2}^{2} (u'_{k_{1}}(\tau_{1})-c.c.)(u'_{k_{2}}(\tau_{2})-c.c.)
(u'_{k_{3}}(\tau_{3})-c.c.)(w_{k_{1}}(\tau_{4})w^{*}_{k_{1}}(\tau_{1})v_{k_{2}}(\tau_{4})\nonumber\\
&~~~v^{*}_{k_{2}}(\tau_{2})w_{k_{3}}(\tau_{4})w^{*}_{k_{3}}(\tau_{3}))
\end{align}
\begin{align}
C_{H^{3}_{33}} &= -\lambda_{3} c_{1} c_{2}^{2} (u'_{k_{1}}(\tau_{1})-c.c.)(u'_{k_{2}}(\tau_{2})-c.c.)
(u'_{k_{3}}(\tau_{3})-c.c.)(w_{k_{1}}(\tau_{4})w^{*}_{k_{1}}(\tau_{1})w_{k_{2}}(\tau_{4})\nonumber\\
&~~~w^{*}_{k_{2}}(\tau_{2})v_{k_{3}}(\tau_{4})v^{*}_{k_{3}}(\tau_{3}))
\end{align}
As we mentioned above the summation of these A, B and C terms give us the final result for the bispestrum in the commutator form.

\subsection{Different terms in the squeezed limit}
\label{AppenB.3}
In the following, we are going to present the whole independent contributions in the squeezed limit in the commutator form.\\
$(2_{A})A_{H^{3}_{2}}$:  Since this term can be easily obtained by replacing $(\nu_{1}, c_{1}, \lambda_{1})$ with $(\nu_{2}, c_{2}, \lambda_{2})$, we skip it now and insert it in the final result.\\
$(3_{A})A_{H^{3}_{3}}$:  This term is related to our new interaction and is divided into three different parts as follows.

$(3_{A1})A_{H^{3}_{31}}$:
\begin{align}
\delta \theta ^{3} (A_{H^{3}_{31}}) &= - \frac{\dot{\theta_{0}}^{3}}{2^{7}} \frac{\lambda_{3}c_{1}c_{2}^{2}}{H\widetilde{R}^{3}}\frac{\pi^{3}}{k_{1}^{4}k_{2}k_{3}} \nonumber \\
& \times {\rm Re} \bigg{[} i\int_{-\infty}^{0} dx_{1}\int_{-\infty}^{x_{1}} dx_{2}\int_{-\infty}^{x_{2}} dx_{3}\int_{-\infty}^{x_{3}} dx_{4} (-x_{1})^{-\frac{1}{2}} (-x_{2})^{\frac{1}{2}}(-x_{3})^{-\frac{1}{2}}(-x_{4})^{-\frac{1}{2}} \nonumber\\
& \times \sin(-x_{1})\left( H^{(1)}_{\nu_{1}}(-x_{1})H^{(2)}_{\nu_{1}}(-x_{2}) - c.c. \right) \left( H^{(2)}_{\nu_{2}}(-\frac{k_{3}}{k_{1}}x_{2})H^{(1)}_{\nu_{2}}(-\frac{k_{3}}{k_{1}}x_{4})e^{-i\frac{k_{3}}{k_{1}}x_{4}} - c.c. \right) \nonumber\\
& \times H^{(1)}_{\nu_{2}}(-\frac{k_{2}}{k_{1}}x_{2})H^{(2)}_{\nu_{2}}(-\frac{k_{2}}{k_{1}}x_{3})e^{i\frac{k_{2}}{k_{1}}x_{3}}\bigg{]} \, .
\end{align}
Again by approximating the small limit of  $-\frac{k_{3}}{k_{1}}x_{2}$ in the third line we have,
\begin{align}
\delta \theta ^{3} (A_{H^{3}_{31}}) &= - \frac{\dot{\theta_{0}}^{3}}{2^{5-\nu_{2}}} \frac{\lambda_{3}c_{1}c_{2}^{2}}{H\widetilde{R}^{3}}\frac{\pi^{2}\Gamma{(\nu_{2})}}{k_{1}^{\frac{7}{2}-\nu_{2}}k_{2}k_{3}^{\frac{3}{2}+\nu_{2}}} \nonumber \\
& \times \int_{-\infty}^{0} dx_{1}\int_{-\infty}^{x_{1}} dx_{2}\int_{-\infty}^{x_{2}} dx_{3} (-x_{1})^{-\frac{1}{2}} (-x_{2})^{\frac{1}{2}-\nu_{2}}(-x_{3})^{-\frac{1}{2}} \nonumber\\
& \times \sin(-x_{1}){\rm Im}\left( H^{(1)}_{\nu_{1}}(-x_{1})H^{(2)}_{\nu_{1}}(-x_{2})\right) {\rm Im}\left( H^{(1)}_{\nu_{2}}(-x_{2})H^{(2)}_{\nu_{2}}(-x_{3})e^{ix_{3}} \right) \nonumber\\
& \times \int_{-\infty}^{0} dy_{4} (-y_{4})^{-\frac{1}{2}} {\rm Re}\left(  H^{(1)}_{\nu_{2}}(-y_{4})e^{-iy_{4}}\right) \, .
\end{align}
We next look at the term with the permutation $k_{1}\leftrightarrow k_{3} $. Again, the scaling behavior of this term is
\begin{align}
\delta \theta ^{3} (A_{H^{3}_{31}})\sim \frac{1}{k_{1}^{5}k_{2}}
\end{align}
Which, as we argued before, is negligible.

Also for the term with the permutation $k_{2}\leftrightarrow k_{3} $ and $\nu_{2} > \frac{1}{2}$ we have,
\begin{align}
\delta \theta ^{3} (A_{H^{3}_{31}}) &= - \frac{\dot{\theta_{0}}^{3}}{2^{5-2\nu_{2}}} \frac{\lambda_{3}c_{1}c_{2}^{2}}{H\widetilde{R}^{3}}\frac{\pi}{k_{1}^{5-2\nu_{2}}k_{2}k_{3}^{2\nu_{2}}}(\Gamma{(\nu_{2})})^{2} \nonumber \\
& \times \int_{-\infty}^{0} dx_{1}\int_{-\infty}^{x_{1}} dx_{2}\int_{-\infty}^{x_{2}} dx_{3}\int_{-\infty}^{x_{3}} dx_{4} (-x_{1})^{-\frac{1}{2}} (-x_{2})^{\frac{1}{2}-\nu_{2}}(-x_{3})^{\frac{1}{2}-\nu_{2}}(-x_{4})^{-\frac{1}{2}} \nonumber\\
& \times \sin(-x_{1}){\rm Im}\left( H^{(1)}_{\nu_{1}}(-x_{1})H^{(2)}_{\nu_{1}}(-x_{2})\right) {\rm Im}\left( H^{(2)}_{\nu_{2}}(-x_{2})H^{(1)}_{\nu_{2}}(-x_{4})e^{-ix_{4}} \right) \, .
\end{align}

$(3_{A2})A_{H^{3}_{32}}$: For this term we can use the results of the above case, $A_{H^{3}_{31}}$:
\begin{align}
\delta \theta ^{3} (A_{H^{3}_{32}}) &= - \frac{\dot{\theta_{0}}^{3}}{2^{5-\nu_{2}}} \frac{\lambda_{3}c_{1}c_{2}^{2}}{H\widetilde{R}^{3}}\frac{\pi^{2}\Gamma{(\nu_{2})}}{k_{1}^{\frac{7}{2}-\nu_{2}}k_{2}k_{3}^{\frac{3}{2}+\nu_{2}}} \nonumber \\
& \times \int_{-\infty}^{0} dx_{1}\int_{-\infty}^{x_{1}} dx_{2}\int_{-\infty}^{x_{2}} dx_{3} (-x_{1})^{-\frac{1}{2}} (-x_{2})^{\frac{1}{2}-\nu_{2}}(-x_{3})^{-\frac{1}{2}} \nonumber\\
& \times \sin(-x_{1}){\rm Im}\left( H^{(1)}_{\nu_{2}}(-x_{1})H^{(2)}_{\nu_{2}}(-x_{2})\right) {\rm Im}\left( H^{(1)}_{\nu_{1}}(-x_{2})H^{(2)}_{\nu_{1}}(-x_{3})e^{ix_{3}} \right) \nonumber\\
& \times \int_{-\infty}^{0} dy_{4} (-y_{4})^{-\frac{1}{2}} {\rm Re}\left(  H^{(1)}_{\nu_{2}}(-y_{4})e^{-iy_{4}}\right) \, .
\end{align}
As we can see the scaling behavior of this term is as $N_{2}$,  which means that this term is not negligible.

We next look at the term with the permutation $k_{1}\leftrightarrow k_{3} $. Again the scaling behavior of this term is
\begin{align}
\delta \theta ^{3} (A_{H^{3}_{32}})\sim \frac{1}{k_{1}^{5}k_{2}} \, ,
\end{align}
which, as we argued before, is negligible.

Also for the term with the permutation $k_{2}\leftrightarrow k_{3} $ and $\nu_{1} > \frac{1}{2}$, we have
\begin{align}
\delta \theta ^{3} (A_{H^{3}_{32}}) &= - \frac{\dot{\theta_{0}}^{3}}{2^{5-2\nu_{1}}} \frac{\lambda_{3}c_{1}c_{2}^{2}}{H\widetilde{R}^{3}}\frac{\pi}{k_{1}^{5-2\nu_{1}}k_{2}k_{3}^{2\nu_{1}}}(\Gamma{(\nu_{1})})^{2} \nonumber \\
& \times \int_{-\infty}^{0} dx_{1}\int_{-\infty}^{x_{1}} dx_{2}\int_{-\infty}^{x_{2}} dx_{3}\int_{-\infty}^{x_{3}} dx_{4} (-x_{1})^{-\frac{1}{2}} (-x_{2})^{\frac{1}{2}-\nu_{1}}(-x_{3})^{\frac{1}{2}-\nu_{1}}(-x_{4})^{-\frac{1}{2}} \nonumber\\
& \times \sin(-x_{1}){\rm Im}\left( H^{(1)}_{\nu_{2}}(-x_{1})H^{(2)}_{\nu_{2}}(-x_{2})\right) {\rm Im}\left( H^{(2)}_{\nu_{2}}(-x_{2})H^{(1)}_{\nu_{2}}(-x_{4})e^{-ix_{4}} \right) \, .
\end{align}

$(3_{A3})A_{H^{3}_{33}}$:  For this term we can use the results of the above case, $A_{H^{3}_{31}}$:
\begin{align}
\delta \theta ^{3} (A_{H^{3}_{33}}) &= - \frac{\dot{\theta_{0}}^{3}}{2^{5-\nu_{1}}} \frac{\lambda_{3}c_{1}c_{2}^{2}}{H\widetilde{R}^{3}}\frac{\pi^{2}\Gamma{(\nu_{1})}}{k_{1}^{\frac{7}{2}-\nu_{1}}k_{2}k_{3}^{\frac{3}{2}+\nu_{1}}} \nonumber \\
& \times \int_{-\infty}^{0} dx_{1}\int_{-\infty}^{x_{1}} dx_{2}\int_{-\infty}^{x_{2}} dx_{3} (-x_{1})^{-\frac{1}{2}} (-x_{2})^{\frac{1}{2}-\nu_{1}}(-x_{3})^{-\frac{1}{2}} \nonumber\\
& \times \sin(-x_{1}){\rm Im}\left( H^{(1)}_{\nu_{2}}(-x_{1})H^{(2)}_{\nu_{2}}(-x_{2})\right) {\rm Im}\left( H^{(1)}_{\nu_{2}}(-x_{2})H^{(2)}_{\nu_{2}}(-x_{3})e^{ix_{3}} \right) \nonumber\\
& \times \int_{-\infty}^{0} dy_{4} (-y_{4})^{-\frac{1}{2}} {\rm Re}\left(  H^{(1)}_{\nu_{1}}(-y_{4})e^{-iy_{4}}\right)
\end{align}
As we can see the scaling behavior of this term is as $N_{1}$  which means that, depending on the parameter space, this term can be significant.
We next look at the term with the permutation $k_{1}\leftrightarrow k_{3} $. Again, the scaling behavior of this term is
\begin{align}
\delta \theta ^{3} (A_{H^{3}_{33}})\sim \frac{1}{k_{1}^{5}k_{2}} \, ,
\end{align}
which, as we argued before, is negligible.

Also for the term with the permutation $k_{2}\leftrightarrow k_{3} $ and $\nu_{2} > \frac{1}{2}$ we have
\begin{align}
\delta \theta ^{3} (A_{H^{3}_{33}}) &= - \frac{\dot{\theta_{0}}^{3}}{2^{5-2\nu_{2}}} \frac{\lambda_{3}c_{1}c_{2}^{2}}{H\widetilde{R}^{3}}\frac{\pi}{k_{1}^{5-2\nu_{2}}k_{2}k_{3}^{2\nu_{2}}}(\Gamma{(\nu_{2})})^{2} \nonumber \\
& \times \int_{-\infty}^{0} dx_{1}\int_{-\infty}^{x_{1}} dx_{2}\int_{-\infty}^{x_{2}} dx_{3}\int_{-\infty}^{x_{3}} dx_{4} (-x_{1})^{-\frac{1}{2}} (-x_{2})^{\frac{1}{2}-\nu_{2}}(-x_{3})^{\frac{1}{2}-\nu_{2}}(-x_{4})^{-\frac{1}{2}} \nonumber\\
& \times \sin(-x_{1}){\rm Im}\left( H^{(1)}_{\nu_{2}}(-x_{1})H^{(2)}_{\nu_{2}}(-x_{2})\right) {\rm Im}\left( H^{(2)}_{\nu_{1}}(-x_{2})H^{(1)}_{\nu_{1}}(-x_{4})e^{-ix_{4}} \right) \, .
\end{align}

$(4_{A})A_{H^{3}_{4}}$:  Again since this term can be obtained from $A_{H^{3}_{3}}$, we skip it now and insert it in the final result.

$(1_{B})B_{H^{3}_{1}}$:  This part is very similar to \cite{Chen:2009zp}:
\begin{align}
\delta \theta ^{3} (B_{H^{3}_{1}}) &= - \frac{\dot{\theta_{0}}^{3}}{2^{6}} \frac{\lambda_{1}c_{1}^{3}}{H\widetilde{R}^{3}}\frac{\pi^{3}}{k_{1}^{4}k_{2}k_{3}} \nonumber \\
& \times {\rm Re} \bigg{[} i\int_{-\infty}^{0} dx_{1}\int_{-\infty}^{x_{1}} dx_{2}\int_{-\infty}^{x_{2}} dx_{3}\int_{-\infty}^{x_{3}} dx_{4} (-x_{1})^{-\frac{1}{2}} (-x_{2})^{-\frac{1}{2}}(-x_{3})^{\frac{1}{2}}(-x_{4})^{-\frac{1}{2}} \nonumber\\
& \times \sin(-x_{1})\sin(-\frac{k_{2}}{k_{1}} x_{2})\left( H^{(2)}_{\nu_{1}}(-x_{1})H^{(2)}_{\nu_{1}}(-\frac{k_{2}}{k_{1}}x_{2})H^{(1)}_{\nu_{1}}(-x_{3})H^{(1)}_{\nu_{1}}(-\frac{k_{2}}{k_{1}}x_{3}) - c.c. \right)\nonumber\\
&\times\left( H^{(1)}_{\nu_{1}}(-\frac{k_{3}}{k_{1}}x_{3})H^{(2)}_{\nu_{1}}(-\frac{k_{3}}{k_{1}}x_{4})e^{i\frac{k_{3}}{k_{1}}x_{4}}\right) \bigg{]} \, .
\end{align}
Now as in \cite{Chen:2009zp}, the term $H^{(1)}_{\nu_{1}}(-\frac{k_{3}}{k_{1}}x_{3})$ in the 4th line  can be approximated in the small $-k_{3}/k_{1}x_{3}$ limit. However, the term $H^{(2)}_{\nu_{1}}(-\frac{k_{3}}{k_{1}}x_{4})$ in the 4th line can not be  can approximated. Then by redefining $y_{4} \equiv k_{3}/k_{1}x_{4}$, we get
\begin{align}
\delta \theta ^{3} (B_{H^{3}_{1}}) &= - \frac{\dot{\theta_{0}}^{3}}{2^{5-\nu_{1}}} \frac{\lambda_{1}c_{1}^{3}}{H\widetilde{R}^{3}}\frac{\pi^{2}\Gamma{(\nu_{1})}}{k_{1}^{\frac{7}{2}-\nu_{1}}k_{2}k_{3}^{\frac{3}{2}+\nu_{1}}} \nonumber \\
& \times \int_{-\infty}^{0} dx_{1}\int_{-\infty}^{x_{1}} dx_{2}\int_{-\infty}^{x_{2}} dx_{3} (-x_{1})^{-\frac{1}{2}}(-x_{2})^{-\frac{1}{2}} (-x_{3})^{\frac{1}{2}-\nu_{1}} \nonumber\\
& \times \sin(-x_{1})\sin(-x_{2}){\rm Im}\left( H^{(2)}_{\nu_{1}}(-x_{1})H^{(2)}_{\nu_{1}}(-x_{2}) \left(H^{(1)}_{\nu_{1}}(-x_{3})\right)^{2}\right) \nonumber\\
&\times \int_{-\infty}^{0} dy_{4} (-y_{4})^{-\frac{1}{2}}{\rm Re}\left(  H^{(2)}_{\nu_{1}}(-y_{4})e^{iy_{4}}\right) \, .
\end{align}
The scaling behavior of the above term is as $N_{1}$.

We next look at the term with the permutation $k_{1}\leftrightarrow k_{3} $:
\begin{align}
\delta \theta ^{3} (B_{H^{3}_{1}}) &= - \frac{\dot{\theta_{0}}^{3}}{2^{6}} \frac{\lambda_{1}c_{1}^{3}}{H\widetilde{R}^{3}}\frac{\pi^{3}}{k_{1}^{4}k_{2}k_{3}} \nonumber \\
& \times {\rm Re} \bigg{[} i\int_{-\infty}^{0} dx_{1}\int_{-\infty}^{x_{1}} dx_{2}\int_{-\infty}^{x_{2}} dx_{3}\int_{-\infty}^{x_{3}} dx_{4} (-x_{1})^{-\frac{1}{2}} (-x_{2})^{-\frac{1}{2}}(-x_{3})^{\frac{1}{2}}(-x_{4})^{-\frac{1}{2}} \nonumber\\
& \times \sin(-\frac{k_{3}}{k_{1}}x_{1})\sin(-\frac{k_{2}}{k_{1}} x_{2})\left( H^{(2)}_{\nu_{1}}(-\frac{k_{3}}{k_{1}}x_{1})H^{(2)}_{\nu_{1}}(-\frac{k_{2}}{k_{1}}x_{2})H^{(1)}_{\nu_{1}}(-\frac{k_{3}}{k_{1}}x_{3})H^{(1)}_{\nu_{1}}(-\frac{k_{2}}{k_{1}}x_{3}) - c.c. \right)\nonumber\\
&\times\left( H^{(1)}_{\nu_{1}}(-x_{3})H^{(2)}_{\nu_{1}}(-x_{4})e^{ix_{4}}\right) \bigg{]}
\end{align}
Again as in \cite{Chen:2009zp}, the terms containing $-k_{3}/k_{1}x_{i} (i= 1,3)$ can be approximated in the small argument limit. The reason for $i=1$ is the following. In the integrand we have the factor $H^{(2)}_{\nu_{1}}(-\frac{k_{2}}{k_{1}}x_{2})$ so if $|x_{2}| \gg 1$, this term becomes fast-oscillating and hence suppresses the integration. On the other hand, since the upper bound of the integral of $x_{2}$ is $x_{1}$, $|x_{1}|<|x_{2}| $. So the terms containing $-k_{3}/k_{1}x_{1}$ is small. In addition, due tot he term $H^{(1)}_{\nu_{1}}(-x_{3})$,
the smallness of the term containing $-k_{3}/k_{1}x_{3}$ is somewhat more clear. Therefore,  we have
\begin{align}
\delta \theta ^{3} (B_{H^{3}_{1}}) &= - \frac{\dot{\theta_{0}}^{3}}{2^{5-2\nu_{1}}} \frac{\lambda_{1}c_{1}^{3}}{H\widetilde{R}^{3}}\frac{\pi }{k_{1}^{5-2\nu_{1}}k_{2}k_{3}^{2\nu_{1}}}(\Gamma{(\nu_{1})})^{2} \nonumber \\
& \times \int_{-\infty}^{0} dx_{1}\int_{-\infty}^{x_{1}} dx_{2}\int_{-\infty}^{x_{2}} dx_{3}\int_{-\infty}^{x_{3}} dx_{4} (-x_{1})^{\frac{1}{2}-\nu_{1}}(-x_{2})^{-\frac{1}{2}} (-x_{3})^{\frac{1}{2}-\nu_{1}}(-x_{4})^{-\frac{1}{2}} \nonumber\\
& \times \sin(-x_{2}){\rm Im}\left( H^{(2)}_{\nu_{1}}(-x_{2})H^{(1)}_{\nu_{1}}(-x_{3})\right) \nonumber\\
&\times {\rm Im}\left( H^{(1)}_{\nu_{1}}(-x_{3})H^{(2)}_{\nu_{1}}(-x_{4})e^{ix_{4}}\right) \, .
\end{align}
Then we look at the term with the permutation $k_{2}\leftrightarrow k_{3}$ :
\begin{align}
\delta \theta ^{3} (B_{H^{3}_{1}}) &= - \frac{\dot{\theta_{0}}^{3}}{2^{6}} \frac{\lambda_{1}c_{1}^{3}}{H\widetilde{R}^{3}}\frac{\pi^{3}}{k_{1}^{4}k_{2}k_{3}} \nonumber \\
& \times {\rm Re} \bigg{[} i\int_{-\infty}^{0} dx_{1}\int_{-\infty}^{x_{1}} dx_{2}\int_{-\infty}^{x_{2}} dx_{3}\int_{-\infty}^{x_{3}} dx_{4} (-x_{1})^{-\frac{1}{2}} (-x_{2})^{-\frac{1}{2}}(-x_{3})^{\frac{1}{2}}(-x_{4})^{-\frac{1}{2}} \nonumber\\
& \times \sin(-x_{1})\sin(-\frac{k_{3}}{k_{1}} x_{2})\left( H^{(2)}_{\nu_{1}}(-x_{1})H^{(2)}_{\nu_{1}}(-\frac{k_{3}}{k_{1}}x_{2})H^{(1)}_{\nu_{1}}(-x_{3})H^{(1)}_{\nu_{1}}(-\frac{k_{3}}{k_{1}}x_{3}) - c.c. \right)\nonumber\\
&\times\left( H^{(1)}_{\nu_{1}}(-\frac{k_{2}}{k_{1}}x_{3})H^{(2)}_{\nu_{1}}(-\frac{k_{2}}{k_{1}}x_{4})e^{i\frac{k_{2}}{k_{1}}x_{4}}\right) \bigg{]} \, .
\end{align}
Again as in \cite{Chen:2009zp}, the terms containing $-k_{3}/k_{1}x_{i}\, , (i= 2,3)$ can be approximated in the small argument limit. The smallness of the term containing $-k_{3}/k_{1}x_{3}$ is more  clear due to the term $H^{(1)}_{\nu_{1}}(-x_{3})$.  On the other hand, the reason for the smallness for $i=2$ is that
the upper bound of the integral of $x_{3}$ is $x_{2}$. This means that $|x_{2}|<|x_{3}| $. So the terms containing $-k_{3}/k_{1}x_{2}$ is small too. Therefore,  we have
\begin{align}
\delta \theta ^{3} (B_{H^{3}_{1}}) &= - \frac{\dot{\theta_{0}}^{3}}{2^{5-2\nu_{1}}} \frac{\lambda_{1}c_{1}^{3}}{H\widetilde{R}^{3}}\frac{\pi }{k_{1}^{5-2\nu_{1}}k_{2}k_{3}^{2\nu_{1}}}(\Gamma{(\nu_{1})})^{2} \nonumber \\
& \times \int_{-\infty}^{0} dx_{1}\int_{-\infty}^{x_{1}} dx_{2}\int_{-\infty}^{x_{2}} dx_{3}\int_{-\infty}^{x_{3}} dx_{4} (-x_{1})^{-\frac{1}{2}}(-x_{2})^{\frac{1}{2}-\nu_{1}} (-x_{3})^{\frac{1}{2}-\nu_{1}}(-x_{4})^{-\frac{1}{2}} \nonumber\\
& \times \sin(-x_{1}){\rm Im}\left( H^{(2)}_{\nu_{1}}(-x_{1})H^{(1)}_{\nu_{1}}(-x_{3})\right) \nonumber\\
&\times {\rm Im}\left( H^{(1)}_{\nu_{1}}(-x_{3})H^{(2)}_{\nu_{1}}(-x_{4})e^{ix_{4}}\right) \, .
\end{align}

$(2_{B})B_{H^{3}_{2}}$ : Since this term can be obtained from the above case, we skip it now and insert it in the final result.

$(3_{B})B_{H^{3}_{3}}$: This term is related to our new interaction and is divided into three different parts:

$(3_{B1})B_{H^{3}_{31}}$:
\begin{align}
\delta \theta ^{3} (B_{H^{3}_{31}}) &= - \frac{\dot{\theta_{0}}^{3}}{2^{6}} \frac{\lambda_{3}c_{1}c_{2}^{2}}{H\widetilde{R}^{3}}\frac{\pi^{3}}{k_{1}^{4}k_{2}k_{3}} \nonumber \\
& \times {\rm Re} \bigg{[} i\int_{-\infty}^{0} dx_{1}\int_{-\infty}^{x_{1}} dx_{2}\int_{-\infty}^{x_{2}} dx_{3}\int_{-\infty}^{x_{3}} dx_{4} (-x_{1})^{-\frac{1}{2}} (-x_{2})^{-\frac{1}{2}}(-x_{3})^{\frac{1}{2}}(-x_{4})^{-\frac{1}{2}} \nonumber\\
& \times \sin(-x_{1})\sin(-\frac{k_{2}}{k_{1}} x_{2})\left( H^{(2)}_{\nu_{1}}(-x_{1})H^{(1)}_{\nu_{1}}(-x_{3})H^{(2)}_{\nu_{2}}(-\frac{k_{2}}{k_{1}}x_{2})H^{(1)}_{\nu_{2}}(-\frac{k_{2}}{k_{1}}x_{3}) - c.c. \right)\nonumber\\
&\times\left( H^{(1)}_{\nu_{2}}(-\frac{k_{3}}{k_{1}}x_{3})H^{(2)}_{\nu_{2}}(-\frac{k_{3}}{k_{1}}x_{4})e^{i\frac{k_{3}}{k_{1}}x_{4}}\right) \bigg{]} \, .
\end{align}
As in the above cases, for the small value of $\frac{k_{3}}{k_{1}}x_{3}$,  we have
\begin{align}
\delta \theta ^{3} (B_{H^{3}_{31}}) &= - \frac{\dot{\theta_{0}}^{3}}{2^{5-\nu_{2}}} \frac{\lambda_{3}c_{1}c_{2}^{2}}{H\widetilde{R}^{3}}\frac{\pi^{2}\Gamma{(\nu_{2})}}{k_{1}^{\frac{7}{2}-\nu_{2}}k_{2}k_{3}^{\frac{3}{2}+\nu_{2}}} \nonumber \\
& \times \int_{-\infty}^{0} dx_{1}\int_{-\infty}^{x_{1}} dx_{2}\int_{-\infty}^{x_{2}} dx_{3} (-x_{1})^{-\frac{1}{2}}(-x_{2})^{-\frac{1}{2}} (-x_{3})^{\frac{1}{2}-\nu_{2}} \nonumber\\
& \times \sin(-x_{1})\sin(-x_{2}){\rm Im}\left( H^{(2)}_{\nu_{1}}(-x_{1})H^{(1)}_{\nu_{1}}(-x_{3})H^{(2)}_{\nu_{2}}(-x_{2})H^{(1)}_{\nu_{2}}(-x_{3})\right) \nonumber\\
&\times \int_{-\infty}^{0} dy_{4} (-y_{4})^{-\frac{1}{2}}{\rm Re}\left(  H^{(2)}_{\nu_{2}}(-y_{4})e^{iy_{4}}\right)
\end{align}
The scaling behavior of the above term is as $N_{2}$.

We next look at the term with the permutation $k_{1}\leftrightarrow k_{3} $:
\begin{align}
\delta \theta ^{3} (B_{H^{3}_{31}}) &= - \frac{\dot{\theta_{0}}^{3}}{2^{6}} \frac{\lambda_{3}c_{1}c_{2}^{2}}{H\widetilde{R}^{3}}\frac{\pi^{3}}{k_{1}^{4}k_{2}k_{3}} \nonumber \\
& \times {\rm Re} \bigg{[} i\int_{-\infty}^{0} dx_{1}\int_{-\infty}^{x_{1}} dx_{2}\int_{-\infty}^{x_{2}} dx_{3}\int_{-\infty}^{x_{3}} dx_{4} (-x_{1})^{-\frac{1}{2}} (-x_{2})^{-\frac{1}{2}}(-x_{3})^{\frac{1}{2}}(-x_{4})^{-\frac{1}{2}} \nonumber\\
& \times \sin(-\frac{k_{3}}{k_{1}}x_{1})\sin(-\frac{k_{2}}{k_{1}} x_{2})\left( H^{(2)}_{\nu_{1}}(-\frac{k_{3}}{k_{1}}x_{1})
H^{(1)}_{\nu_{1}}(-\frac{k_{3}}{k_{1}}x_{3})
H^{(2)}_{\nu_{2}}(-\frac{k_{2}}{k_{1}}x_{2})H^{(1)}_{\nu_{2}}(-\frac{k_{2}}{k_{1}}x_{3}) - c.c. \right)\nonumber\\
&\times\left( H^{(1)}_{\nu_{2}}(-x_{3})H^{(2)}_{\nu_{2}}(-x_{4})e^{ix_{4}}\right) \bigg{]}
\end{align}
Again, the terms containing $-k_{3}/k_{1}x_{i}, \,  (i= 1,3)$ can be approximated in the small argument limit. The proof is the same as in the above cases:
\begin{align}
\delta \theta ^{3} (B_{H^{3}_{31}}) &= - \frac{\dot{\theta_{0}}^{3}}{2^{5-2\nu_{1}}} \frac{\lambda_{3}c_{1}c_{2}^{2}}{H\widetilde{R}^{3}}\frac{\pi }{k_{1}^{5-2\nu_{1}}k_{2}k_{3}^{2\nu_{1}}}(\Gamma{(\nu_{1})})^{2} \nonumber \\
& \times \int_{-\infty}^{0} dx_{1}\int_{-\infty}^{x_{1}} dx_{2}\int_{-\infty}^{x_{2}} dx_{3}\int_{-\infty}^{x_{3}} dx_{4} (-x_{1})^{\frac{1}{2}-\nu_{1}}(-x_{2})^{-\frac{1}{2}} (-x_{3})^{\frac{1}{2}-\nu_{1}}(-x_{4})^{-\frac{1}{2}} \nonumber\\
& \times \sin(-x_{2}){\rm Im}\left( H^{(2)}_{\nu_{2}}(-x_{2})H^{(1)}_{\nu_{2}}(-x_{3})\right) \nonumber\\
&\times {\rm Im}\left( H^{(1)}_{\nu_{2}}(-x_{3})H^{(2)}_{\nu_{2}}(-x_{4})e^{ix_{4}}\right)
\end{align}
Then we look at the term with the permutation $k_{2}\leftrightarrow k_{3} $:
\begin{align}
\delta \theta ^{3} (B_{H^{3}_{31}}) &= - \frac{\dot{\theta_{0}}^{3}}{2^{6}} \frac{\lambda_{3}c_{1}c_{2}^{2}}{H\widetilde{R}^{3}}\frac{\pi^{3}}{k_{1}^{4}k_{2}k_{3}} \nonumber \\
& \times {\rm Re} \bigg{[} i\int_{-\infty}^{0} dx_{1}\int_{-\infty}^{x_{1}} dx_{2}\int_{-\infty}^{x_{2}} dx_{3}\int_{-\infty}^{x_{3}} dx_{4} (-x_{1})^{-\frac{1}{2}} (-x_{2})^{-\frac{1}{2}}(-x_{3})^{\frac{1}{2}}(-x_{4})^{-\frac{1}{2}} \nonumber\\
& \times \sin(-x_{1})\sin(-\frac{k_{3}}{k_{1}} x_{2})\left( H^{(2)}_{\nu_{1}}(-x_{1})H^{(1)}_{\nu_{1}}(-x_{3})H^{(2)}_{\nu_{2}}(-\frac{k_{3}}{k_{1}}x_{2})H^{(1)}_{\nu_{2}}(-\frac{k_{3}}{k_{1}}x_{3}) - c.c. \right)\nonumber\\
&\times\left( H^{(1)}_{\nu_{2}}(-\frac{k_{2}}{k_{1}}x_{3})H^{(2)}_{\nu_{2}}(-\frac{k_{2}}{k_{1}}x_{4})e^{i\frac{k_{2}}{k_{1}}x_{4}}\right) \bigg{]}
\end{align}
Again, the terms containing $-k_{3}/k_{1}x_{i}, \,  (i= 2,3)$ can be approximated in the small argument limit:
\begin{align}
\delta \theta ^{3} (B_{H^{3}_{31}}) &= - \frac{\dot{\theta_{0}}^{3}}{2^{5-2\nu_{2}}} \frac{\lambda_{3}c_{1}c_{2}^{2}}{H\widetilde{R}^{3}}\frac{\pi }{k_{1}^{5-2\nu_{2}}k_{2}k_{3}^{2\nu_{2}}}(\Gamma{(\nu_{2})})^{2} \nonumber \\
& \times \int_{-\infty}^{0} dx_{1}\int_{-\infty}^{x_{1}} dx_{2}\int_{-\infty}^{x_{2}} dx_{3}\int_{-\infty}^{x_{3}} dx_{4} (-x_{1})^{-\frac{1}{2}}(-x_{2})^{\frac{1}{2}-\nu_{2}} (-x_{3})^{\frac{1}{2}-\nu_{2}}(-x_{4})^{-\frac{1}{2}} \nonumber\\
& \times \sin(-x_{1}){\rm Im}\left( H^{(2)}_{\nu_{1}}(-x_{1})H^{(1)}_{\nu_{1}}(-x_{3})\right) \nonumber\\
&\times {\rm Im}\left( H^{(1)}_{\nu_{2}}(-x_{3})H^{(2)}_{\nu_{2}}(-x_{4})e^{ix_{4}}\right) \, .
\end{align}

$(3_{B2})B_{H^{3}_{32}}$:  Since the analysis in this case is very similar to the above case for $B_{H^{3}_{31}}$, we just mention the final result:
\begin{align}
\delta \theta ^{3} (B_{H^{3}_{32}}) &= - \frac{\dot{\theta_{0}}^{3}}{2^{5-\nu_{2}}} \frac{\lambda_{3}c_{1}c_{2}^{2}}{H\widetilde{R}^{3}}\frac{\pi^{2}\Gamma{(\nu_{2})}}{k_{1}^{\frac{7}{2}-\nu_{2}}k_{2}k_{3}^{\frac{3}{2}+\nu_{2}}} \nonumber \\
& \times \int_{-\infty}^{0} dx_{1}\int_{-\infty}^{x_{1}} dx_{2}\int_{-\infty}^{x_{2}} dx_{3} (-x_{1})^{-\frac{1}{2}}(-x_{2})^{-\frac{1}{2}} (-x_{3})^{\frac{1}{2}-\nu_{2}} \nonumber\\
& \times \sin(-x_{1})\sin(-x_{2}){\rm Im}\left( H^{(2)}_{\nu_{2}}(-x_{1})H^{(1)}_{\nu_{2}}(-x_{3})H^{(2)}_{\nu_{1}}(-x_{2})H^{(1)}_{\nu_{1}}(-x_{3})\right) \nonumber\\
&\times \int_{-\infty}^{0} dy_{4} (-y_{4})^{-\frac{1}{2}}{\rm Re}\left(  H^{(2)}_{\nu_{2}}(-y_{4})e^{iy_{4}}\right)
\end{align}
The scaling behavior of the above term is as $N_{2}$.

We next look at the term with the permutation $k_{1}\leftrightarrow k_{3} $:
\begin{align}
\delta \theta ^{3} (B_{H^{3}_{32}}) &= - \frac{\dot{\theta_{0}}^{3}}{2^{5-2\nu_{2}}} \frac{\lambda_{3}c_{1}c_{2}^{2}}{H\widetilde{R}^{3}}\frac{\pi }{k_{1}^{5-2\nu_{2}}k_{2}k_{3}^{2\nu_{2}}}(\Gamma{(\nu_{2})})^{2} \nonumber \\
& \times \int_{-\infty}^{0} dx_{1}\int_{-\infty}^{x_{1}} dx_{2}\int_{-\infty}^{x_{2}} dx_{3}\int_{-\infty}^{x_{3}} dx_{4} (-x_{1})^{\frac{1}{2}-\nu_{2}}(-x_{2})^{-\frac{1}{2}} (-x_{3})^{\frac{1}{2}-\nu_{2}}(-x_{4})^{-\frac{1}{2}} \nonumber\\
& \times \sin(-x_{2}){\rm Im}\left( H^{(2)}_{\nu_{1}}(-x_{2})H^{(1)}_{\nu_{1}}(-x_{3})\right) \nonumber\\
&\times {\rm Im}\left( H^{(1)}_{\nu_{2}}(-x_{3})H^{(2)}_{\nu_{2}}(-x_{4})e^{ix_{4}}\right) \, .
\end{align}
Then we look at the term with the permutation $k_{2}\leftrightarrow k_{3} $:
\begin{align}
\delta \theta ^{3} (B_{H^{3}_{32}}) &= - \frac{\dot{\theta_{0}}^{3}}{2^{5-2\nu_{1}}} \frac{\lambda_{3}c_{1}c_{2}^{2}}{H\widetilde{R}^{3}}\frac{\pi }{k_{1}^{5-2\nu_{1}}k_{2}k_{3}^{2\nu_{1}}}(\Gamma{(\nu_{1})})^{2} \nonumber \\
& \times \int_{-\infty}^{0} dx_{1}\int_{-\infty}^{x_{1}} dx_{2}\int_{-\infty}^{x_{2}} dx_{3}\int_{-\infty}^{x_{3}} dx_{4} (-x_{1})^{-\frac{1}{2}}(-x_{2})^{\frac{1}{2}-\nu_{1}} (-x_{3})^{\frac{1}{2}-\nu_{1}}(-x_{4})^{-\frac{1}{2}} \nonumber\\
& \times \sin(-x_{1}){\rm Im}\left( H^{(2)}_{\nu_{2}}(-x_{1})H^{(1)}_{\nu_{2}}(-x_{3})\right) \nonumber\\
&\times {\rm Im}\left( H^{(1)}_{\nu_{2}}(-x_{3})H^{(2)}_{\nu_{2}}(-x_{4})e^{ix_{4}}\right) \, .
\end{align}

$(3_{B3})B_{H^{3}_{33}}$: Since the analysis is very similar to the previous terms, we just mention the final result,
\begin{align}
\delta \theta ^{3} (B_{H^{3}_{33}}) &= - \frac{\dot{\theta_{0}}^{3}}{2^{5-\nu_{1}}} \frac{\lambda_{3}c_{1}c_{2}^{2}}{H\widetilde{R}^{3}}\frac{\pi^{2}\Gamma{(\nu_{1})}}{k_{1}^{\frac{7}{2}-\nu_{1}}k_{2}k_{3}^{\frac{3}{2}+\nu_{1}}} \nonumber \\
& \times \int_{-\infty}^{0} dx_{1}\int_{-\infty}^{x_{1}} dx_{2}\int_{-\infty}^{x_{2}} dx_{3} (-x_{1})^{-\frac{1}{2}}(-x_{2})^{-\frac{1}{2}} (-x_{3})^{\frac{1}{2}-\nu_{1}} \nonumber\\
& \times \sin(-x_{1})\sin(-x_{2}){\rm Im}\left( H^{(2)}_{\nu_{2}}(-x_{1})H^{(2)}_{\nu_{2}}(-x_{2})(H^{(1)}_{\nu_{2}}(-x_{3}))^{2}\right) \nonumber\\
&\times \int_{-\infty}^{0} dy_{4} (-y_{4})^{-\frac{1}{2}}{\rm Re}\left(  H^{(2)}_{\nu_{1}}(-y_{4})e^{iy_{4}}\right) \, .
\end{align}
Then we look at the term with the permutation $k_{1}\leftrightarrow k_{3} $:
\begin{align}
\delta \theta ^{3} (B_{H^{3}_{33}}) &= - \frac{\dot{\theta_{0}}^{3}}{2^{5-2\nu_{2}}} \frac{\lambda_{3}c_{1}c_{2}^{2}}{H\widetilde{R}^{3}}\frac{\pi }{k_{1}^{5-2\nu_{2}}k_{2}k_{3}^{2\nu_{2}}}(\Gamma{(\nu_{2})})^{2} \nonumber \\
& \times \int_{-\infty}^{0} dx_{1}\int_{-\infty}^{x_{1}} dx_{2}\int_{-\infty}^{x_{2}} dx_{3}\int_{-\infty}^{x_{3}} dx_{4} (-x_{1})^{\frac{1}{2}-\nu_{2}}(-x_{2})^{-\frac{1}{2}} (-x_{3})^{\frac{1}{2}-\nu_{2}}(-x_{4})^{-\frac{1}{2}} \nonumber\\
& \times \sin(-x_{2}){\rm Im}\left( H^{(2)}_{\nu_{2}}(-x_{2})H^{(1)}_{\nu_{2}}(-x_{3})\right) \nonumber\\
&\times {\rm Im}\left( H^{(1)}_{\nu_{1}}(-x_{3})H^{(2)}_{\nu_{1}}(-x_{4})e^{ix_{4}}\right) \, .
\end{align}
Finally we look at the term with the permutation $k_{2}\leftrightarrow k_{3} $:
\begin{align}
\delta \theta ^{3} (B_{H^{3}_{33}}) &= - \frac{\dot{\theta_{0}}^{3}}{2^{5-2\nu_{2}}} \frac{\lambda_{3}c_{1}c_{2}^{2}}{H\widetilde{R}^{3}}\frac{\pi }{k_{1}^{5-2\nu_{2}}k_{2}k_{3}^{2\nu_{2}}}(\Gamma{(\nu_{2})})^{2} \nonumber \\
& \times \int_{-\infty}^{0} dx_{1}\int_{-\infty}^{x_{1}} dx_{2}\int_{-\infty}^{x_{2}} dx_{3}\int_{-\infty}^{x_{3}} dx_{4} (-x_{1})^{-\frac{1}{2}}(-x_{2})^{\frac{1}{2}-\nu_{2}} (-x_{3})^{\frac{1}{2}-\nu_{2}}(-x_{4})^{-\frac{1}{2}} \nonumber\\
& \times \sin(-x_{1}){\rm Im}\left( H^{(2)}_{\nu_{2}}(-x_{1})H^{(1)}_{\nu_{2}}(-x_{3})\right) \nonumber\\
&\times {\rm Im}\left( H^{(1)}_{\nu_{1}}(-x_{3})H^{(2)}_{\nu_{1}}(-x_{4})e^{ix_{4}}\right) \, .
\end{align}

$(4_{B})B_{H^{3}_{4}}$: Again, since this term can be obtained from $B_{H^{3}_{3}}$, we skip it now and insert it in the final result.

Now we move on to the final case, $C$.

$(1_{C})C_{H^{3}_{1}}$: Unlike the case \cite{Chen:2009zp}, this term can be  relevant in our case. The reason is that we have two different $\nu_{i}$ and we can choose our parameter space such that the related term can be large enough. So we should consider all of the terms related to this part carefully.
\begin{align}
\delta \theta ^{3} (C_{H^{3}_{1}}) &= \frac{\dot{\theta_{0}}^{3}}{2^{5}} \frac{\lambda_{1}c_{1}^{3}}{H\widetilde{R}^{3}}\frac{\pi^{3}}{k_{1}^{4}k_{2}k_{3}} \nonumber \\
& \times {\rm Re} \bigg{[} i\int_{-\infty}^{0} dx_{1}\int_{-\infty}^{x_{1}} dx_{2}\int_{-\infty}^{x_{2}} dx_{3}\int_{-\infty}^{x_{3}} dx_{4} (-x_{1})^{-\frac{1}{2}} (-x_{2})^{-\frac{1}{2}}(-x_{3})^{-\frac{1}{2}}(-x_{4})^{\frac{1}{2}} \nonumber\\
& \times \sin(-x_{1})\sin(-\frac{k_{2}}{k_{1}} x_{2})\sin(-\frac{k_{3}}{k_{1}} x_{3})\bigg{(}
H^{(1)}_{\nu_{1}}(-x_{4})H^{(1)}_{\nu_{1}}(-\frac{k_{2}}{k_{1}}x_{4})H^{(1)}_{\nu_{1}}(-\frac{k_{3}}{k_{1}}x_{4})\nonumber\\
&\times H^{(2)}_{\nu_{1}}(-x_{1})H^{(2)}_{\nu_{1}}(-\frac{k_{2}}{k_{1}}x_{2})H^{(2)}_{\nu_{1}}(-\frac{k_{3}}{k_{1}}x_{3})\bigg{)} \bigg{]} \, .
\end{align}
One can easily see that the terms containing $-k_{3}/k_{1}x_{i} (i= 3,4)$ can be approximated in the small argument limit. The result is
\begin{align}
\delta \theta ^{3} (C_{H^{3}_{1}}) &= \frac{\dot{\theta_{0}}^{3}}{2^{5-2\nu_{1}}} \frac{\lambda_{1}c_{1}^{3}}{H\widetilde{R}^{3}}\frac{\pi }{k_{1}^{5-2\nu_{1}}k_{2}k_{3}^{2\nu_{1}}}(\Gamma{(\nu_{1})})^{2} \nonumber \\
& \times \int_{-\infty}^{0} dx_{1}\int_{-\infty}^{x_{1}} dx_{2}\int_{-\infty}^{x_{2}} dx_{3}\int_{-\infty}^{x_{3}} dx_{4} (-x_{1})^{-\frac{1}{2}}(-x_{2})^{-\frac{1}{2}} (-x_{3})^{\frac{1}{2}-\nu_{1}}(-x_{4})^{\frac{1}{2}-\nu_{1}} \nonumber\\
& \times \sin(-x_{1})\sin(-x_{2}){\rm Im}\left( H^{(2)}_{\nu_{1}}(-x_{1})H^{(2)}_{\nu_{1}}(-x_{2})(H^{(1)}_{\nu_{1}}(-x_{4}))^{2}\right) \, .
\end{align}
We next look at the term with the permutation $k_{1}\leftrightarrow k_{3} $:
\begin{align}
\delta \theta ^{3} (C_{H^{3}_{1}}) &= \frac{\dot{\theta_{0}}^{3}}{2^{5-2\nu_{1}}} \frac{\lambda_{1}c_{1}^{3}}{H\widetilde{R}^{3}}\frac{\pi }{k_{1}^{5-2\nu_{1}}k_{2}k_{3}^{2\nu_{1}}}(\Gamma{(\nu_{1})})^{2} \nonumber \\
& \times \int_{-\infty}^{0} dx_{1}\int_{-\infty}^{x_{1}} dx_{2}\int_{-\infty}^{x_{2}} dx_{3}\int_{-\infty}^{x_{3}} dx_{4} (-x_{1})^{\frac{1}{2}-\nu_{1}}(-x_{2})^{-\frac{1}{2}} (-x_{3})^{-\frac{1}{2}}(-x_{4})^{\frac{1}{2}-\nu_{1}} \nonumber\\
& \times \sin(-x_{2})\sin(-x_{3}){\rm Im}\left( H^{(2)}_{\nu_{1}}(-x_{2})H^{(2)}_{\nu_{1}}(-x_{3})(H^{(1)}_{\nu_{1}}(-x_{4}))^{2}\right)
\end{align}
Then we look at the term with the permutation $k_{2}\leftrightarrow k_{3} $:
\begin{align}
\delta \theta ^{3} (C_{H^{3}_{1}}) &= \frac{\dot{\theta_{0}}^{3}}{2^{5-2\nu_{1}}} \frac{\lambda_{1}c_{1}^{3}}{H\widetilde{R}^{3}}\frac{\pi }{k_{1}^{5-2\nu_{1}}k_{2}k_{3}^{2\nu_{1}}}(\Gamma{(\nu_{1})})^{2} \nonumber \\
& \times \int_{-\infty}^{0} dx_{1}\int_{-\infty}^{x_{1}} dx_{2}\int_{-\infty}^{x_{2}} dx_{3}\int_{-\infty}^{x_{3}} dx_{4} (-x_{1})^{-\frac{1}{2}}(-x_{2})^{\frac{1}{2}-\nu_{1}} (-x_{3})^{-\frac{1}{2}}(-x_{4})^{\frac{1}{2}-\nu_{1}} \nonumber\\
& \times \sin(-x_{1})\sin(-x_{3}){\rm Im}\left( H^{(2)}_{\nu_{1}}(-x_{1})H^{(2)}_{\nu_{1}}(-x_{3})(H^{(1)}_{\nu_{1}}(-x_{4}))^{2}\right) \, .
\end{align}

$(2_{C})C_{H^{3}_{2}}$: Since this term is very similar to the above case, $C_{H^{3}_{1}}$ , we skip it at the moment and insert it in the final result.

$(3_{C})C_{H^{3}_{3}}$: This part is related to our new interaction and is divided in three terms.

$(3_{C1})C_{H^{3}_{31}}$: Again,  we just mention the final result:
\begin{align}
\delta \theta ^{3} (C_{H^{3}_{31}}) &= \frac{\dot{\theta_{0}}^{3}}{2^{5-2\nu_{2}}} \frac{\lambda_{3}c_{1}c_{2}^{2}}{H\widetilde{R}^{3}}\frac{\pi }{k_{1}^{5-2\nu_{2}}k_{2}k_{3}^{2\nu_{2}}}(\Gamma{(\nu_{2})})^{2} \nonumber \\
& \times \int_{-\infty}^{0} dx_{1}\int_{-\infty}^{x_{1}} dx_{2}\int_{-\infty}^{x_{2}} dx_{3}\int_{-\infty}^{x_{3}} dx_{4} (-x_{1})^{-\frac{1}{2}}(-x_{2})^{-\frac{1}{2}} (-x_{3})^{\frac{1}{2}-\nu_{2}}(-x_{4})^{\frac{1}{2}-\nu_{2}} \nonumber\\
& \times \sin(-x_{1})\sin(-x_{2}){\rm Im}\left( H^{(2)}_{\nu_{1}}(-x_{1})H^{(2)}_{\nu_{2}}(-x_{2})H^{(1)}_{\nu_{1}}(-x_{4})H^{(1)}_{\nu_{2}}(-x_{4})\right) \, .
\end{align}
We next look at the term with the permutation $k_{1}\leftrightarrow k_{3} $:
\begin{align}
\delta \theta ^{3} (C_{H^{3}_{31}}) &= \frac{\dot{\theta_{0}}^{3}}{2^{5-2\nu_{1}}} \frac{\lambda_{3}c_{1}c_{2}^{2}}{H\widetilde{R}^{3}}\frac{\pi }{k_{1}^{5-2\nu_{1}}k_{2}k_{3}^{2\nu_{1}}}(\Gamma{(\nu_{1})})^{2} \nonumber \\
& \times \int_{-\infty}^{0} dx_{1}\int_{-\infty}^{x_{1}} dx_{2}\int_{-\infty}^{x_{2}} dx_{3}\int_{-\infty}^{x_{3}} dx_{4} (-x_{1})^{\frac{1}{2}-\nu_{1}}(-x_{2})^{-\frac{1}{2}} (-x_{3})^{-\frac{1}{2}}(-x_{4})^{\frac{1}{2}-\nu_{1}} \nonumber\\
& \times \sin(-x_{2})\sin(-x_{3}){\rm Im}\left( H^{(2)}_{\nu_{2}}(-x_{2})H^{(2)}_{\nu_{2}}(-x_{3})(H^{(1)}_{\nu_{2}}(-x_{4}))^{2}\right)
\end{align}
Then we look at the term with the permutation $k_{2}\leftrightarrow k_{3} $:
\begin{align}
\delta \theta ^{3} (C_{H^{3}_{31}}) &= \frac{\dot{\theta_{0}}^{3}}{2^{5-2\nu_{2}}} \frac{\lambda_{3}c_{1}c_{2}^{2}}{H\widetilde{R}^{3}}\frac{\pi }{k_{1}^{5-2\nu_{2}}k_{2}k_{3}^{2\nu_{2}}}(\Gamma{(\nu_{2})})^{2} \nonumber \\
& \times \int_{-\infty}^{0} dx_{1}\int_{-\infty}^{x_{1}} dx_{2}\int_{-\infty}^{x_{2}} dx_{3}\int_{-\infty}^{x_{3}} dx_{4} (-x_{1})^{-\frac{1}{2}}(-x_{2})^{\frac{1}{2}-\nu_{2}} (-x_{3})^{-\frac{1}{2}}(-x_{4})^{\frac{1}{2}-\nu_{2}} \nonumber\\
& \times \sin(-x_{1})\sin(-x_{3}){\rm Im}\left( H^{(2)}_{\nu_{1}}(-x_{1})H^{(2)}_{\nu_{2}}(-x_{3})H^{(1)}_{\nu_{1}}(-x_{4})H^{(1)}_{\nu_{2}}(-x_{4})\right) \, .
\end{align}

$(3_{C2})C_{H^{3}_{32}}$: Again we just mention the final result:
\begin{align}
\delta \theta ^{3} (C_{H^{3}_{32}}) &= \frac{\dot{\theta}_{0}^{3}}{2^{5-2\nu_{2}}} \frac{\lambda_{3}c_{1}c_{2}^{2}}{H\widetilde{R}^{3}}\frac{\pi }{k_{1}^{5-2\nu_{2}}k_{2}k_{3}^{2\nu_{2}}}(\Gamma{(\nu_{2})})^{2} \nonumber \\
& \times \int_{-\infty}^{0} dx_{1}\int_{-\infty}^{x_{1}} dx_{2}\int_{-\infty}^{x_{2}} dx_{3}\int_{-\infty}^{x_{3}} dx_{4} (-x_{1})^{-\frac{1}{2}}(-x_{2})^{-\frac{1}{2}} (-x_{3})^{\frac{1}{2}-\nu_{2}}(-x_{4})^{\frac{1}{2}-\nu_{2}} \nonumber\\
& \times \sin(-x_{1})\sin(-x_{2}){\rm Im}\left( H^{(2)}_{\nu_{2}}(-x_{1})H^{(2)}_{\nu_{1}}(-x_{2})H^{(1)}_{\nu_{1}}(-x_{4})H^{(1)}_{\nu_{2}}(-x_{4})\right) \, .
\end{align}
We next look at the term with the permutation $k_{1}\leftrightarrow k_{3} $:
\begin{align}
\delta \theta ^{3} (C_{H^{3}_{32}}) &= \frac{\dot{\theta}_{0}^{3}}{2^{5-2\nu_{2}}} \frac{\lambda_{3}c_{1}c_{2}^{2}}{H\widetilde{R}^{3}}\frac{\pi }{k_{1}^{5-2\nu_{2}}k_{2}k_{3}^{2\nu_{2}}}(\Gamma{(\nu_{2})})^{2} \nonumber \\
& \times \int_{-\infty}^{0} dx_{1}\int_{-\infty}^{x_{1}} dx_{2}\int_{-\infty}^{x_{2}} dx_{3}\int_{-\infty}^{x_{3}} dx_{4} (-x_{1})^{\frac{1}{2}-\nu_{2}}(-x_{2})^{-\frac{1}{2}} (-x_{3})^{-\frac{1}{2}}(-x_{4})^{\frac{1}{2}-\nu_{2}} \nonumber\\
& \times \sin(-x_{2})\sin(-x_{3}){\rm Im}\left( H^{(2)}_{\nu_{1}}(-x_{2})H^{(2)}_{\nu_{2}}(-x_{3})H^{(1)}_{\nu_{1}}(-x_{4})H^{(1)}_{\nu_{2}}(-x_{4})\right)
\end{align}
Then we look at the term with the permutation $k_{2}\leftrightarrow k_{3} $:
\begin{align}
\delta \theta ^{3} (C_{H^{3}_{32}}) &= \frac{\dot{\theta}_{0}^{3}}{2^{5-2\nu_{1}}} \frac{\lambda_{3}c_{1}c_{2}^{2}}{H\widetilde{R}^{3}}\frac{\pi }{k_{1}^{5-2\nu_{1}}k_{2}k_{3}^{2\nu_{1}}}(\Gamma{(\nu_{1})})^{2} \nonumber \\
& \times \int_{-\infty}^{0} dx_{1}\int_{-\infty}^{x_{1}} dx_{2}\int_{-\infty}^{x_{2}} dx_{3}\int_{-\infty}^{x_{3}} dx_{4} (-x_{1})^{-\frac{1}{2}}(-x_{2})^{\frac{1}{2}-\nu_{1}} (-x_{3})^{-\frac{1}{2}}(-x_{4})^{\frac{1}{2}-\nu_{1}} \nonumber\\
& \times \sin(-x_{1})\sin(-x_{3}){\rm Im}\left( H^{(2)}_{\nu_{2}}(-x_{1})H^{(2)}_{\nu_{2}}(-x_{3})(H^{(1)}_{\nu_{2}}(-x_{4}))^{2}\right) \, .
\end{align}

$(3_{C3})C_{H^{3}_{33}}$:  Again we just mention the final result:
\begin{align}
\delta \theta ^{3} (C_{H^{3}_{33}}) &= \frac{\dot{\theta}_{0}^{3}}{2^{5-2\nu_{1}}} \frac{\lambda_{3}c_{1}c_{2}^{2}}{H\widetilde{R}^{3}}\frac{\pi }{k_{1}^{5-2\nu_{1}}k_{2}k_{3}^{2\nu_{1}}}(\Gamma{(\nu_{1})})^{2} \nonumber \\
& \times \int_{-\infty}^{0} dx_{1}\int_{-\infty}^{x_{1}} dx_{2}\int_{-\infty}^{x_{2}} dx_{3}\int_{-\infty}^{x_{3}} dx_{4} (-x_{1})^{-\frac{1}{2}}(-x_{2})^{-\frac{1}{2}} (-x_{3})^{\frac{1}{2}-\nu_{1}}(-x_{4})^{\frac{1}{2}-\nu_{1}} \nonumber\\
& \times \sin(-x_{1})\sin(-x_{2}){\rm Im}\left( H^{(2)}_{\nu_{2}}(-x_{1})H^{(2)}_{\nu_{2}}(-x_{2})(H^{(1)}_{\nu_{2}}(-x_{4}))^{2}\right) \, .
\end{align}
We next look at the term with the permutation $k_{1}\leftrightarrow k_{3} $:
\begin{align}
\delta \theta ^{3} (C_{H^{3}_{33}}) &= \frac{\dot{\theta}_{0}^{3}}{2^{5-2\nu_{2}}} \frac{\lambda_{3}c_{1}c_{2}^{2}}{H\widetilde{R}^{3}}\frac{\pi }{k_{1}^{5-2\nu_{2}}k_{2}k_{3}^{2\nu_{2}}}(\Gamma{(\nu_{2})})^{2} \nonumber \\
& \times \int_{-\infty}^{0} dx_{1}\int_{-\infty}^{x_{1}} dx_{2}\int_{-\infty}^{x_{2}} dx_{3}\int_{-\infty}^{x_{3}} dx_{4} (-x_{1})^{\frac{1}{2}-\nu_{2}}(-x_{2})^{-\frac{1}{2}} (-x_{3})^{-\frac{1}{2}}(-x_{4})^{\frac{1}{2}-\nu_{2}} \nonumber\\
& \times \sin(-x_{2})\sin(-x_{3}){\rm Im}\left( H^{(2)}_{\nu_{2}}(-x_{2})H^{(2)}_{\nu_{1}}(-x_{3})H^{(1)}_{\nu_{1}}(-x_{4})H^{(1)}_{\nu_{2}}(-x_{4})\right) \, .
\end{align}
Then we look at the term with the permutation $k_{2}\leftrightarrow k_{3} $:
\begin{align}
\delta \theta ^{3} (C_{H^{3}_{33}}) &= \frac{\dot{\theta}_{0}^{3}}{2^{5-2\nu_{2}}} \frac{\lambda_{3}c_{1}c_{2}^{2}}{H\widetilde{R}^{3}}\frac{\pi }{k_{1}^{5-2\nu_{2}}k_{2}k_{3}^{2\nu_{2}}}(\Gamma{(\nu_{2})})^{2} \nonumber \\
& \times \int_{-\infty}^{0} dx_{1}\int_{-\infty}^{x_{1}} dx_{2}\int_{-\infty}^{x_{2}} dx_{3}\int_{-\infty}^{x_{3}} dx_{4} (-x_{1})^{-\frac{1}{2}}(-x_{2})^{\frac{1}{2}-\nu_{2}} (-x_{3})^{-\frac{1}{2}}(-x_{4})^{\frac{1}{2}-\nu_{2}} \nonumber\\
& \times \sin(-x_{1})\sin(-x_{3}){\rm Im}\left( H^{(2)}_{\nu_{2}}(-x_{1})H^{(2)}_{\nu_{1}}(-x_{3})H^{(1)}_{\nu_{1}}(-x_{4})H^{(1)}_{\nu_{2}}(-x_{4})\right) \, .
\end{align}

$(4_{C})C_{H^{3}_{4}}$:  Again we skip this part and insert it in the final result.

Finally we note that the other permutation $k_{1}\leftrightarrow k_{2} $ gives each term a factor 2.
\subsection{Squeezed limit amplitudes}
\label{AppenB.4}
In the following, we present the details of $S_{i}$'s in the squeezed limit,
\ba
S_{1}(\nu_{1} , \nu_{2}) &\equiv& \frac{\pi^{2}\Gamma{(\nu_{1})}}{2^{4-\nu_{1}}} \int_{-\infty}^{0} dx_{1}\int_{-\infty}^{x_{1}} dx_{2}\int_{-\infty}^{x_{2}} dx_{3}\nonumber \\
&& \times \bigg{[} \bigg{(}  (-x_{1})^{-\frac{1}{2}} (-x_{2})^{\frac{1}{2}-\nu_{1}}(-x_{3})^{-\frac{1}{2}}
\sin(-x_{1})  \bigg{(}  \lambda_{1}c_{1}^{3} {\rm Im}\left( H^{(1)}_{\nu_{1}}(-x_{1})H^{(2)}_{\nu_{1}}(-x_{2})\right) \nonumber \\
&& \times {\rm Im}\left( H^{(1)}_{\nu_{1}}(-x_{2})H^{(2)}_{\nu_{1}}(-x_{3})e^{ix_{3}}\right) + \lambda_{4}c_{1}^{2}c_{2}{\rm Im}\left( H^{(1)}_{\nu_{2}}(-x_{1})H^{(2)}_{\nu_{2}}(-x_{2})\right) \nonumber\\
&&\times {\rm Im}\left( H^{(1)}_{\nu_{1}}(-x_{2})H^{(2)}_{\nu_{1}}(-x_{3})e^{ix_{3}} \right) + \lambda_{4}c_{1}^{2}c_{2}{\rm Im}\left( H^{(1)}_{\nu_{1}}(-x_{1})H^{(2)}_{\nu_{1}}(-x_{2})\right) \nonumber\\
&&\times {\rm Im}\left( H^{(1)}_{\nu_{2}}(-x_{2})H^{(2)}_{\nu_{2}}(-x_{3})e^{ix_{3}} \right) + \lambda_{3}c_{1}c_{2}^{2}{\rm Im}\left( H^{(1)}_{\nu_{2}}(-x_{1})H^{(2)}_{\nu_{2}}(-x_{2})\right) \nonumber \\
&& \times {\rm Im}\left( H^{(1)}_{\nu_{2}}(-x_{2})H^{(2)}_{\nu_{2}}(-x_{3})e^{ix_{3}} \right)
\bigg{)}+ (-x_{1})^{-\frac{1}{2}}(-x_{2})^{-\frac{1}{2}}(-x_{3})^{\frac{1}{2}-\nu_{1}} \nonumber\\
&& \times \sin(-x_{1})\sin(-x_{2}) \bigg{(}\lambda_{1}c_{1}^{3} {\rm Im}\left( H^{(2)}_{\nu_{1}}(-x_{1})H^{(2)}_{\nu_{1}}(-x_{2}) \left(H^{(1)}_{\nu_{1}}(-x_{3})\right)^{2}\right) \nonumber \\
&&+\lambda_{4}c_{1}^{2}c_{2}{\rm Im}\left( H^{(2)}_{\nu_{2}}(-x_{1})H^{(1)}_{\nu_{2}}(-x_{3})H^{(2)}_{\nu_{1}}(-x_{2})H^{(1)}_{\nu_{1}}(-x_{3})\right)\nonumber\\
&&+\lambda_{4}c_{1}^{2}c_{2}{\rm Im}\left( H^{(2)}_{\nu_{1}}(-x_{1})H^{(1)}_{\nu_{1}}(-x_{3})H^{(2)}_{\nu_{2}}(-x_{2})H^{(1)}_{\nu_{2}}(-x_{3})\right)\nonumber\\
&&+\lambda_{3}c_{1}c_{2}^{2}{\rm Im}\left( H^{(2)}_{\nu_{2}}(-x_{1})H^{(2)}_{\nu_{2}}(-x_{2})(H^{(1)}_{\nu_{2}}(-x_{3}))^{2}\right)
\bigg{)}  \bigg{)} \nonumber\\
&& \times \int_{-\infty}^{0} dy_{4} (-y_{4})^{-\frac{1}{2}}{\rm Re}\left(  H^{(1)}_{\nu_{1}}(-y_{4})e^{-iy_{4}}\right) \bigg{]}
\ea
Similarly, $S_2(\nu_1, \nu_2) $ is easily obtained by replacing $\nu_1 \longleftrightarrow \nu_2$, $\lambda_{1}\longleftrightarrow\lambda_{2} $,  $\lambda_{3}\longleftrightarrow\lambda_{4} $  and $c_{1}\longleftrightarrow c_{2} $   in $S_1(\nu_1, \nu_2)$.\\

\ba
S_{3}(\nu_{1} , \nu_{2}) &\equiv& \frac{\pi (\Gamma{(\nu_{1})})^{2}}{2^{4-\nu_{1}}}\int_{-\infty}^{0} dx_{1}\int_{-\infty}^{x_{1}} dx_{2}\int_{-\infty}^{x_{2}} dx_{3}\int_{-\infty}^{x_{3}} dx_{4} \bigg{[}(-x_{1})^{-\frac{1}{2}} (-x_{2})^{\frac{1}{2}-\nu_{1}} (-x_{3})^{\frac{1}{2}-\nu_{1}}(-x_{4})^{-\frac{1}{2}} \sin(-x_{1}) \times\nonumber \\&& \hspace{-0.4cm}\bigg{(} \lambda_{1}c_{1}^{3} {\rm Im}\left( H^{(1)}_{\nu_{1}}(-x_{1})H^{(2)}_{\nu_{1}}(-x_{2})\right)
{\rm Im}\left( H^{(2)}_{\nu_{1}}(-x_{2})H^{(1)}_{\nu_{1}}(-x_{4})e^{-ix_{4}} \right) + \lambda_{4}c_{1}^{2}c_{2} {\rm Im}\left( H^{(1)}_{\nu_{2}}(-x_{1})H^{(2)}_{\nu_{2}}(-x_{2})\right) \times \nonumber\\
&& \hspace{-0.2cm}
{\rm Im}\left( H^{(2)}_{\nu_{1}}(-x_{2})H^{(1)}_{\nu_{1}}(-x_{4})e^{-ix_{4}} \right)+ \lambda_{3}c_{1}c_{2}^{2}{\rm Im}\left( H^{(1)}_{\nu_{2}}(-x_{1})H^{(2)}_{\nu_{2}}(-x_{2})\right){\rm Im}\left( H^{(2)}_{\nu_{2}}(-x_{2})H^{(1)}_{\nu_{2}}(-x_{4})e^{-ix_{4}} \right)+ \nonumber \\
&&  \hspace{-0.2cm} \lambda_{4}c_{1}^{2}c_{2}{\rm Im}\left( H^{(1)}_{\nu_{1}}(-x_{1})H^{(2)}_{\nu_{1}}(-x_{2})\right) {\rm Im}\left( H^{(2)}_{\nu_{2}}(-x_{2})H^{(1)}_{\nu_{2}}(-x_{4})e^{-ix_{4}} \right) + \lambda_{1}c_{1}^{3} {\rm Im}\left( H^{(2)}_{\nu_{1}}(-x_{1})H^{(1)}_{\nu_{1}}(-x_{3})\right)\times \nonumber\\
&&    \hspace{-0.2cm}  {\rm Im}\left( H^{(1)}_{\nu_{1}}(-x_{3})H^{(2)}_{\nu_{1}}(-x_{4})e^{ix_{4}}\right)+\lambda_{4}c_{1}^{2}c_{2}{\rm Im}\left( H^{(2)}_{\nu_{2}}(-x_{1})H^{(1)}_{\nu_{2}}(-x_{3})\right) {\rm Im}\left( H^{(1)}_{\nu_{1}}(-x_{3})H^{(2)}_{\nu_{1}}(-x_{4})e^{ix_{4}}\right) + \nonumber\\
&& \hspace{-0.2cm} \lambda_{3}c_{1}c_{2}^{2} {\rm Im}\left( H^{(2)}_{\nu_{2}}(-x_{1})H^{(1)}_{\nu_{2}}(-x_{3})\right){\rm Im}\left( H^{(1)}_{\nu_{2}}(-x_{3})H^{(2)}_{\nu_{2}}(-x_{4})e^{ix_{4}}\right)+\lambda_{4}c_{1}^{2}c_{2} {\rm Im}\left( H^{(2)}_{\nu_{1}}(-x_{1})H^{(1)}_{\nu_{1}}(-x_{3})\right) \times   \nonumber\\
&&  \hspace{-0.2cm} {\rm Im}\left( H^{(1)}_{\nu_{2}}(-x_{3})H^{(2)}_{\nu_{2}}(-x_{4})e^{ix_{4}}\right)\bigg{)} + (-x_{1})^{\frac{1}{2}-\nu_{1}}(-x_{2})^{-\frac{1}{2}} (-x_{3})^{\frac{1}{2}-\nu_{1}}(-x_{4})^{-\frac{1}{2}} \sin(-x_{2}) \times  \nonumber\\
&&  \hspace{-0.3cm}\bigg{(} \lambda_{1}c_{1}^{3} {\rm Im}\left( H^{(2)}_{\nu_{1}}(-x_{2})H^{(1)}_{\nu_{1}}(-x_{3})\right)
{\rm Im}\left( H^{(1)}_{\nu_{1}}(-x_{3})H^{(2)}_{\nu_{1}}(-x_{4})e^{ix_{4}}\right) + \lambda_{3}c_{1}c_{2}^{2} {\rm Im}\left( H^{(2)}_{\nu_{2}}(-x_{2})H^{(1)}_{\nu_{2}}(-x_{3})\right)\times \nonumber\\
&& \hspace{-0.2cm} {\rm Im}\left( H^{(1)}_{\nu_{2}}(-x_{3})H^{(2)}_{\nu_{2}}(-x_{4})e^{ix_{4}}\right) + \lambda_{4}c_{1}^{2}c_{2}  {\rm Im}\left( H^{(2)}_{\nu_{2}}(-x_{2})H^{(1)}_{\nu_{2}}(-x_{3})\right){\rm Im}\left( H^{(1)}_{\nu_{1}}(-x_{3})H^{(2)}_{\nu_{1}}(-x_{4})e^{ix_{4}}\right) + \nonumber\\
&& \hspace{-0.2cm} \lambda_{4}c_{1}^{2}c_{2}{\rm Im}\left( H^{(2)}_{\nu_{1}}(-x_{2})H^{(1)}_{\nu_{1}}(-x_{3})\right){\rm Im}\left( H^{(1)}_{\nu_{2}}(-x_{3})H^{(2)}_{\nu_{2}}(-x_{4})e^{ix_{4}}\right)\bigg{)} -  (-x_{1})^{-\frac{1}{2}}(-x_{2})^{-\frac{1}{2}} (-x_{3})^{\frac{1}{2}-\nu_{1}} \times \nonumber\\
&&  \hspace{-0.2cm} (-x_{4})^{\frac{1}{2}-\nu_{1}}\sin(-x_{1})\sin(-x_{2}) \bigg{(} \lambda_{1}c_{1}^{3} {\rm Im}\left( H^{(2)}_{\nu_{1}}(-x_{1})H^{(2)}_{\nu_{1}}(-x_{2})(H^{(1)}_{\nu_{1}}(-x_{4}))^{2}\right) + \lambda_{4}c_{1}^{2}c_{2} \times \nonumber\\
&&  \hspace{-0.2cm} {\rm Im}\left( H^{(2)}_{\nu_{2}}(-x_{1})H^{(2)}_{\nu_{1}}(-x_{2})H^{(1)}_{\nu_{2}}(-x_{4})H^{(1)}_{\nu_{1}}(-x_{4})\right) + \lambda_{4}c_{1}^{2}c_{2} {\rm Im}\bigg{(} H^{(2)}_{\nu_{1}}(-x_{1})H^{(2)}_{\nu_{2}}(-x_{2})H^{(1)}_{\nu_{2}}(-x_{4}) \times \nonumber\\
&&  \hspace{-0.2cm} H^{(1)}_{\nu_{1}}(-x_{4})\bigg{)} +\lambda_{3}c_{1}c_{2}^{2}{\rm Im}\left( H^{(2)}_{\nu_{2}}(-x_{1})H^{(2)}_{\nu_{2}}(-x_{2})(H^{(1)}_{\nu_{2}}(-x_{4}))^{2}\right) \bigg{)} - (-x_{1})^{\frac{1}{2}-\nu_{1}}(-x_{2})^{-\frac{1}{2}} (-x_{3})^{-\frac{1}{2}}\times \nonumber\\
&& \hspace{-0.2cm}(-x_{4})^{\frac{1}{2}-\nu_{1}}\sin(-x_{2}) \sin(-x_{3})\bigg{(}\lambda_{1}c_{1}^{3}{\rm Im}\left( H^{(2)}_{\nu_{1}}(-x_{2})H^{(2)}_{\nu_{1}}(-x_{3})(H^{(1)}_{\nu_{1}}(-x_{4}))^{2}\right)+  \lambda_{3}c_{1}c_{2}^{2} \times\nonumber\\
&& \hspace{-0.2cm}{\rm Im}\left( H^{(2)}_{\nu_{2}}(-x_{2})H^{(2)}_{\nu_{2}}(-x_{3})(H^{(1)}_{\nu_{2}}(-x_{4}))^{2}\right)+ \lambda_{4}c_{1}^{2}c_{2} {\rm Im}\left( H^{(2)}_{\nu_{2}}(-x_{2})H^{(2)}_{\nu_{1}}(-x_{3})H^{(1)}_{\nu_{2}}(-x_{4})H^{(1)}_{\nu_{1}}(-x_{4})\right)+ \nonumber\\
&& \hspace{-0.2cm} \lambda_{4}c_{1}^{2}c_{2}{\rm Im}\left( H^{(2)}_{\nu_{1}}(-x_{2})H^{(2)}_{\nu_{2}}(-x_{3})H^{(1)}_{\nu_{2}}(-x_{4})H^{(1)}_{\nu_{1}}(-x_{4})\right) \bigg{)}- (-x_{1})^{-\frac{1}{2}}(-x_{2})^{\frac{1}{2}-\nu_{1}} (-x_{3})^{-\frac{1}{2}}(-x_{4})^{\frac{1}{2}-\nu_{1}}\times \nonumber\\
&& \sin(-x_{1}) \sin(-x_{3})\bigg{(}\lambda_{1}c_{1}^{3} {\rm Im}\left( H^{(2)}_{\nu_{1}}(-x_{1})H^{(2)}_{\nu_{1}}(-x_{3})(H^{(1)}_{\nu_{1}}(-x_{4}))^{2}\right) + \lambda_{4}c_{1}^{2}c_{2}{\rm Im}\bigg{(} H^{(2)}_{\nu_{2}}(-x_{1})H^{(2)}_{\nu_{1}}(-x_{3}) \times\nonumber\\
&& \hspace{-0.2cm}  H^{(1)}_{\nu_{2}}(-x_{4})H^{(1)}_{\nu_{1}}(-x_{4})\bigg{)}+
\lambda_{3}c_{1}c_{2}^{2}{\rm Im}\left( H^{(2)}_{\nu_{2}}(-x_{1})H^{(2)}_{\nu_{2}}(-x_{3})(H^{(1)}_{\nu_{2}}(-x_{4}))^{2}\right)+ \lambda_{4}c_{1}^{2}c_{2}{\rm Im}\bigg{(} H^{(2)}_{\nu_{1}}(-x_{1})\nonumber\\
&& \hspace{-0.2cm}H^{(2)}_{\nu_{2}}(-x_{3})H^{(1)}_{\nu_{2}}(-x_{4})H^{(1)}_{\nu_{1}}(-x_{4})\bigg{)}\bigg{)}
\bigg{]}
\ea
Similarly, $S_4(\nu_1, \nu_2) $ is easily obtained by replacing $\nu_1 \longleftrightarrow \nu_2$, $\lambda_{1}\longleftrightarrow\lambda_{2} $,  $\lambda_{3}\longleftrightarrow\lambda_{4} $  and $c_{1}\longleftrightarrow c_{2} $   in $S_3(\nu_1, \nu_2)$.

\end{document}